\PassOptionsToPackage{table}{xcolor}
\documentclass[onecolumn]{IEEEtran}

\usepackage{cite}
\usepackage{graphicx}
\usepackage{tikz}
\usetikzlibrary{arrows.meta, positioning}
\usetikzlibrary{positioning,calc}
\usetikzlibrary{positioning,fit}
\usepackage{xcolor}
\usepackage{subcaption}
\usepackage{tabularx}
\usepackage{carlito}
\fontfamily{Carlito-TLF}\selectfont
\usepackage{booktabs}
\usepackage{multirow}
\usepackage[T1]{fontenc}
\usepackage[utf8]{inputenc}
\usepackage{longtable}
\usepackage{array}
\usepackage{stfloats}
\usepackage{tcolorbox}
\usepackage{hhline}
\usepackage{enumitem}
\usepackage{tcolorbox}
\usepackage{hyperref}
\usepackage{titlesec}

\tcbset{
    colback=white,
    colframe=black,
    boxrule=0.5pt,
    arc=2pt,
    left=4pt,
    right=4pt,
    top=4pt,
    bottom=4pt
}
\usepackage{amsmath,amssymb}
\usepackage{hyperref}
\usepackage{enumitem}
\usepackage{pgfplots}
\pgfplotsset{compat=1.18}
\usepackage{comment}
\usepackage{comment}
\usepackage{titlesec}

\renewcommand{\thesection}{\arabic{section}}

\titleformat{\section}
  {\normalfont\large\bfseries}
  {\thesection}
  {1em}
  {}

\makeatletter

\renewenvironment{abstract}
{\par\noindent\textbf{Abstract}\par\smallskip}
{\par}

\makeatother

\begin{document}

\title{Psychological Safety in Software Engineering Teams: A Systematic Mapping Study of Team Processes and Performance}

\author{
\IEEEauthorblockN{
Adnan Manzoor$^{a}$,
Muhammad Azeem Akbar$^{a}$,
Jussi Kasurinen$^{a}$,
Muhammad Hamza$^{a}$,
Aakash Ahmad$^{c}$, 
Abdul Razzaq$^{d}$
}

\IEEEauthorblockA{
$^{a}$Department of Software Engineering, LUT University, Lappeenranta, Finland.\\
$^{c}$Data Science Research Centre, School of Computing, University of Derby, Derby, UK.\\
$^{d}$ Department of Electronics and Computer Engineering, University of Limerick, Limerick, Ireland.
}
}

\maketitle

\begin{abstract}
\noindent
\textit{Context:}
Psychological safety is an important human and behavioral factor in software development teams, where collaboration, knowledge sharing, interpersonal risk-taking, and continuous learning are central to effective teamwork. However, evidence on psychological safety in software engineering remains fragmented across team processes, performance outcomes, contextual conditions, barriers, and improvement practices.
\textit{Objective:}
This study systematically maps and synthesizes the literature on psychological safety in software engineering, examining its conceptualization, associations with team performance and team processes, contextual influences, barriers, and strategies for strengthening it.
\textit{Method:}
We conducted a systematic mapping study following established guidelines for systematic studies in software engineering. A total of 112 primary studies published between 2006 and June 2026 were analyzed using descriptive statistics and thematic analysis across six research questions.
\textit{Results:}
Psychological safety emerges as a multidimensional and context-dependent team-level construct characterized by interpersonal risk-taking, open communication, collaboration, learning, trust, and supportive social interaction. It is predominantly associated with productivity, software quality, innovation, and team satisfaction, as well as collaboration, communication, and decision-making. These relationships are shaped by leadership, organizational culture, work environment, trust, and sociocultural and individual conditions. Recurring barriers include communication problems, organizational and structural constraints, leadership issues, diversity and inclusion challenges, and workload and resource pressures. Although diverse practices are reported for strengthening psychological safety, evidence regarding their effectiveness and long-term sustainability remains limited.
\textit{Conclusion:}
Psychological safety is a multidimensional and context-dependent team-level human and behavioral construct associated with team interaction and performance. This study consolidates fragmented evidence and highlights the need for longitudinal, context-sensitive research evaluating psychological safety interventions across contemporary software development environments.
\end{abstract}

\textbf{keywords:}Software Development Teams, Psychological Safety, Human Factors, Team Performance, systematic mapping studies.

%-------------Introduction---------------------------

\section{Introduction}

Software engineering has evolved into a fundamentally collaborative and socio-technical discipline, where successful project outcomes depend not only on technical expertise but also on effective human interaction within development teams~\cite{p129, p130}. Contemporary software projects are characterized by increasing system complexity, rapid release cycles, distributed collaboration, and continuous integration practices~\cite{p131, p132}, all of which intensify the reliance on coordination, communication, and collective decision-making.As a result, human factors, including social, cognitive, and organizational aspects, have become increasingly important for understanding team performance in software engineering environments.~\cite{p130}.\vspace{5pt}

Within this broader domain of human factors, psychological safety has emerged as a key socio-behavioral construct that shapes how individuals engage in team interactions. Psychological safety refers to a shared belief among team members that the work environment supports interpersonal risk-taking, enabling individuals to express ideas, ask questions, admit mistakes, and challenge existing practices without fear of negative consequences~\cite{p133}. In software engineering contexts, where knowledge sharing, rapid feedback, and collaborative problem-solving are essential, such conditions are particularly important~\cite{p134, p130}. Prior research suggests that psychologically safe teams exhibit improved collaboration, more effective communication, and stronger learning behaviors, which have been associated with outcomes such as productivity, software quality, innovation, and team satisfaction.~\cite{p133, p134}.\vspace{5pt}

Despite this growing body of work, the current state of knowledge on psychological safety in software engineering remains fragmented and lacks conceptual coherence. Existing studies are dispersed across multiple disciplines, including organizational behavior, management science, and software engineering, often addressing isolated aspects such as communication, leadership, or collaboration rather than providing an integrated perspective~\cite{p134, p130}. Furthermore, prior research demonstrates substantial variation in how psychological safety is defined, operationalized, and measured, leading to inconsistencies in findings and limiting the ability to build cumulative knowledge~\cite{p134}. Importantly, many studies examine psychological safety in general organizational settings, with limited attention to the unique characteristics of software engineering environments, such as agile practices, cross-functional teams, distributed development, and high cognitive workload~\cite{p130, p131}. Although previous reviews have synthesized psychological safety in broader organizational contexts, they do not provide a software-engineering-specific mapping of how psychological safety relates to team performance, team processes, contextual factors, barriers, and improvement strategies. The distinctive characteristics of software development, including agile practices, cross-functional collaboration, distributed work, and rapid feedback cycles, therefore warrant a focused synthesis of the available evidence.\vspace{5pt}

This fragmentation introduces both theoretical and practical challenges. From a theoretical perspective, the lack of a unified conceptualization hinders the development of a coherent understanding of how psychological safety interacts with broader human factors to influence team performance in software engineering contexts~\cite{p134, p130}. From a practical perspective, the absence of synthesized and structured evidence limits practitioners’ ability to design and implement effective interventions for fostering psychologically safe and high-performing software engineering teams~\cite{p133, p134}. Consequently, there is a need for a systematic and comprehensive synthesis that consolidates existing knowledge and provides a structured foundation for both research and practice.\vspace{5pt}

To address this need, this study conducts a systematic mapping study (SMS) to synthesize existing research on psychological safety in software development teams. The objective (see Table~\ref{tab:rq}) of this study is to consolidate dispersed findings and develop a structured understanding of how psychological safety is conceptualized in software engineering, how it relates to team performance and team processes, which contextual factors shape these relationships, what barriers hinder its establishment, and which strategies have been reported to strengthen it.\vspace{5pt}

This study makes the following contributions to software engineering research:

\begin{itemize}
    \item It provides a comprehensive synthesis of how psychological safety is defined and conceptualized in software engineering research.
    
    \item It provides an integrated perspective on the reported relationships among psychological safety, team performance outcomes, team processes, and contextual factors.
    
    \item It identifies reported barriers to psychological safety and synthesizes strategies for fostering psychologically safe software development teams, while highlighting gaps for future research.
\end{itemize}

\begin{figure*}[ht]
    \centering
    \includegraphics[width=0.7\linewidth]{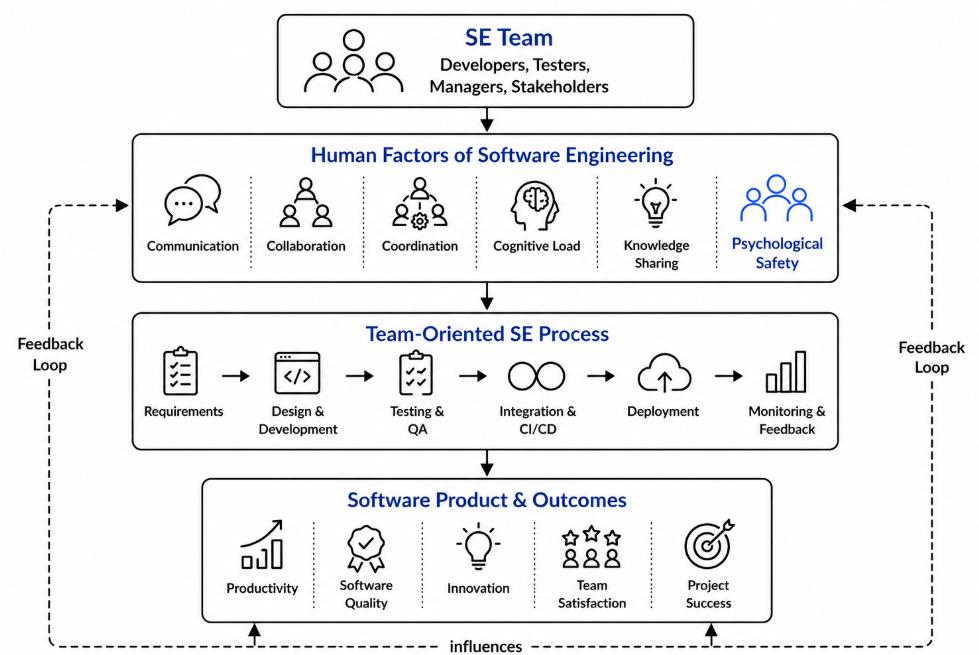}
    \caption{Conceptual Framework of Psychological Safety in Team-Oriented Software Engineering}
    \label{fig:1}
\end{figure*}

Together, these contributions provide a structured foundation for future research and practical guidance for software engineering teams seeking to foster psychologically safe work environments. Figure ~\ref{fig:1} provides a conceptual overview of the relationships examined in this study, linking psychological safety with contextual human factors, team processes, and team performance outcomes.

%----------------Fig. Schematic representation of the research m ethodology---
%-------------------------------
\section{Background}
\subsection{\textbf{Software Engineering Teams and Their Performance}}
Software engineering is inherently team-oriented, involving professionals such as developers, testers, architects, and project managers who coordinate interdependent development activities. Modern software teams frequently operate in agile, cross-functional, and distributed environments, making effective teamwork an important component of successful software development
 ~\cite{p136, p137}.\vspace{5pt}

Team performance in software engineering is commonly evaluated through multiple dimensions, including productivity, software quality, innovation, and team satisfaction ~\cite{p138, p139}. These dimensions are particularly relevant to the present study because they capture both technical outcomes, such as productivity and software quality, and human-centered outcomes, such as innovation and team satisfaction.\vspace{5pt}

Software development is therefore considered a socio-technical activity, in which both technical and social factors jointly influence team outcomes ~\cite{p140}. Understanding how teams function and perform requires attention to both process-level coordination and interpersonal dynamics within SE environments. Accordingly, understanding software team performance requires consideration not only of technical processes and capabilities but also of the human factors that shape interaction, coordination, and collective work.\vspace{5pt}

\subsection{\textbf{Human Factors in Software Engineering Teams}}
Human factors refer to the behavioral, cognitive, and social aspects that influence how individuals interact within teams and perform their tasks. In software engineering, these factors include communication, leadership, trust, coordination, and cognitive load, all of which play a significant role in shaping team dynamics and performance ~\cite{p140,p142}.\vspace{5pt}

In this study, human factors are treated as the broader set of interpersonal, behavioral, cognitive, and organizational conditions surrounding teamwork, while psychological safety represents a specific team-level construct concerned with interpersonal risk-taking. Human factors such as leadership, trust, communication, and team climate may therefore act as antecedents, contextual influences, or related mechanisms through which psychological safety develops and affects team functioning.
Effective communication enables team members to exchange information, clarify requirements, and resolve issues efficiently. Leadership influences team motivation, direction, and the creation of supportive environments, while coordination ensures alignment of tasks and dependencies across team members. Contextual conditions such as workload, work pressure, and resource constraints can further shape team interaction, decision-making, and overall effectiveness.\vspace{5pt}

Prior software engineering research has demonstrated that human factors influence both development processes and team outcomes, particularly in collaborative and agile environments ~\cite{p143,p144}. Communication, trust, leadership, and supportive team conditions can shape how effectively team members coordinate, exchange knowledge, respond to problems, and adapt their working practices. Among these human factors, psychological safety has received increasing attention because it directly concerns whether team members feel able to participate openly in these interactions.

\subsection{\textbf{Psychological Safety in Software Engineering Teams}}
Psychological safety refers to a shared belief that the team is safe for interpersonal risk-taking, allowing members to express ideas, ask questions, acknowledge mistakes, and raise concerns without fear of negative interpersonal consequences ~\cite{p133}. It is widely recognized as a key factor influencing team learning, collaboration, and performance in organizational and team-based settings ~\cite{p145,p146}.\vspace{5pt}

Within software engineering teams, psychological safety plays a critical role in enabling open communication, knowledge sharing, and collaborative problem-solving. In software development settings, psychological safety can be reflected in behaviors such as speaking up, discussing errors openly, seeking feedback, and engaging in learning-oriented interactions.\vspace{5pt}

Psychologically safe environments encourage individuals to engage in constructive discussions, share feedback, and learn from mistakes, which are essential for managing the complexity and uncertainty of software development tasks ~\cite{p147}. Moreover, psychological safety is closely linked to broader human factors such as leadership, trust, and team climate, and it serves as a mechanism through which these factors influence team processes and performance outcomes ~\cite{p148}.\vspace{5pt}

%000000000000000000 sir azeem 000000000

Despite growing interest in psychological safety, the available evidence in software engineering remains dispersed across different research topics and team contexts. Studies have examined psychological safety in relation to communication, collaboration, leadership, trust, learning, and team outcomes, but these relationships have not been consistently organized within a software-engineering-specific perspective. Furthermore, contextual conditions such as team structure, organizational culture, leadership, and distributed work may shape how psychological safety develops and operates. These limitations motivate a systematic mapping of the available evidence to clarify the current state of knowledge and identify areas requiring further investigation.
%00000000000000000000000000000000000000

%---------------------end bacground----------------
 
%-------------------Research Method----------------
\section{Research Method}
\begin{figure*}[ht]
    \centering
    \includegraphics[width=0.8\linewidth]{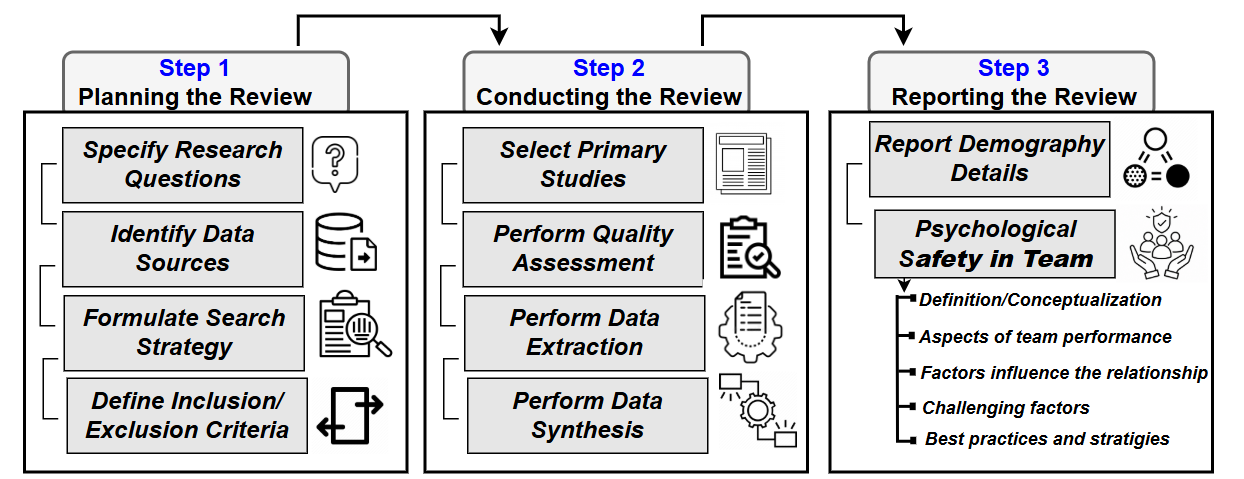}
    \caption{Schematic representation of the research methodology implemented in this study}
    \label{fig:2}
\end{figure*}

This study was conducted as a Systematic Mapping Study (SMS) following established guidelines for systematic studies in software engineering~\cite{p108,p109} and adhering to the recommendations of the ACM SIGSOFT Empirical Standards for Systematic Reviews~\cite{p135}. The study was structured into three phases: (i) Planning the Review, (ii) Conducting the Review, and (iii) Reporting the Review. The overall process is illustrated in Fig.~\ref{fig:2}.\vspace{5pt}
 
To enhance the reliability and consistency of the review process, inter-rater agreement (IRA) was used during study selection, while structured verification procedures were applied during data extraction \cite{p110}. In addition, predefined quality assessment criteria were used to evaluate the selected studies.

%----------------Fig. Schematic representation of the research m ethodology---
%-------------------------------
\subsection{\textbf{Planning the Review}}
The \textbf{Phase 1} begins with the formulation of research questions that define the scope and objectives of the SMS and guide the subsequent search, selection, extraction, and synthesis processes. These questions guide the overall direction, scope, and structure of the study.\vspace{5pt}

\subsubsection{\textbf{Specify Research Questions}}

We formulated six Research Questions (RQs) using a structured approach inspired by the Goal Question Metric (GQM) ~\cite{p119} to guide this Systematic Mapping Study (SMS). This approach ensures alignment between the study objectives, research questions, and data extraction process.\vspace{5pt}

The six research questions are organized around key mapping dimensions. RQ1 examines how psychological safety is defined and conceptualized in software engineering teams. RQ2 investigates its relationship with team performance outcomes, while RQ3 examines its relationship with team processes. RQ4 identifies contextual factors that may shape the relationship between psychological safety and team performance. From a practical perspective, RQ5 investigates reported challenges and barriers, whereas RQ6 examines strategies, practices, and interventions for strengthening psychological safety in software teams (see table ~\ref{tab:rq}).\vspace{5pt}
%-------------GQM------------

%-----------------table RQs ---------------
\renewcommand{\arraystretch}{1.3}

\begin{table*}[t]
\centering
\caption{Research Questions and Rationales}
\label{tab:rq}

\footnotesize
\setlength{\tabcolsep}{5pt}

\begin{tabular}{c p{5.5cm} p{8.5cm}}
\toprule

\rowcolor{gray!40}
\textbf{RQ} & \textbf{Research Question} & \textbf{Rationale} \\

\midrule

RQ1 & How is psychological safety defined and conceptualized in the context of software engineering teams? &
Psychological safety and human factors are interpreted differently across studies and disciplines. Some researchers define psychological safety in terms of interpersonal risk-taking, while others emphasize communication, trust, or learning behavior. Without a clear understanding of how these concepts are defined in software engineering contexts, it is difficult to compare findings or build cumulative knowledge. Therefore, RQ1 aims to clarify and synthesize how psychological safety and human factors are conceptualized in relation to team performance. \\

RQ2 &  What aspects of team performance are associated with psychological safety in software engineering teams?
\begin{itemize}
\item Productivity
\item Software Quality
\item Innovation
\item Satisfaction
\end{itemize}
&
Team performance in software development can be measured through various dimensions such as productivity, software quality, innovation, and team satisfaction. However, different studies emphasize different performance aspects. To understand the practical impact of psychological safety, it is necessary to identify which performance dimensions are most frequently associated with it. RQ2 addresses this need by categorizing performance-related outcomes reported in the literature. \\

RQ3 & How is psychological safety associated with team processes in software engineering teams?
\begin{itemize}
\item Collaboration
\item Communication
\item Decision Making
\end{itemize}
&
Team performance is often influenced indirectly through team processes such as collaboration, communication, decision-making, and knowledge sharing. Psychological safety may enhance these processes, which in turn affect performance outcomes. Understanding this process-level mechanism is important for explaining how psychological safety operates within software teams. Therefore, RQ3 explores the influence of psychological safety on core team processes. \\

RQ4 & What factors influence the relationship between psychological safety and team performance? 
\begin{itemize}
\item Leadership
\item Culture
\item Work Environment
\item Trust
\end{itemize}
&
The relationship between psychological safety and team performance may not be uniform across all contexts. Factors such as leadership style, organizational culture, work environment, and trust may strengthen or weaken this relationship. Identifying these moderating variables is essential for understanding contextual dependencies and for designing effective interventions. RQ4 aims to identify these influencing factors. \\

RQ5 & What are the challenges or barriers to establishing psychological safety in software development teams? &
Despite its importance, establishing psychological safety in software development teams can be difficult. Organizational hierarchies, fear of blame, time pressure, and cultural resistance may prevent open communication. Understanding these barriers is crucial for both researchers and practitioners. RQ5 seeks to identify commonly reported challenges in creating psychologically safe software development environments. \\

RQ6 & What strategies, practices, or interventions are reported in literature for enhancing psychological safety in software teams? &
Beyond identifying problems, it is important to examine how psychological safety can be improved. Many studies propose leadership practices, communication strategies, training programs, or agile practices to enhance team safety. RQ6 investigates the strategies, practices, and interventions reported in the literature for strengthening psychological safety in software teams. \\

\bottomrule
\end{tabular}
\end{table*}
%-------------end table RQs------------

\subsubsection{\textbf{Identify data sources}}
Electronic Data Sources (EDS) were utilized to conduct a systematic and automated search using predefined search strings. In systematic studies, EDS enable the efficient identification of relevant literature while minimizing selection bias and ensuring reproducibility of the search process ~\cite{p125}. Prior empirical research has emphasized the importance of selecting appropriate digital libraries to achieve comprehensive coverage of the domain under investigation ~\cite{p126,p127}.\vspace{5pt}

Based on these recommendations, we selected five widely recognized digital libraries: ACM Digital Library, IEEE Xplore, ScienceDirect, SpringerLink, and Wiley Online Library. These databases are considered authoritative sources for software engineering and interdisciplinary research, making them particularly suitable for identifying studies related to psychological safety and human factors in software development teams.\vspace{5pt}

Although the selected EDS do not guarantee exhaustive coverage of all existing literature, prior studies have demonstrated that these sources are sufficient and appropriate for capturing a representative set of relevant publications ~\cite{p125,p126}. Therefore, their use supports the reliability and validity of this Systematic Mapping Study.\vspace{5pt}

\subsubsection{\textbf{Formulate of Search Query}}
\label{sec:search_query}
At the initial stage, the PICO framework~\cite{p120} was explored to guide the construction of the search string. However, the multidisciplinary nature and broad scope of the topic required a complex combination of keywords and Boolean operators, which exceeded the maximum query length permitted by several digital libraries (see Table~\ref{string}). Consequently, the search string was developed using an iterative approach based on domain expertise and pilot searches~\cite{p120}. This process allowed the refinement of keywords and ensured adequate coverage of relevant literature within the operational limitations of the selected databases i.e. (IEEE, ACM, Wiley Online Library, Science Direct, Springer Link). A structured Boolean search query was developed to capture variations of key concepts. The search string combined three main components:\vspace{5pt}
\begin{itemize}
    \item 	Psychological and human-factor constructs: \textit{“psychological safety” OR “team psychological safety” OR “human factors” OR “behavioral factors” OR “social factors”}
\end{itemize}
\begin{itemize}
    \item 	Team performance and process dimensions: \textit{(“team performance” OR “team effectiveness” OR productivity OR “software quality” OR innovation OR collaboration OR communication OR “decision making” OR “knowledge sharing”}
\end{itemize}
\begin{itemize}
    \item 	Software engineering context: \textit{“software engineering” OR “software development” OR “software teams” OR “development teams” OR “agile teams” OR “devops” OR “software projects”}
\end{itemize}\vspace{5pt}
The use of Boolean operators (AND/OR) ensured that retrieved studies were directly related to both psychological and performance-related aspects within software engineering contexts. The final database searches were conducted in 06/2026, covering publications from 2006 through 2026.\vspace{5pt}

%-------------Table search string----------
\begin{table*}[t]
\centering
\caption{Search Strategy and Data Sources}
\label{string}
\footnotesize
\setlength{\tabcolsep}{4pt}

% --- Search String ---
\begin{tabular}{p{10.5cm}}
\toprule
\rowcolor{gray!40}
\textbf{Search String} \\
\midrule
("psychological safety" OR "team psychological safety" OR "human factors" OR "behavioral factors" OR "social factors")\\
{\centering \textbf{AND}\par}
("team performance" OR "team effectiveness" OR productivity OR "software quality" OR innovation OR collaboration OR communication OR "decision making" OR "knowledge sharing")\\
{\centering \textbf{AND}\par}
("software engineering" OR "software development" OR "software teams" OR "development teams" OR "agile teams" OR "devops" OR "software projects")
\\
\bottomrule
\end{tabular}

% --- Databases ---
\begin{tabular}{l l}
\toprule
\rowcolor{gray!40}
\textbf{Database} & \textbf{Search Scope} \\
\midrule
IEEE & Keywords, title, and abstract \\
ScienceDirect & Keywords, title, and abstract \\
ACM & Keywords, title, and abstract \\
Wiley Online Library & Keywords, title, and abstract \\
Springer Link & Keywords, title, and abstract \\
\bottomrule
\end{tabular}

\end{table*}
%-------end table search string---------

\subsubsection{\textbf{Define Inclusion and Exclusion Criteria}}
To ensure relevance and methodological rigor, explicit inclusion and exclusion criteria were defined prior to the study selection process, following established guidelines for systematic studies ~\cite{p120} (see table ~\ref{inclusion and exclusion}). These criteria were designed to filter the search results by removing duplicate, irrelevant, non-peer-reviewed, and out-of-scope studies. Methodological quality was assessed separately during the subsequent quality-assessment stage.\vspace{5pt}

The inclusion criteria ensure that only studies relevant to psychological safety and human factors within software engineering contexts are considered. In contrast, the exclusion criteria eliminate studies that do not meet the required scope, such as non-peer-reviewed work, non-English publications, and studies outside the software engineering domain.\vspace{5pt}

The criteria were applied consistently across all stages of the screening process, including title, abstract, and full-text review, to ensure transparency, consistency, and reproducibility.
%-----------------Table Inclusion/Exclusion---------------
\begin{table}[h]
\centering
\footnotesize
\caption{Inclusion and Exclusion Criteria}
\label{inclusion and exclusion}
\begin{tabular}{ll|ll}
\toprule
\rowcolor{gray!40}
\textbf{Code} & \textbf{Inclusion Criteria} & \textbf{Code} & \textbf{Exclusion Criteria} \\
\midrule

Inc1 & Studies published from 2006 to 06/2026 
& Exc1 & Duplicate studies \\

Inc2 & Studies indexed in major digital libraries 
& Exc2 & Non-English publications \\

Inc3 & Studies focusing on psychological safety and/or human factors 
& Exc3 & Studies outside the software engineering domain \\

Inc4 & Studies conducted within a software engineering context 
& Exc4 & Studies not focused on team-level aspects \\

Inc5 & Studies reporting empirical evidence or relevant findings 
& Exc5 & Non-peer-reviewed studies \\

\bottomrule
\end{tabular}
\end{table}

%----------------- end Table Inclusion/Exclusion---------------

These criteria were applied consistently at all stages of the study selection process to ensure transparency, consistency, and reproducibility.

\subsection{\textbf{Conducting the Review}}
The \textbf{Phase 2} of the study involves conducting the review, which is guided by the protocol defined during the planning phase (see Fig ~\ref{fig:2}). This phase systematically applies the predefined procedures to identify, select, and analyze relevant studies. The key steps involved in this phase are described as follows:\vspace{5pt}

\subsubsection{\textbf{Select Primary studies}}
The search for primary studies commenced by querying the selected digital repositories using the search string described in Section~\ref{sec:search_query}.The selection of the primary study was conducted in two parts. These two parts helped to ensure comprehensive coverage of the relevant literature.\vspace{5pt}

\textbf{Part 1:}
The study selection process began with a systematic search of multiple electronic databases using the predefined search string ~\cite{p120},~\cite{p121} (see Table~\ref{tab:2}). The search was conducted across five major digital libraries, as illustrated in Table~\ref{tab:2}, covering the period from 2006 to 06/2026. The overall study selection procedure is described in the following steps.\vspace{5pt}

%-------table 2---------------------------------------
\renewcommand{\arraystretch}{1.2}

\begin{table*}[h]
\centering
\caption{Study Selection Process Across Databases}
\label{tab:2}
\footnotesize
\setlength{\tabcolsep}{4pt}

\begin{tabular}{c l c c c c c c}
\toprule

% -------- Step Row --------

\rowcolor{gray!40}
 &  & \textbf{Step 1} & \textbf{Step 2} & \textbf{Step 3} 
& \textbf{Step 4} & \textbf{Step 5} & \textbf{Step 6} \\

% -------- Header Row --------
\rowcolor{gray!20}
\textbf{\#} & \textbf{Database} 
& \textbf{Initial Search} 
& \textbf{Irrelevant Excluded} 
& \textbf{Title-based Selection} 
& \textbf{Abstract-based Selection} 
& \textbf{Final Selection} 
& \textbf{After Snowballing} \\

\midrule
1 & Science Direct & 250 & 100 & 150 & 53 & 18 & 21 \\
2 & ACM & 220 & 111 & 109 & 43 & 14 & 14 \\
3 & IEEE & 658 & 408 & 250 & 112 & 61 & 66 \\
4 & Wiley Online Library & 32 & 14 & 16 & 8 & 6 & 6 \\
5 & Springer Link & 24 & 14 & 10 & 8 & 5 & 5 \\

\midrule

\rowcolor{gray!20}
 & \textbf{Total} & \textbf{1184} & \textbf{647} & \textbf{535} 
& \textbf{224} & \textbf{104} & \textbf{112} \\

\bottomrule
\end{tabular}
\end{table*}
%-------end table 2---------------------------------------

\begin{itemize}
    \item \textbf{Step 1: Data Collection and Extraction:} 
    Customized search strings were executed across the selected digital databases (see Table~\ref{tab:2}) to retrieve study metadata, including titles, author names, publication years, venues, publication types, and abstracts. This initial search yielded a total of 1184 studies from five databases.
\end{itemize}\vspace{1pt}

\begin{itemize}
    \item \textbf{Step 2: Relevance Screening: } In this step,the collected studies were reviewed and categorized into relevant, partially relevant, and irrelevant based on their titles, keywords, and research focus. Studies that were not related to the research topic were removed at this stage. After this filtering process, 647 irrelevant studies were excluded, and the remaining studies were considered for further screening. To ensure the reliability and consistency of the relevance screening process, Inter-rater agreement was measured using Cohen’s Kappa statistic~\cite{p123}. A random sample of 10 percent of the studies \textit{(n = 68)} was independently reviewed by two reviewers (R1 and R2), who classified each study as either included or excluded based on the inclusion and exclusion criteria. The agreement between the reviewers is presented in Table~\ref{tab:kappa}. The calculated Cohen’s Kappa value was \textit{kappa(k) = 0.77 }which indicates substantial agreement between the reviewers. This result demonstrates that the inclusion and exclusion criteria were applied consistently and that the relevance screening process was reliable. Any disagreements between the reviewers were discussed and resolved through mutual agreement.
\end{itemize}\vspace{1pt}

\begin{itemize}
    \item \textbf{Step 3: Title-Based Screening: }
In this stage, the titles of the remaining studies were reviewed to determine their relevance to the research objectives. Studies that did not clearly relate to the research topic based on their titles were excluded. After title-based screening, 535 studies were retained for abstract screening.  To ensure the reliability of the title screening process, inter-rater agreement was measured using Cohen’s Kappa statistics ~\cite{p123}. A random sample of 10 percent of the studies  \textit{(n = 54)} was independently reviewed by two reviewers (R1 and R2), who applied the inclusion and exclusion criteria based on the titles of the studies. The agreement between the reviewers is presented in Table~\ref{tab:kappa}. The calculated Cohen’s Kappa value was \textit{kappa(k) = 0.73 }, which indicates substantial agreement between the reviewers. Any disagreements were discussed and resolved through mutual agreement to reach a final decision.
    \end{itemize}\vspace{1pt}
  
\begin{itemize}
    \item \textbf{Step 4: Abstract-Based Screening: } In this step, the abstracts of the selected studies were carefully reviewed to further assess their relevance. Studies that did not meet the inclusion criteria based on their abstracts were excluded. After abstract screening, 224 studies remained for full-text review. To evaluate the consistency of the abstract screening stage, inter-rater agreement was assessed using Cohen’s Kappa statistic ~\cite{p123}. A randomly selected subset of 10 percent of the studies \textit{(n = 22)} was independently evaluated by two reviewers (R1 and R2), who applied the predefined inclusion and exclusion criteria based on the abstracts. The level of agreement between the reviewers is shown in Table~\ref{tab:kappa} Computed Cohen’s Kappa value was \textit{kappa(k) = 0.68 }, indicating a substantial level of agreement between the reviewers. Any discrepancies in the reviewers’ decisions were resolved through discussion until a consensus was reached.
    \end{itemize}\vspace{1pt}
 
\begin{itemize}
    \item \textbf{Step 5: Full-Text Screening (Final Selection): } In the final stage, the full texts of the remaining studies were reviewed in detail to ensure that they met the inclusion criteria and were directly relevant to the research questions. After full-text screening, 104 studies were selected as the final set of primary studies for data extraction and analysis. To measure the level of agreement between the reviewers during the full-text screening stage, Cohen’s Kappa statistics ~\cite{p123} were used. A random sample of 10 percent of the studies \textit{(n = 10)} was independently evaluated by two reviewers (R1 and R2). The inter-rater agreement results are presented in Table~\ref{tab:kappa}. The calculated Cohen’s Kappa value was \textit{kappa(k) = 0.74 }, indicating a substantial level of agreement between the reviewers. Any discrepancies between the reviewers’ decisions were resolved through discussion to reach a final consensus.   
    \end{itemize}\vspace{1pt}

% --------table 3--------

\renewcommand{\arraystretch}{1.2}

\begin{table*}[t]
\centering
\caption{Inter-rater Agreement Results Across Study Selection Stages}
\label{tab:kappa}

% -------- (a) --------
\begin{subtable}[t]{0.45\textwidth}
\centering
\caption{Relevant/Irrelevant Screening}
\begin{tabular}{lccc}
\toprule
\rowcolor{gray!40}
 & \textbf{R2 Include} & \textbf{R2 Exclude} & \textbf{Total} \\
\midrule
\textbf{R1 Include} & 46 & 4 & 50 \\
\textbf{R1 Exclude} & 3 & 15 & 18 \\
\midrule
\rowcolor{gray!20}
\textbf{Total} & 49 & 19 & 68 \\
\bottomrule
\end{tabular}

\vspace{1mm}
{\footnotesize $\kappa = 0.77$}
\end{subtable}
\hfill
% -------- (b) --------
\begin{subtable}[t]{0.45\textwidth}
\centering
\caption{Title-based Screening}
\begin{tabular}{lccc}
\toprule
\rowcolor{gray!40}
 & \textbf{R2 Include} & \textbf{R2 Exclude} & \textbf{Total} \\
\midrule
\textbf{R1 Include} & 34 & 3 & 37 \\
\textbf{R1 Exclude} & 3 & 14 & 17 \\
\midrule
\rowcolor{gray!20}
\textbf{Total} & 37 & 17 & 54 \\
\bottomrule
\end{tabular}

\vspace{1mm}
{\footnotesize $\kappa = 0.73$}
\end{subtable}

\vspace{0.4cm}

% -------- (c) --------
\begin{subtable}[t]{0.45\textwidth}
\centering
\caption{Abstract Screening}
\begin{tabular}{lccc}
\toprule
\rowcolor{gray!40}
 & \textbf{R2 Include} & \textbf{R2 Exclude} & \textbf{Total} \\
\midrule
\textbf{R1 Include} & 14 & 1 & 15 \\
\textbf{R1 Exclude} & 2 & 5 & 7 \\
\midrule
\rowcolor{gray!20}
\textbf{Total} & 16 & 6 & 22 \\
\bottomrule
\end{tabular}

\vspace{1mm}
{\footnotesize $\kappa = 0.68$}
\end{subtable}
\hfill
% -------- (d) --------
\begin{subtable}[t]{0.45\textwidth}
\centering
\caption{Full-text Screening}
\begin{tabular}{lccc}
\toprule
\rowcolor{gray!40}
 & \textbf{R2 Include} & \textbf{R2 Exclude} & \textbf{Total} \\
\midrule
\textbf{R1 Include} & 7 & 0 & 7 \\
\textbf{R1 Exclude} & 1 & 2 & 3 \\
\midrule
\rowcolor{gray!20}
\textbf{Total} & 8 & 2 & 10 \\
\bottomrule
\end{tabular}

\vspace{1mm}
{\footnotesize $\kappa = 0.74$}
\end{subtable}

\end{table*}
%-------end table 3-------------------------------------

\textbf{Part 2:}
In the second phase, backward and forward snowballing were performed on the 104 primary studies identified through database searching and full-text screening to locate additional relevant publications ~\cite{p122}, ~\cite{p118}. The same inclusion and exclusion criteria were applied to candidate studies identified through reference lists and citation tracking. Snowballing identified eight additional eligible studies, resulting in a final dataset of 112 primary studies. The process continued until no additional relevant studies were identified (see Table~\ref{tab:2}).\vspace{5pt}

%--------------- IRA---------------------

\subsubsection{\textbf{Perform a Quality Assessment}}
The quality assessment was conducted to evaluate the methodological quality, relevance, and evidential adequacy of the included studies, following the guidelines proposed by ~\cite{p120}. This process aimed to reduce potential bias and ensure that only high-quality and relevant studies were included in the analysis.\vspace{5pt}

Each study was assessed using five quality assessment (QA) criteria (QA1–QA5), as summarized in Table~\ref{tab:qality assement}. A three-point scoring scale was applied, where a score of 1 indicates that the criterion is fully satisfied, 0.5 indicates partial satisfaction, and 0 indicates that the criterion is not addressed, as adopted in prior studies ~\cite{p128}. The total quality score for each study was calculated as the sum of scores across all QA criteria.\vspace{5pt}

The assessment was conducted by the primary researcher and independently verified by co-authors to ensure consistency and reduce subjectivity. A minimum threshold score of 2.5 was defined to filter out low-quality studies, and studies scoring below this threshold were excluded from further analysis. All 112 studies met the minimum quality threshold and were therefore retained for data extraction and synthesis.\vspace{5pt}

In total, 112 studies were evaluated using this process. The results were systematically recorded to ensure transparency and traceability. Due to space limitations, only a subset of the quality assessment results is presented in Table~\ref{tab:qality assement}. The complete results for all selected studies (P1–P112) are provided as supplementary material.\vspace{5pt}

This quality assessment process enhances the reliability, validity, and consistency of the study findings and supports rigorous data extraction and synthesis.\vspace{5pt}

\begin{table*}[htbp]
\caption{Quality Assessment Criteria and Sample Results}
\label{tab:qality assement}
\centering
\begin{tabular}{lcccccc}
\toprule
\rowcolor{gray!40}
\multicolumn{7}{l}{\textbf{Quality Assessment Criteria}} \\
QA1: Psychological safety relevance \\
QA2: Software engineering context \\
QA3: Clear research methodology \\
QA4: Relevant findings on team performance \\
QA5: Evidence-supported findings \\
\midrule
\rowcolor{gray!40}
 \textbf{Study ID} & \textbf{QA1} & \textbf{QA2} & \textbf{QA3} & \textbf{QA4} & \textbf{QA5} & \textbf{Total} \\
P1 & 1 & 1 & 0.5 & 1 & 1 & 4.5 \\
P2 & 1 & 1 & 1 & 0.5 & 1 & 4.5 \\
P3 & 0.5 & 1 & 0.5 & 1 & 0.5 & 3.5 \\
P4 & 1 & 0.5 & 1 & 1 & 1 & 4.5 \\
P5 & 1 & 1 & 1 & 1 & 1 & 5.0 \\
\midrule
\rowcolor{gray!20}
\multicolumn{7}{l}{\textbf{Scoring:} 1 = Yes, 0.5 = Partially, 0 = No} \\
\bottomrule
\end{tabular}
\end{table*}

\subsubsection{\textbf{Perform Data Extraction}} 
A structured data extraction form was developed to systematically collect relevant information from the selected studies in accordance with the research questions of this study~\cite{p120}. To improve the reliability and consistency of the extraction process, a pilot data extraction was conducted on a subset of the included studies to refine the extraction form and resolve ambiguities before performing the full data extraction~\cite{p120}. Based on the pilot extraction results, the data extraction form was refined and finalized for the formal data extraction process. To minimize bias and ambiguity, the extracted data were reviewed and discussed during the extraction process.\vspace{5pt}

The data extraction items were organized into two main parts, as presented in Table~\ref{tab:data_extraction}. Part 1 consists of demographic information of the selected studies, while Part 2 includes research questions related data items. Each data item was assigned a unique code (S1, S2, S3, …) to ensure consistency and easy identification during data analysis.\vspace{5pt}

\textbf{Part 1 (S1–S8)} includes demographic information such as publication year, publisher, publication venue, publication type, authors’ affiliation, study aim, and study contribution. This information provides an overview of the general characteristics of the selected studies.\vspace{5pt}

\textbf{Part 2 (S9–S14)} contains RQ-specific data items covering the conceptualization of psychological safety, team performance aspects, team processes, contextual or moderating factors, challenges and barriers, and strategies or practices for strengthening psychological safety.\vspace{5pt}

All extracted data were recorded and organized using Microsoft Excel to facilitate further analysis. The use of coded data items helped ensure consistency in data extraction and supported systematic data analysis.\vspace{5pt}

%-----------table 4------------

\renewcommand{\arraystretch}{1.2}

\begin{table*}[t]
\centering
\caption{Data Extraction Form}
\label{tab:data_extraction}

\footnotesize
\setlength{\tabcolsep}{5pt}

\begin{tabular}{c l p{8cm} c}
\toprule

\rowcolor{gray!40}
\textbf{ID} & \textbf{Data Item} & \textbf{Description} & \textbf{RQs} \\

\midrule

% -------- Demographics Group --------
\multirow{8}{*}{\textbf{S1--S8}} 
& Index & Paper ID & \multirow{8}{*}{\textit{Demographics}} \\
& Publication Year & Publication year of the study & \\
& Publisher & The publisher of the study & \\
& Venue & Publishing venue name & \\
& Publication Type & Journal or conference & \\
& Authors Affiliation & University or industry & \\
& Study Aim & Aim of the paper & \\
& Study Contribution & Contributions made by the paper & \\

\midrule

% -------- RQs --------
S9 & Definition of Psychological Safety 
& Definition or conceptualization used in the study 
& R1 \\

S10 & Team Performance Aspects 
& Aspects of team performance studied (e.g., productivity, software quality, innovation, team satisfaction)
& R2 \\

S11 & Team Processes 
& Team processes influenced by psychological safety (e.g., communication, collaboration, decision-making) 
& R3 \\

S12 & Moderating Factors 
& Factors influencing the relationship between psychological safety and team performance (e.g., leadership style, culture, work environment, trust) 
& R4 \\

S13 & Challenges/Barriers 
&Challenges or barriers to establishing psychological safety in software development teams
& R5 \\

S14 & Strategies/Practices 
& Strategies, practices, or interventions used to improve psychological safety 
& R6 \\

% S15 & Recommendations 
% & Lessons and recommendations for improving team performance 
% & R7 \\

\bottomrule
\end{tabular}

\end{table*}
%-----------------end table 4 ---------------

\subsubsection{\textbf{Perform Data synthesis}} 
The extracted data were analyzed using both descriptive statistics and thematic analysis. Descriptive statistics were used to analyze the quantitative data, while thematic analysis~\cite{p115} was used to analyze the qualitative data extracted from the selected studies.\vspace{5pt}

\textbf{Descriptive statistics:}
Descriptive statistics were used to analyze the demographic data of the selected studies. This included data items such as publication year (S2), publisher (S3), publication venue (S4), publication type (S5), authors’ affiliation (S6), study aim (S7), and study contribution (S8). The purpose of this analysis was to provide an overview of the distribution and general characteristics of the selected studies. The results of this analysis are presented using tables and charts, such as the number of publications per year, distribution of studies by publication type, and distribution by publisher.\vspace{5pt}

\textbf{Thematic Analysis} A predefined technique ~\cite{p115} was employed to systematically analyze the qualitative data extracted from the selected studies, particularly those related to the research questions (S9--S14), including definitions, team performance aspects, effects on team performance, team processes, moderating factors, challenges, strategies, and recommendations. 
The analysis was conducted in several steps:\vspace{5pt}

\begin{itemize}
\item \textbf{Step 1 Data Familiarization:}
In the first step, the selected studies were carefully reviewed multiple times to become familiar with the extracted data. We read the relevant sections of each study and noted important information related to definitions, team performance aspects, team processes, moderating factors, challenges, strategies, and recommendations. This step helped in developing a clear understanding of the data ~\cite{p115}.
\end{itemize}\vspace{1pt}

\begin{itemize}
\item \textbf{Step 2 Initial Coding: }
In this step, descriptive coding was applied to the extracted qualitative data. We carefully reviewed the extracted text from the selected studies and assigned short descriptive labels (codes) to meaningful text segments. These codes were used to summarize the main idea or concept represented in the data. For example, codes such as “trust”, “communication”, “leadership support”, “team collaboration”, and “fear of speaking” were assigned to relevant text segments. The purpose of this step was to organize the data into meaningful groups and to identify important concepts that appeared frequently across the studies. These initial codes served as the foundation for the development of sub-themes and main themes in the later stages of thematic analysis~\cite{p115}.
\end{itemize}\vspace{1pt}

\begin{itemize}
\item \textbf{Step 3 Code Refinement: }
 After descriptive coding, the codes were further refined by breaking down the text into smaller meaningful units. Short and precise codes were assigned to represent key concepts and ideas found in the data. This step helped in identifying different concepts related to psychological safety, team processes, challenges, and strategies \cite{p115}.
\end{itemize}\vspace{1pt}

\begin{itemize}
\item \textbf{Step 4 Sub-Categories/Sub-Themes: }
 In this step, the initial codes were examined and compared to identify similarities and relationships between them. Codes that shared common meanings were grouped together into sub-categories or sub-themes. These sub-themes represented more specific concepts within the data and helped organize the large number of codes into manageable groups. The formation of sub-themes provided a structured foundation for developing broader themes in the next stage of the thematic analysis \cite{p115}.
\end{itemize}\vspace{1pt}

\begin{itemize}
\item \textbf{Step 5 Themes: }
In this step, the sub-themes developed in the previous stage were further analyzed and grouped into broader themes. These themes represented the major concepts and patterns identified in the data. The purpose of this step was to reduce the data into meaningful categories that represent the main findings of the study. Each theme included multiple related sub-themes and provided a structured way to present the results of the thematic analysis. These themes were then used to answer the research questions and present the study findings \cite{p115}.
\end{itemize}\vspace{1pt}

\begin{itemize}
\item \textbf{Step 6 Review and Refinement of Themes: }
In this stage, the initially developed themes and sub-themes were reviewed and refined to ensure that they accurately represented the data and were aligned with the research objectives. We re-examined the codes, sub-themes, and themes to identify similarities, overlaps, and inconsistencies. Some themes were merged where they represented similar concepts, while others were renamed or removed to improve clarity and consistency. This refinement process helped ensure that the final themes were coherent, meaningful, and clearly represented the patterns identified in the data \cite{p115}.
\end{itemize}\vspace{1pt}

\begin{itemize}
\item \textbf{Step 7 Defining and Naming Themes: }
In this final stage of thematic analysis, each identified theme was carefully defined and given a clear and meaningful name. The purpose of this step was to ensure that each theme accurately represented the underlying data and was clearly distinct from other themes. The definitions of the themes were written to explain the scope and meaning of each theme. These final themes were then used to structure the results and to address the research 
of the study \cite{p115}.
\end{itemize}\vspace{5pt}

\begin{figure*}[ht]
    \centering
    \includegraphics[width=0.8\linewidth]{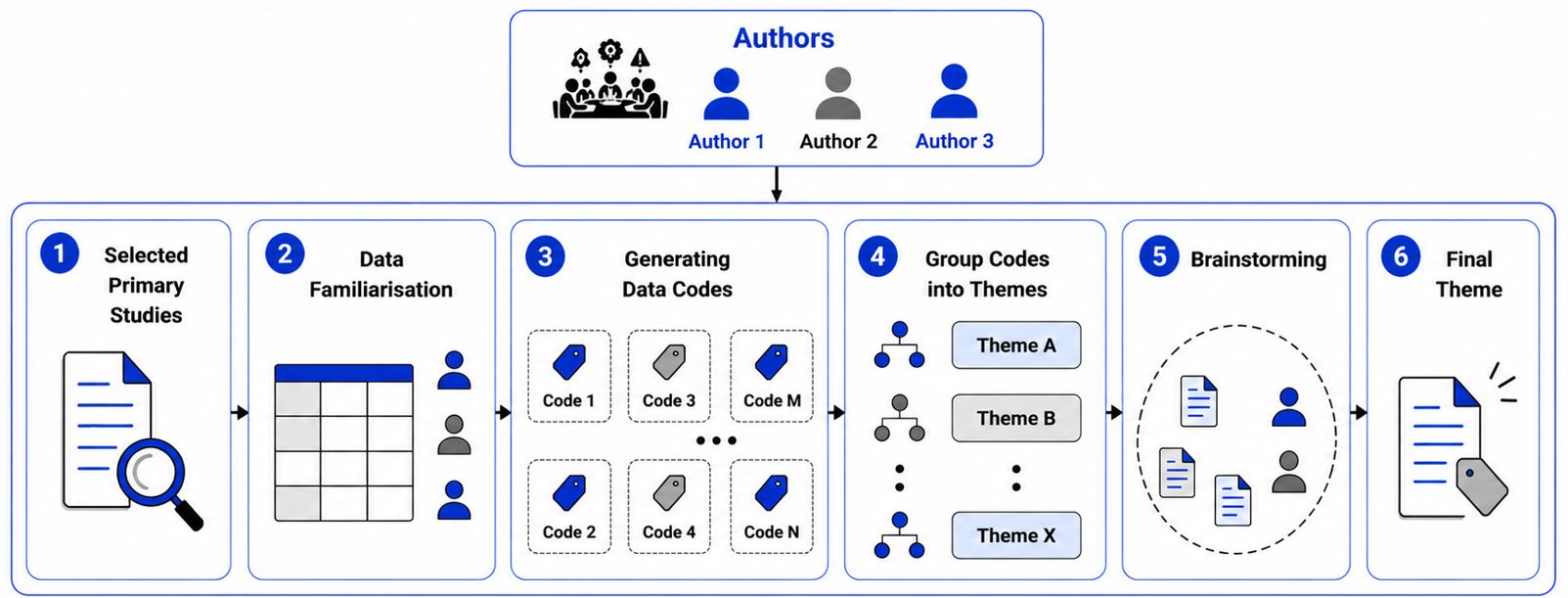}
    \caption{Thematic analysis process.}
    \label{fig:themes coding}
\end{figure*}

The thematic analysis process of this SMS is illustrated in Fig.~\ref{fig:themes coding}. All authors collaboratively participated in the brainstorming phase to minimize bias and ensure consistency in the identification, refinement, and naming of the key themes. This collaborative approach enhances the reliability of thematic classification and aligns with recommended practices for systematic mapping studies in software engineering \cite{p121,p120}.\vspace{5pt}

\textbf{Replication Package:}
The complete dataset used in this study, including the selected primary
studies, extracted data, initial coding, open codes, themes, and subthemes,
is available in the accompanying replication package at
Zenodo~\cite{dataset2026}.\vspace{5pt}

To further support transparency and reproducibility, a replication package is provided. It includes details of the selected primary studies based on the defined search strategy, the quality assessment scores of the included studies, and the extracted data corresponding to each research question (RQ1–RQ6). Such practices are consistent with established guidelines emphasizing traceability and replicability in systematic studies \cite{p120}.

\subsection{\textbf{Reporting the Review}}
We report the results of the systematic mapping study (SMS) in a structured manner, organized according to the defined research questions (RQs). Specifically, (i) the thematic findings addressing RQ1 to RQ6 covering the conceptualization of psychological safety, its influence on team performance and processes, moderating factors, challenges, and improvement strategies are presented in Section 4; (ii) the interpretation and in-depth discussion of these findings are provided in Section 5, where results are analyzed in relation to existing literature and research implications.\vspace{5pt}

Furthermore, a structured comparison with existing studies is presented in Section 6, followed by the synthesis of research gaps in Section 7. Finally, threats to validity are discussed in Section 8, and the overall conclusions and key contributions of the SMS are summarized in Section 9.\vspace{5pt}
%---------Result----------------
\section{Results}
This section presents the findings of the systematic mapping study, structured around the six predefined research questions (RQ1–RQ6). The results first provide an overview of publication trends and the distribution of the included studies, followed by the findings for each research question. Together, these findings provide a structured understanding of how psychological safety is conceptualized and associated with team performance, team processes, contextual factors, barriers, and improvement strategies in software engineering teams. Because individual studies could contribute evidence to multiple themes or subthemes, the reported thematic frequencies are non-mutually exclusive and may therefore exceed the total number of studies addressing a research question.

\begin{figure*}[t]
\centering

\begin{minipage}{0.95\textwidth} % ⭐ controls full layout width
\centering

%---------------- TOP ROW ----------------
\begin{minipage}[t]{0.48\textwidth}
\centering
\begin{tikzpicture}
\begin{axis}[
    ybar,
    bar width=4pt, 
    width=\linewidth,
    height=4.8cm,
    ymin=0,
    ymax=25,
    symbolic x coords={,2006,2007,2008,2009,2010,2011,2012,2013,2014,2015,2016,2017,2018,2019,2020,2021,2022,2023,2024,2025,2026},
    nodes near coords,
    nodes near coords style={font=\small},
    xtick=data,
    xticklabel style={font=\scriptsize, rotate=90},
    ylabel={Studies},
]
\addplot[fill=blue] coordinates {
(2006,3) (2007,0) (2008,2) (2009,1)
(2010,3) (2011,4) (2012,4) (2013,1)
(2014,2) (2015,1) (2016,8) (2017,1)
(2018,8) (2019,6) (2020,5) (2021,5)
(2022,8) (2023,10) (2024,16) (2025,17) (2026,7) 
};
\node[anchor=north east] at (rel axis cs:1,1) {\small Total = 112};
\end{axis}
\end{tikzpicture}

\small (a) Yearly distribution
\end{minipage}
\hfill
\begin{minipage}[t]{.48\textwidth}
\centering
\begin{tikzpicture}
\begin{axis}[
    ybar,
    bar width=4pt, 
    width=\linewidth,
    height=4.8cm,
    ymin=0,
    ymax=80,
    symbolic x coords={IEEE, SciD, ACM, wiley, Springer},
    nodes near coords,
    nodes near coords style={font=\small},
    xtick=data,
    xticklabel style={font=\scriptsize, rotate=15},
    ylabel={Studies},
]
\addplot[fill=blue] coordinates {
(IEEE,66)
(SciD,21)
(ACM,14)
(wiley,6)
(Springer,5)
};
\node[anchor=north east] at (rel axis cs:1,1) {\small Total = 112};
\end{axis}
\end{tikzpicture}

\small (b) Distribution by Digital Library
\end{minipage}

\vspace{0.3cm} % ⭐ tighter spacing

%---------------- BOTTOM CENTER ----------------
\begin{minipage}{0.6\textwidth}
\centering
\begin{tikzpicture}
\begin{axis}[
    ybar,
    bar width=4pt, 
    width=\linewidth,
    height=4.8cm,
    ymin=0,
    ymax=130,
    symbolic x coords={RQ1,RQ2,RQ3,RQ4,RQ5,RQ6},
    nodes near coords,
    nodes near coords style={font=\small},
    xtick=data,
    xticklabel style={font=\scriptsize, rotate=15},
    ylabel={Studies},
]
\addplot[fill=blue] coordinates {
(RQ1,106)
(RQ2,80)
(RQ3,58)
(RQ4,74)
(RQ5,63)
(RQ6,44)
};
\end{axis}
\end{tikzpicture}

\small (c) RQ distribution
\end{minipage}

\end{minipage}

\caption{Overview of publication trends and study distribution}
\label{fig:demo}

\end{figure*}
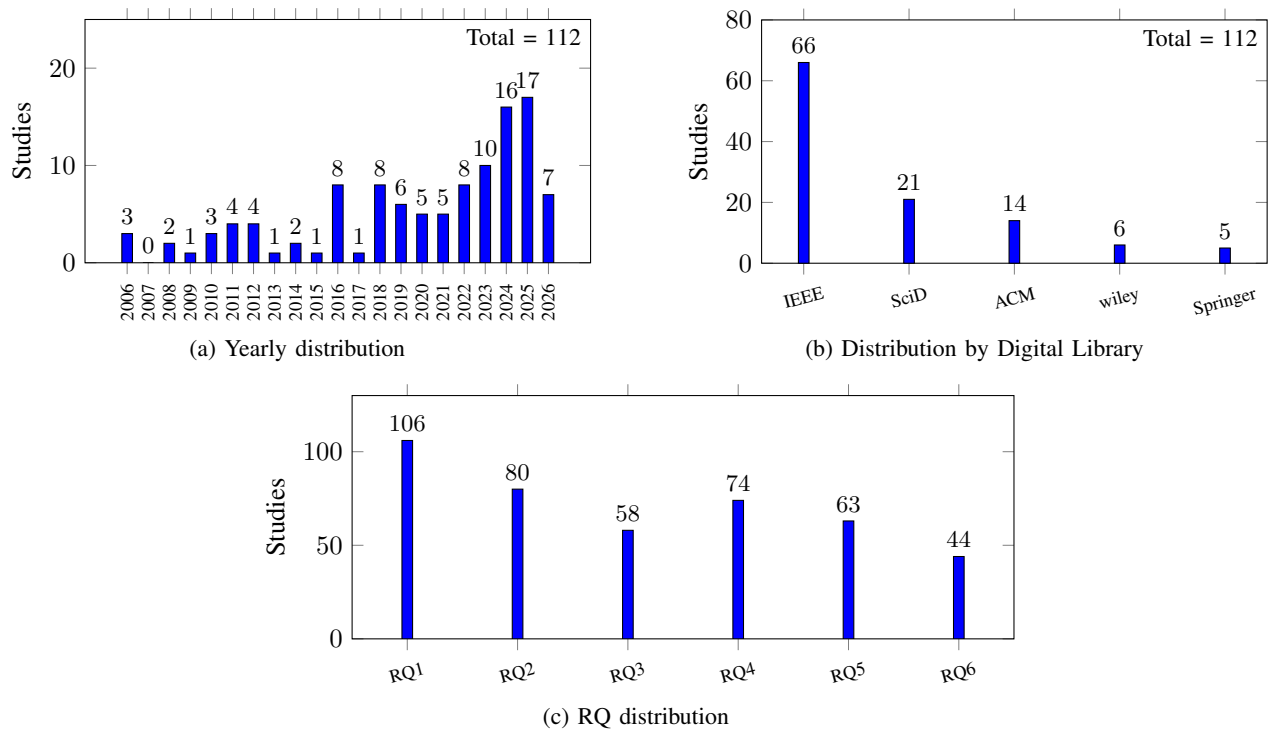
\subsection{\textbf{Overview of Included Studies:}}
\textbf{Yearly Publication Trends:} Figure~\ref{fig:demo} illustrates the yearly distribution of the included studies published between 2006 and june 2026. The figure shows the number of studies published per year and the overall research trend over time. The earliest studies included in this review were published in 2006, with only a few studies identified, followed by relatively low publication activity between 2007 and 2015, indicating limited research activity in the early years. The number of publications began to increase gradually from 2016 onwards, with a noticeable rise in 2018 and 2019, suggesting growing research interest in this area.\vspace{5pt}

Publication activity remained relatively limited during the earlier years but increased more noticeably from 2016 onward. The highest numbers of included studies were observed in 2024 and 2025, indicating growing recent research attention to psychological safety and related human factors in software engineering. Overall, the trend shows a gradual increase in publications over time, with a notable surge in recent years, suggesting that this research area is gaining increasing attention from the research community.\vspace{5pt}

\textbf{Distribution by Digital Library:}
Figure ~\ref{fig:demo} presents the distribution of the 112 included studies across the five digital libraries. IEEE contributed the largest number of studies, with 66 papers (58.9\%), followed by ScienceDirect with 21 (18.8\%), ACM with 14 (12.5\%), Wiley with 6 (5.4\%), and SpringerLink with 5 (4.5\%). The distribution shows that a substantial proportion of the included literature was retrieved from IEEE, while relevant studies were also identified across the other four sources.\vspace{5pt}

It also suggests that most of the relevant literature is concentrated in a limited number of major digital libraries, which can guide researchers on where to search for and publish studies related to this topic.\vspace{5pt}

%---------table 5---------------

The included studies contributed to the six research questions to varying degrees, Table~\ref{tab:rq}. RQ1 received the broadest coverage, with 106 studies (94.6\%), followed by RQ2 with 80 studies (71.4\%), RQ4 with 74 studies (66.1\%), RQ5 with 63 studies (56.3\%), RQ3 with 58 studies (51.8\%), and RQ6 with 44 studies (39.3\%). These results indicate that the conceptualization of psychological safety (RQ1) has received comparatively greater attention in the literature, whereas strategies and interventions for enhancing psychological safety (RQ6) remain less frequently investigated as shown in Figure~\ref{fig:demo}.

%-----------RQ1-----------------

\subsection{\textbf{Conceptualization of Psychological Safety in Software Engineering Teams (RQ1):}}
RQ1 investigates how psychological safety is defined and conceptualized in software engineering teams. The included studies were examined to identify the key conceptual characteristics, interpersonal conditions, and team-level mechanisms through which psychological safety is described in software engineering contexts.\vspace{5pt}

Of the 112 included studies, 106 (94.6\%) contributed evidence addressing RQ1. The thematic analysis organized the evidence into seven major themes and twenty-five subthemes, as illustrated in Table ~\ref{tab:conceptualization}. Communication-related conceptualizations received the greatest coverage, with 37 studies (33.0\%), followed by collaboration with 18 studies (16.1\%). The remaining themes capture different conceptual dimensions of psychological safety, including team dynamics and performance factors, social environment, learning and knowledge sharing, interpersonal openness, and individual security and well-being.\vspace{5pt}

By conducting thematic analysis, the literature was organized into seven \textit{major themes and twenty-five sub-themes. The identified themes include Collaboration Shapes Psychological Safety in Team Performance, Conceptualizations of Communication’s Role in Psychological Safety and Performance, Team Dynamics and Performance Factors, Social Environment Factors in Team Psychological Safety and Performance, Psychological Safety as an Enabler of Learning and Knowledge Sharing, Psychological Safety and Interpersonal Openness and Communication, and Psychological Safety and Individual Security and Well-Being,} as illustrated in Table~\ref{tab:conceptualization}.\vspace{5pt}

\textbf{Collaboration Shapes Psychological Safety in Team Performance} examines how collaboration contributes to psychological safety and team performance, supported by 18 studies (16.1\%).The sub-theme Psychological Safety and Collaboration is supported by nine studies (8.0\%), highlighting that collaborative practices encourage knowledge sharing and collective problem solving \cite{p7,p23,p25,p27,p39,p60,p87,p88,p94} . Communication Supporting Psychological Safety in Collaborative Teams appears in five studies (4.5\%), emphasizing open communication and constructive feedback \cite{p2,p29,p42,p59,p82}. Three studies (2.7\%) identify Trust and Communication in Team Performance, demonstrating that trust-based communication strengthens teamwork \cite{p57,p71,p109a}. One study (0.9\%) highlights Psychological Safety Supporting Development, Motivation, and Openness, showing that safe environments encourage participation \cite{p75}. Overall, the evidence characterizes collaboration, supported by communication and trust, as an important condition associated with psychologically safe team environments.\vspace{5pt}

\textbf{Conceptualizations of communication’s role in psychological safety and team performance} captures how communication-related behaviors and conditions are associated with the conceptualization and expression of psychological safety in software teams., identified in 37 studies (33.0\%). The sub-theme \textit{Interpersonal Risk-Taking and Safety} appears in eight studies (7.1\%), indicating that psychological safety enables behaviors such as asking questions and admitting mistakes \cite{p3,p10,p24,p34,p61,p64,p77,p91}. Seven studies (6.3\%) highlight \textit{Shared Belief, Speak-Up Behavior, and Identity}, emphasizing shared beliefs that speaking up is safe \cite{p11,p32,p76,p89,p97,p103,p104}. Five studies (4.5\%) identify \textit{Trust, Respect, and Team Climate}, showing that respectful relationships support open dialogue.
\cite{p38,p56,p70,p73,p92}. Four studies (3.6\%) emphasize \textit{Supervisor Support and Freedom of Expression}, highlighting the role of supportive leadership in encouraging open communication \cite{p21,p80,p86,p108a}. In summary, communication practices and supportive leadership are central mechanisms for fostering psychological safety.\vspace{5pt}

The theme \textbf{Team Dynamics and Performance Factors} investigates how team dynamics influence psychological safety and performance, supported by 15 studies (13.4\%). Five studies (4.5\%) highlight \textit{Human Factors Influencing Team Performance}, emphasizing interpersonal interactions and collaboration patterns \cite{p4,p5,p9,p74,p102}. Six studies (5.4\%) discuss \textit{Psychological Safety and Team Performance}, showing that safe teams engage more actively in knowledge sharing and collaboration \cite{p22,p79,p83,p96,p99,p100}. One study (0.9\%) identifies \textit{Individual Attributes and Personality Factors}, indicating the influence of personal characteristics on team participation \cite{p6}. Three studies (2.7\%) highlight \textit{Leadership Factors Supporting Psychological Safety}, where leaders encourage open communication and participation \cite{p1,p13,p93}. Together, these studies associate psychological safety with team dynamics, leadership behaviors, interpersonal interactions, and individual characteristics.\vspace{5pt}

The theme \textbf{Social Environment Factors in Team Psychological Safety and Performance} examines the role of social and organizational environments in psychological safety, supported by 13 studies (11.6\%). Nine studies (8.0\%) highlight \textit{Psychological Safety and Trust-Based Team Climate}, indicating that trust promotes open communication and collaboration \cite{p20,p31,p33,p40,p53,p65,p85,p98,p101}. Three studies (2.7\%) discuss \textit{Psychological Safety and Supportive Team Climate}, emphasizing supportive organizational environments \cite{p30,p37,p108a}. One study (0.9\%) identifies \textit{Team Cohesion and Boundary-Spanning}, linking cohesive teams with improved coordination and knowledge sharing \cite{p66}. Overall, the evidence highlights supportive social environments and trust-based climates as recurring conditions associated with psychological safety.\vspace{5pt}

\begin{table*}[t]
\centering
\caption{Conceptualization of Psychological Safety in Software Engineering Teams (RQ1)}
\label{tab:conceptualization}

{
\fontfamily{Carlito-TLF}\selectfont
\fontsize{7.5}{8.5}\selectfont
\renewcommand{\arraystretch}{1.05}
\setlength{\tabcolsep}{2pt}

% Clear table borders
\setlength{\arrayrulewidth}{0.7pt}
\arrayrulecolor{black}

\begin{tabular}{|
>{\raggedright\arraybackslash}m{0.27\textwidth}|
>{\raggedright\arraybackslash}m{0.35\textwidth}|
>{\raggedright\arraybackslash}m{0.30\textwidth}|}
\hline

% ==================================================
% RQ HEADING
% ==================================================

\rowcolor{yellow!25}
\multicolumn{3}{|l|}{%
\color{black}\bfseries
RQ1: Definition/Conceptualization of Psychological Safety in team performance
}
\\
\hline

% ==================================================
% COLUMN HEADINGS
% ==================================================

\rowcolor{black}
\color{white}\bfseries Themes &
\color{white}\bfseries Sub-Themes &
\color{white}\bfseries Paper IDs
\\
\hline

% ==================================================
% THEME 1 — LIGHT GRAY
% ==================================================

\cellcolor{gray!15}
&
\cellcolor{gray!15}
1.1 Psychological Safety \& Collaboration
&
\cellcolor{gray!15}
P7, P23, P25, P27, P39, P60, P87, P88, P94
\\

\hhline{|>{\arrayrulecolor{gray!15}}->{\arrayrulecolor{black}}|-|-|}

\cellcolor{gray!15}
&
\cellcolor{gray!15}
1.2 Psychological Safety \& Communication
&
\cellcolor{gray!15}
P2, P29, P42, P59, P82
\\

\hhline{|>{\arrayrulecolor{gray!15}}->{\arrayrulecolor{black}}|-|-|}

\cellcolor{gray!15}
&
\cellcolor{gray!15}
1.3 Trust, Communication \& Team Performance
&
\cellcolor{gray!15}
P57, P71, P109
\\

\hhline{|>{\arrayrulecolor{gray!15}}->{\arrayrulecolor{black}}|-|-|}

\cellcolor{gray!15}
\multirow[c]{-4}{0.27\textwidth}[2.0ex]{%
\raggedright
1. Collaboration Shapes Psychological Safety in Team Performance
}
&
\cellcolor{gray!15}
1.4 Development, Motivation \& Openness
&
\cellcolor{gray!15}
P75
\\

\arrayrulecolor{black}
\hline

% ==================================================
% THEME 2 — WHITE
% ==================================================

&
2.1 Interpersonal Risk-Taking \& Safety
&
P3, P10, P24, P34, P61, P64, P77, P91
\\
\cline{2-3}

&
2.2 Shared Belief, Speak-Up \& Identity
&
P11, P32, P76, P89, P97, P103, P104
\\
\cline{2-3}

&
2.3 Trust, Respect \& Team Climate
&
P38, P56, P70, P73, P92
\\
\cline{2-3}

\multirow[c]{-4}{0.27\textwidth}[2.0ex]{%
\raggedright
2. Conceptualizations of communication’s role in psychological safety \& team performance 
}
&
2.4 Supervisor Support \& Freedom of Expression
&
P21, P80, P86, P90
\\
\hline

% ==================================================
% THEME 3 — LIGHT GRAY
% ==================================================

\cellcolor{gray!15}
&
\cellcolor{gray!15}
3.1 Human Factors Influencing Team Performance
&
\cellcolor{gray!15}
P4, P5, P9, P74, P102
\\

\hhline{|>{\arrayrulecolor{gray!15}}->{\arrayrulecolor{black}}|-|-|}

\cellcolor{gray!15}
&
\cellcolor{gray!15}
3.2 Psychological Safety \& Team Performance
&
\cellcolor{gray!15}
P22, P79, P83, P96, P99, P100
\\

\hhline{|>{\arrayrulecolor{gray!15}}->{\arrayrulecolor{black}}|-|-|}

\cellcolor{gray!15}
&
\cellcolor{gray!15}
3.3 Individual Attributes \& Personality Factors
&
\cellcolor{gray!15}
P6
\\

\hhline{|>{\arrayrulecolor{gray!15}}->{\arrayrulecolor{black}}|-|-|}

\cellcolor{gray!15}
\multirow[c]{-4}{0.27\textwidth}[2.0ex]{%
\raggedright
3. Team Dynamics and Performance Factors
}
&
\cellcolor{gray!15}
3.4 Leadership Factors Supporting Psychological Safety
&
\cellcolor{gray!15}
P1, P13, P93
\\

\arrayrulecolor{black}
\hline

% ==================================================
% THEME 4 — WHITE
% ==================================================

&
4.1 Psychological Safety \& Trust-Based Team Climate
&
P20, P31, P33, P40, P53, P65, P85, P98, P101
\\
\cline{2-3}

&
4.2 Psychological Safety \& Supportive / Safety-Supportive
&
P30, P37, P108
\\
\cline{2-3}

\multirow[c]{-3}{0.27\textwidth}[-0.2ex]{%
\raggedright
4. Social Environment Factors in Team Psychological Safety and Performance
}
&
4.3 Cohesion, Boundary Spanning \& Psychological Safety
&
P66
\\
\hline

% ==================================================
% THEME 5 — LIGHT GRAY
% ==================================================

\cellcolor{gray!15}
&
\cellcolor{gray!15}
5.1 Psychological Safety \& Learning / Knowledge Sharing
&
\cellcolor{gray!15}
P8, P12, P14, P15, P16, P17, P18, P78
\\

\hhline{|>{\arrayrulecolor{gray!15}}->{\arrayrulecolor{black}}|-|-|}

\cellcolor{gray!15}
&
\cellcolor{gray!15}
5.2 Psychological Safety as Foundation for Learning
&
\cellcolor{gray!15}
P41, P43, P44, P26
\\

\hhline{|>{\arrayrulecolor{gray!15}}->{\arrayrulecolor{black}}|-|-|}

\cellcolor{gray!15}
\multirow[c]{-3}{0.27\textwidth}[-0.2ex]{%
\raggedright
5. Psychological Safety as an Enabler of Learning \& Knowledge Sharing
}
&
\cellcolor{gray!15}
5.3 Psychological Safety \& Team-Level Comfort
&
\cellcolor{gray!15}
P46
\\

\arrayrulecolor{black}
\hline

% ==================================================
% THEME 6 — WHITE
% ==================================================

&
6.1 Psychological Safety \& Shared Feedback / Team Goals / Improvement
&
P28, P35, P45
\\
\cline{2-3}

&
6.2 Psychological Safety \& Shared Team Consensus
&
P48, P49, P50
\\
\cline{2-3}

\multirow[c]{-3}{0.27\textwidth}[2.2ex]{%
\raggedright
6. Psychological Safety and Interpersonal Openness \& Communication
}
&
6.3 Psychological Safety \& Openness / Idea Contribution
&
P36
\\
\hline

% ==================================================
% THEME 7 — LIGHT GRAY
% ==================================================

\cellcolor{gray!15}
&
\cellcolor{gray!15}
7.1 Psychological Safety \& Acceptance
&
\cellcolor{gray!15}
P19
\\

\hhline{|>{\arrayrulecolor{gray!15}}->{\arrayrulecolor{black}}|-|-|}

\cellcolor{gray!15}
&
\cellcolor{gray!15}
7.2 Psychological Safety \& Role Comfort \& Career Security
&
\cellcolor{gray!15}
P51, P52, P54, P58, P63, P67, P68, P69, P111
\\

\hhline{|>{\arrayrulecolor{gray!15}}->{\arrayrulecolor{black}}|-|-|}

\cellcolor{gray!15}
&
\cellcolor{gray!15}
7.3 Psychological Safety \& Role Comfort (Self-Image / Status / Career Protection)
&
\cellcolor{gray!15}
P95, P105, P106, P109
\\

\hhline{|>{\arrayrulecolor{gray!15}}->{\arrayrulecolor{black}}|-|-|}

\cellcolor{gray!15}
\multirow[c]{-4}{0.27\textwidth}[4.0ex]{%
\raggedright
7. Psychological Safety and Individual Security \& Well-Being
}
&
\cellcolor{gray!15}
7.4 Psychological Safety \& Well-Being / Retention / Competitive Advantage
&
\cellcolor{gray!15}
P81 P84
\\

\arrayrulecolor{black}
\hline

\end{tabular}

}

\end{table*}

\begin{comment}
\begin{figure*}[!t]
    \centering
    \includegraphics[width=\textwidth]{RQ1.png}
    \caption{Conceptualization of human factors and psychological safety in team performance.}
    \label{fig:4}
\end{figure*}
\end{comment}

The theme \textbf{Psychological Safety as an Enabler of Learning and Knowledge Sharing} explores how psychological safety facilitates learning and knowledge sharing, identified in 13 studies (11.6\%). Eight studies (7.1\%) highlight \textit{Psychological Safety and Learning or Knowledge Sharing}, showing that safe environments encourage information exchange and collective learning \cite{p8,p12,p14,p15,p16,p17,p18,p78}. Four studies (3.6\%) emphasize \textit{Psychological Safety as a Foundation for Team Learning}, allowing individuals to admit mistakes and seek feedback \cite{p41,p43,p44,p26}. One study (0.9\%) identifies \textit{Team-Level Comfort in Collaborative Work}, linking comfort with active participation in team discussions \cite{p46}.Overall, psychological safety is consistently associated with learning behaviors, knowledge exchange, feedback seeking, and participation in collaborative learning.\vspace{5pt}

The theme \textbf{Psychological Safety and Interpersonal Openness and Communication} examines how psychological safety promotes interpersonal openness, supported by 7 studies (6.3\%). Three studies (2.7\%) highlight \textit{Shared Feedback, Team Goals, and Improvement}, where safe environments enable constructive feedback \cite{p28,p35,p45}. Three studies (2.7\%) identify \textit{Psychological Safety and Shared Team Consensus} \cite{p48,p49,p50}. One study (0.9\%) highlights \textit{Openness and Idea Contribution}, showing that individuals freely contribute ideas in psychologically safe environments \cite{p36}. Considering this discussion, psychological safety promotes open dialogue and collaborative team interaction.\vspace{5pt}

The theme \textbf{Psychological Safety and Individual Security and Well-Being} examines how psychological safety contributes to individual security and well-being, supported by 16 studies (14.3\%). One study (0.9\%) highlights \textit{Psychological Safety and Acceptance within Teams}, emphasizing the importance of social acceptance \cite{p19}. Nine studies (8.0\%) discuss \textit{Role Comfort and Career Security}, indicating that psychologically safe environments reduce fear of criticism \cite{p51,p52,p54,p58,p63,p67,p68,p69,p111a}. Four studies (3.6\%) identify \textit{Self-Image, Status, and Career Protection}, showing that psychological safety reduces concerns about professional reputation \cite{p95,p105,p106,p109a}. Two studies (1.8\%) highlight \textit{Employee Well-Being and Organizational Outcomes}, linking psychological safety with employee satisfaction and retention \cite{p81,p84}. Overall, the studies associate psychological safety with greater role comfort, reduced concerns about interpersonal and professional consequences, and improved individual well-being and engagement.\vspace{5pt}

Overall, the findings show that psychological safety is conceptualized in software engineering primarily as a shared team-level condition that enables interpersonal risk-taking, open communication, idea expression, feedback, and discussion of mistakes without fear of negative consequences. Across the identified themes, this conceptualization is closely associated with collaboration, trust, supportive team dynamics, learning and knowledge sharing, interpersonal openness, and individual security and well-being. The evidence therefore suggests that psychological safety is represented not only through perceptions of interpersonal safety but also through the communication and interaction behaviors that characterize psychologically safe software teams.\vspace{5pt}

%------------------Takeaway RQ1
\begin{tcolorbox}[
    colback=gray!5,
    colframe=black,
    title=Takeaways (RQ1)
]

\begin{itemize}[
    label=\raisebox{-0.12\height}{\includegraphics[width=0.42cm]{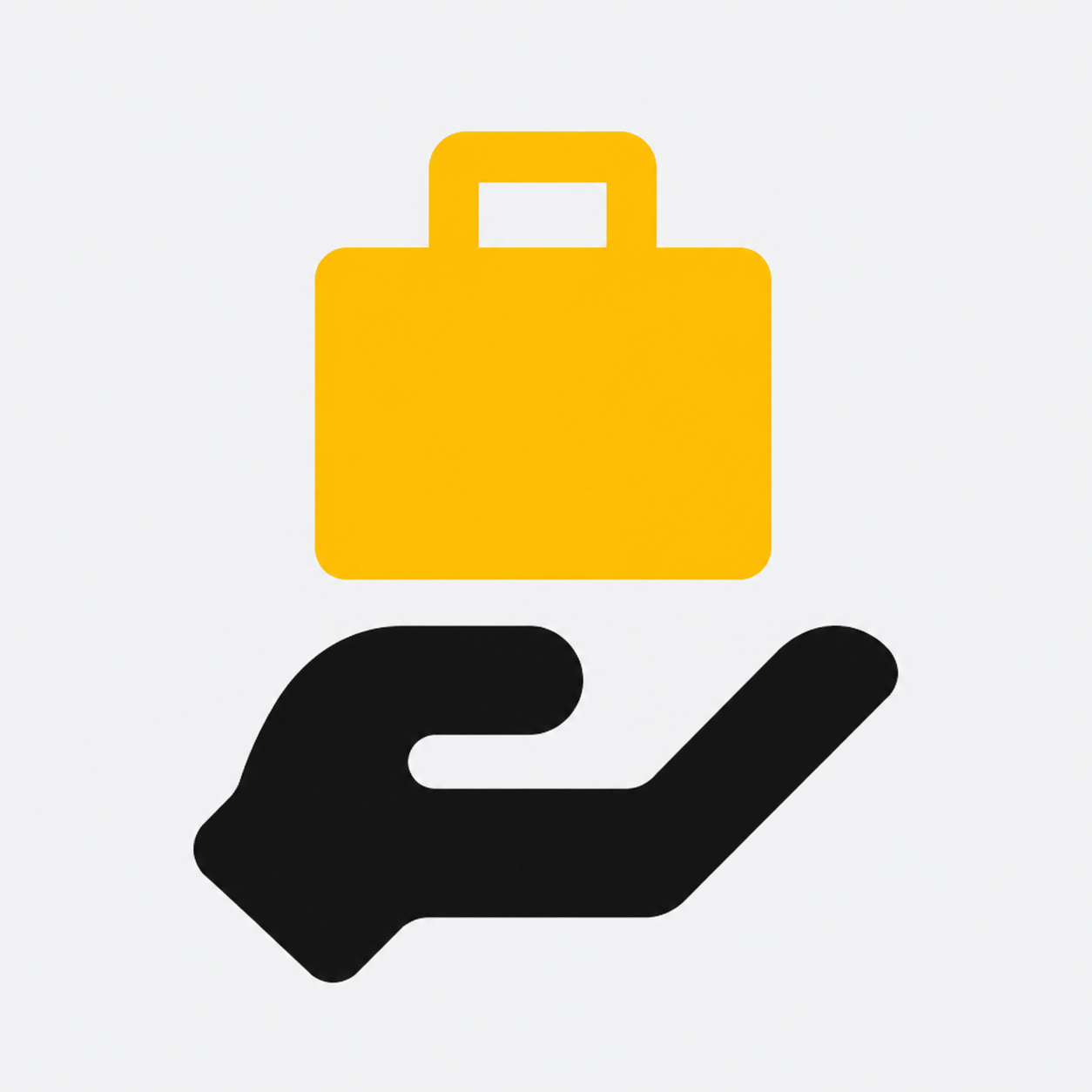}},
    leftmargin=0.70cm,
    labelsep=0.12cm,
    itemsep=-1pt,
parsep=0pt,
topsep=0pt,
partopsep=0pt
]

\item Psychological safety in software development teams is primarily conceptualized as a shared belief enabling open communication and interpersonal risk-taking.

\item Communication-related behaviors and conditions, including speak-up behavior, interpersonal risk-taking, and leadership-supported dialogue, emerge as central characteristics of psychological safety.

\item Collaboration, trust-based interactions, and supportive team dynamics form the structural conditions that sustain psychologically safe team environments.

\item Psychological safety enables learning, knowledge sharing, and individual well-being, thereby strengthening both team interaction and overall team performance.

\end{itemize}

\end{tcolorbox}
%----------- RQ2---------------------------------------------------
\subsection{\textbf{Aspects of Team Performance Associated with Psychological Safety (e.g. productivity, software quality, innovation, satisfaction) (RQ2):}}
RQ2 examines the aspects of team performance associated with psychological safety in software engineering teams. The thematic synthesis identified four major performance aspects reported in the literature: productivity, software quality, innovation, and team satisfaction. The findings for each performance aspect are presented in the following subsections.\vspace{5pt}

\subsubsection{\textbf{Psychological Safety and Team Productivity}}
Of the 112 included studies, 67 (59.8\%) reported evidence related to psychological safety and productivity, while 45 studies (40.2\%) did not directly address productivity as a team performance outcome.\vspace{5pt}

The thematic analysis of productivity-related evidence identified four major themes and eight subthemes, as illustrated in Table.~\ref{tab:RQ2}. The four themes capture the roles of collaboration and communication; learning, feedback, and knowledge processes; emotional climate and team functioning; and creativity, motivation, and proactive behavior in productivity-related outcomes.\vspace{5pt}

A total of 16 studies (14.3\%) highlight the role of \textbf{\textit{Psychological Safety in Collaboration and Communication for Improving Team Productivity}}. The first subtheme, \textit{Collaboration Strengthening Psychological Safety and Enhancing Team Productivity}, states that psychologically safe environments encourage collaboration and collective problem solving, which improves team productivity \cite{p1,p7,p16,p70,p71,p72,p74,p112a}. The second subtheme, \textit{Effective Communication Increasing Psychological Safety and Improving Coordination Efficiency}, states that open communication supported by psychological safety improves coordination and information sharing among team members, leading to better productivity \cite{p11,p19,p20,p25,p27,p29,p73,p108a}.\vspace{5pt}

A total of 24 studies (21.4\%) highlight the role of \textbf{\textit{PS-Enabled Learning, Feedback, and Knowledge Processes for Productivity}}. The first subtheme, \textit{Learning Behaviors Enhancing Psychological Safety and Boosting Team Capability and Productivity}, shows that learning behaviors supported by psychological safety improve team capability and productivity \cite{p10,p15,p21,p22,p26,p31,p32,p111a}. The second subtheme, \textit{Reflective Practices Raising Psychological Safety and Improving R\&D Productivity}, indicates that reflective practices strengthen psychological safety and enhance R\&D productivity \cite{p37,p40,p46,p100,p103,p96,p79,p109a}. The third subtheme, \textit{Feedback and Problem-Solving Behaviors Reinforcing Psychological Safety and Improving Knowledge-Driven Productivity}, demonstrates that feedback and problem-solving behaviors reinforce psychological safety and improve knowledge-driven productivity \cite{p55,p59,p62,p82,p85,p101,p88,p108a}.\vspace{5pt}

A total of 16 studies (14.3\%) highlight the role of \textbf{\textit{Psychological Safety and Human Factors in Emotional Climate and Team Functioning}}. The first subtheme, \textit{Trust and Emotional Climate Strengthening Psychological Safety and Improving Team Functioning}, shows that trust and a positive emotional climate strengthen psychological safety and support effective team functioning \cite{p14,p17,p18,p67,p68,p64,p65,p86}. The second subtheme, \textit{Psychological Safety Directly Increasing Team Productivity and Performance}, indicates that psychologically safe environments enable team members to perform effectively and directly improve team productivity \cite{p24,p30,p33,p89,p90,p92,p97,p112a}.\vspace{5pt}

A total of 8 studies (7.1\%) highlight the role of \textbf{\textit{Psychological Safety in Creativity, Motivation and Proactive Productivity}}. The first subtheme, \textit{Psychological Safety-Enabled Creativity and Learning for Productivity}, shows that psychologically safe environments encourage creativity and learning behaviors that support team productivity \cite{p23,p28,p39,p81,p110a}. The second subtheme, \textit{Psychological Safety-Driven Motivation and Initiative for Productivity}, indicates that psychological safety motivates team members to take initiative and contribute proactively to improving team productivity \cite{p36,p88,p109a}.\vspace{5pt}

Overall, the productivity-related evidence indicates that psychological safety is associated with productivity both directly and through team-level mechanisms involving collaboration, communication, learning, feedback, knowledge sharing, trust, and proactive behavior. The findings suggest that productivity-related benefits are most frequently reported when psychological safety supports effective interaction and learning within software teams.\vspace{5pt}

\subsubsection{\textbf{Psychological Safety and Software Quality}}
This subsection examines the relationship between psychological safety and software quality as an aspect of team performance. The reviewed studies associate software quality with interpersonal and team-level conditions such as collaboration, communication, trust, learning, error management, and innovation. Based on the selected set of 112 literature studies, 25 studies (22.3\%) were found to contain relevant information related to software quality, while 82 studies (73.2\%) did not explicitly discuss this aspect. Through thematic analysis, the relevant studies were organized into two main themes and five subthemes. The identified themes include \textit{Psychological Safety and Human Interaction Factors Influencing Software Quality} and \textit{PS-Enabled Learning, Error Management and Innovation Affecting Software Quality}. The first theme consists of three subthemes, while the second theme contains two subthemes, as shown in Table~\ref{tab:RQ2}, highlighting how collaboration, emotional and interpersonal factors, trust, learning behaviors, and innovation supported by psychological safety contribute to improving software quality in team-based development environments.\vspace{5pt}

A total of 18 studies (16.8\%) highlight the role of \textbf{\textit{Psychological Safety, Human Interaction, and Software Quality}}. The first subtheme, \textit{Collaboration Strengthening Psychological Safety and Improving Software Quality}, shows that collaborative environments supported by psychological safety encourage open dialogue, reporting mistakes, and conflict resolution, which helps improve software quality \cite{p7,p11,p25,p27,p28,p79,p111a}. The second subtheme, \textit{Emotional and Interpersonal Factors Increasing Psychological Safety and Improving Communication Quality}, indicates that emotional competence, engagement, and positive interpersonal relations strengthen psychological safety and improve communication quality within teams \cite{p9,p17,p94,p20,p89,p87}. The third subtheme, \textit{Trust, Supportive Practices, and Structure Strengthening Psychological Safety for High Software Quality}, demonstrates that leadership support, DevOps practices, and trust-based team structures reinforce psychological safety and contribute to maintaining high software quality \cite{p14,p16,p71,p62,p56}.\vspace{5pt}

\begin{comment}
\begin{figure*}[t]
    \centering
    \includegraphics[width=0.7\textwidth]{RQ2.2.png}
    \caption{Human factors and psychological safety influencing software quality.}
    \label{fig:6}
\end{figure*}
\end{comment}

A total of 11 studies (10.3\%) highlight the role of \textbf{\textit{Psychological Safety-Enabled Learning, Error Management and Innovation Affecting Software Quality}}. The first subtheme, \textit{Learning, Voice, and Error Prevention Enhancing Psychological Safety and Software Quality}, shows that psychologically safe environments encourage team members to speak up, share knowledge, and identify errors early, which helps improve software quality and problem-solving effectiveness \cite{p10,p12,p13,p20,p34,p74,p102}. The second subtheme, \textit{Psychological Safety Enabling Creativity and Innovation for High-Quality Performance}, indicates that psychological safety with learning behaviors, error reporting, knowledge sharing, and innovation-related practices that contribute to software quality. \cite{p21,p87,p89,p111a}.\vspace{5pt}

Overall, the findings indicate that psychological safety is associated with software quality through two broad mechanisms: supportive human interactions involving collaboration, communication, and trust, and learning-oriented processes involving error reporting, knowledge sharing, feedback, and innovation. These findings suggest that the relationship between psychological safety and software quality is primarily reported through team interaction and learning processes rather than as a direct technical effect.\vspace{5pt}

\subsubsection{\textbf{Psychological Safety and Innovation}}
This section examines how human factors and psychological safety influence innovation in team performance.This subsection examines the relationship between psychological safety and innovation as an aspect of team performance. Of the 112 included studies, 33 studies (29.5\%) reported evidence related to innovation, while 79 studies (70.5\%) did not explicitly address innovation as a performance outcome. Through thematic analysis, the relevant evidence was organized into two major themes and seven subthemes: Innovation Processes Enabled by Psychological Safety and Social and Environmental Conditions Supporting Innovation. These themes capture how idea sharing, creativity, exploration, learning, knowledge sharing, and supportive team environments are associated with innovation in psychologically safe teams, as shown in Table~\ref{tab:RQ2}.\vspace{5pt}

A total of 29 studies (25.9\%) highlight the role of \textbf{\textit{Innovation Processes Enabled by Psychological Safety}}. The first subtheme, \textit{Idea Sharing Builds Psychological Safety and Enhances Innovation}, shows that psychologically safe environments encourage open idea sharing that supports innovation \cite{p3,p9,p11,p15,p16,p17,p19,p21}. The second subtheme, \textit{Creativity and Error Awareness Strengthen Psychological Safety and Improve Innovation Quality}, indicates that creativity and awareness of mistakes improve innovation quality \cite{p23,p24,p26,p27,p29,p111a}. The third subtheme, \textit{Exploration Behaviors Increase Psychological Safety and Improve Innovation Quality}, demonstrates that experimentation and exploration enhance innovation capability \cite{p31,p33,p37,p38,p41}. The fourth subtheme, \textit{Learning and Knowledge Sharing Boost Psychological Safety and Innovation Outcomes}, suggests that knowledge sharing and learning behaviors strengthen innovation outcomes \cite{p46,p60,p75,p77,p78,p83}. The fifth subtheme, \textit{Core Psychological Safety Mechanisms Enable Innovation and High Performance}, shows that trust, openness, and supportive communication enable innovative team performance \cite{p1,p9,p16,p111a}.Overall, the reviewed studies associate psychological safety with innovation through idea sharing, creativity, experimentation, learning, knowledge sharing, and supportive team interactions.\vspace{5pt}

\begin{comment}

\begin{figure*}[t]
    \centering
    \includegraphics[width=0.6\textwidth, keepaspectratio]{RQ2.3.png}
    \caption{Impact of Human Factors and PS on Innovation.}
    \label{fig:7}
\end{figure*}
\end{comment}

A total of 9 studies (8.0\%) highlight the role of \textbf{\textit{Social and Environmental Conditions Supporting Innovation}}. The first subtheme, \textit{Inclusive Climate Enhances Psychological Safety and Drives Innovation}, shows that inclusive and supportive climates encourage idea expression and innovative behaviors \cite{p85,p88,p90,p91,p93}. The second subtheme, \textit{Positive Team Conditions Strengthen Psychological Safety and Improve Creative Performance}, indicates that positive team environments strengthen psychological safety and enhance creative performance \cite{p102,p103,p104,p106}. Overall, supportive team climates and inclusive environments are frequently associated with innovative behaviors in psychologically safe teams.\vspace{5pt}

Overall, the innovation-related evidence associates psychological safety with idea sharing, creativity, experimentation, learning, and knowledge sharing. The findings further highlight supportive and inclusive team environments as important contextual conditions associated with innovative behaviors in software development teams.\vspace{5pt}

\subsubsection{\textbf{Psychological Safety and Team Satisfaction}}
This section examines how human factors and psychological safety influence individual and team satisfaction. Out of the 112 selected studies, 34 studies (30.3\%) were found to contain relevant information related to the satisfaction aspect, while 73 studies (65.1\%) did not explicitly discuss this aspect. Through thematic analysis, the relevant studies were organized into two main themes and five subthemes. The identified themes include \textit{Psychological Safety Shaping Individual and Team Satisfaction} and \textit{Psychological Safety-Enabled Performance, Structure, and Organizational Conditions Influencing Satisfaction}. The first theme consists of three subthemes, while the second theme contains two subthemes, as shown in Table~\ref{tab:RQ2}, highlighting how psychological safety, trust, well-being, and supportive organizational conditions contribute to improving satisfaction within teams.\vspace{5pt}

\begin{table*}[t]
\centering
\caption{Aspects of Psychological Safety on Team Performance}
\label{tab:RQ2}

{
\fontfamily{Carlito-TLF}\selectfont
\fontsize{7.5}{8.5}\selectfont
\renewcommand{\arraystretch}{1.05}
\setlength{\tabcolsep}{2pt}

% Clear table borders
\setlength{\arrayrulewidth}{0.7pt}
\arrayrulecolor{black}

\begin{tabular}{|
>{\raggedright\arraybackslash}m{0.30\textwidth}|
>{\raggedright\arraybackslash}m{0.38\textwidth}|
>{\raggedright\arraybackslash}m{0.24\textwidth}|}
\hline

% ==================================================
% RQ 2.1 — PRODUCTIVITY
% ==================================================

\rowcolor{yellow!25}
\multicolumn{3}{|l|}{%
\color{black}\bfseries
RQ 2.1: Identify that what are the aspects of Psychological Safety
on team performance (Productivity)
}
\\
\hline

\rowcolor{black}
\color{white}\bfseries Themes &
\color{white}\bfseries Sub-Themes &
\color{white}\bfseries Paper IDs
\\
\hline

% ---------------- THEME 1 — WHITE ----------------

\multirow[c]{2}{0.30\textwidth}{%
\raggedright
1. PS in Collaboration and Communication for Team Productivity
}
&
1.1 Collaboration Strengthening PS and Enhancing Team Productivity
&
P1, P7, P16, P70, P71, P72, P74, P112
\\
\cline{2-3}

&
1.2 Effective Communication Increasing PS and Improving Coordination Efficiency
&
P11, P19, P20, P25, P27, P29, P73, P108
\\
\hline

% ---------------- THEME 2 — LIGHT GRAY ----------------

\cellcolor{gray!15}
&
\cellcolor{gray!15}
2.1 Learning Behaviors Enhancing PS and Boosting Team Capability and Productivity
&
\cellcolor{gray!15}
P10, P15, P21, P22, P26, P31, P32, P111
\\

\hhline{|>{\arrayrulecolor{gray!15}}->{\arrayrulecolor{black}}|-|-|}

\cellcolor{gray!15}
&
\cellcolor{gray!15}
2.2 Reflective Practices Raising PS and Improving R\&D Productivity
&
\cellcolor{gray!15}
P37, P40, P46, P100, P103, P96, P79, P109
\\

\hhline{|>{\arrayrulecolor{gray!15}}->{\arrayrulecolor{black}}|-|-|}

\cellcolor{gray!15}
\multirow[c]{-3}{0.30\textwidth}[6.0ex]{%
\raggedright
2. PS-Enabled Learning, Feedback, and Knowledge Processes for Productivity
}
&
\cellcolor{gray!15}
2.3 Feedback and Problem-Solving Behaviors Reinforcing PS and Improving Knowledge-Driven Productivity
&
\cellcolor{gray!15}
P55, P59, P62, P82, P85, P101, P88, P108
\\

\arrayrulecolor{black}
\hline

% ---------------- THEME 3 — WHITE ----------------

\multirow[c]{2}{0.30\textwidth}{%
\raggedright
3. PS and Human Factors in Emotional Climate and Team Functioning
}
&
3.1 Trust and Emotional Climate Strengthening PS and Improving Team Functioning
&
P14, P17, P18, P67, P68, P64, P65, P86
\\
\cline{2-3}

&
3.2 Psychological Safety Directly Increasing Team Productivity and Performance
&
P24, P30, P33, P89, P90, P92, P97, P112
\\
\hline

% ---------------- THEME 4 — LIGHT GRAY ----------------

\cellcolor{gray!15}
&
\cellcolor{gray!15}
4.1 PS-Enabled Creativity and Learning for Productivity
&
\cellcolor{gray!15}
P23, P28, P39, P81, P110
\\

\hhline{|>{\arrayrulecolor{gray!15}}->{\arrayrulecolor{black}}|-|-|}

\cellcolor{gray!15}
\multirow[c]{-2}{0.30\textwidth}{%
\raggedright
4. PS in Creativity, Motivation and Proactive Productivity
}
&
\cellcolor{gray!15}
4.2 PS-Driven Motivation and Initiative for Productivity
&
\cellcolor{gray!15}
P36, P88, P56
\\

\arrayrulecolor{black}
\hline

% ==================================================
% RQ 2.2 — SOFTWARE QUALITY
% ==================================================

\rowcolor{yellow!25}
\multicolumn{3}{|l|}{%
\color{black}\bfseries
RQ 2.2: Identify that what are the aspects of Psychological Safety
on team performance (Software Quality)
}
\\
\hline

\rowcolor{black}
\color{white}\bfseries Themes &
\color{white}\bfseries Sub-Themes &
\color{white}\bfseries Paper IDs
\\
\hline

% ---------------- THEME 1 — LIGHT GRAY ----------------

\cellcolor{gray!15}
\multirow[c]{3}{0.30\textwidth}{%
\raggedright
1. Psychological Safety, Human Interaction, and Software Quality.
}
&
\cellcolor{gray!15}
1.1 Collaboration Strengthening PS and Improving Software Quality
&
\cellcolor{gray!15}
P7, P11, P25, P27, P28, P79, P111
\\

\hhline{|>{\arrayrulecolor{gray!15}}->{\arrayrulecolor{black}}|-|-|}

\cellcolor{gray!15}
&
\cellcolor{gray!15}
1.2 Emotional and Interpersonal Factors Increasing PS and Improving Communication Quality
&
\cellcolor{gray!15}
P9, P17, P94, P20, P89, P87
\\

\hhline{|>{\arrayrulecolor{gray!15}}->{\arrayrulecolor{black}}|-|-|}

\cellcolor{gray!15}
&
\cellcolor{gray!15}
1.3 Trust, Supportive Practices, and Structure Strengthening PS for High Software Quality
&
\cellcolor{gray!15}
P14, P16, P71, P62, P56
\\

\arrayrulecolor{black}
\hline

% ---------------- THEME 2 — WHITE ----------------

\multirow[c]{2}{0.30\textwidth}{%
\raggedright
2. PS Enabled Learning, Error Management, and Innovation Affecting Software Quality
}
&
2.1 Learning, Voice, and Error Prevention Enhancing PS and Software Quality
&
P10, P12, P13, P20, P34, P74, P102
\\
\hhline{|~|-|-|}

&
2.2 Psychological Safety Enabling Creativity and Innovation for High-Quality Performance
&
P21, P87, P89, P111
\\
\hline

% ==================================================
% RQ 2.3 — INNOVATION
% ==================================================

\rowcolor{yellow!25}
\multicolumn{3}{|l|}{%
\color{black}\bfseries
RQ 2.3: Identify that what are the aspects of Psychological Safety
on team performance (Innovation)
}
\\
\hline

\rowcolor{black}
\color{white}\bfseries Themes &
\color{white}\bfseries Sub-Themes &
\color{white}\bfseries Paper IDs
\\
\hline

% ---------------- THEME 1 — LIGHT GRAY ----------------

\cellcolor{gray!15}
&
\cellcolor{gray!15}
1.1 Idea Sharing Builds PS and Enhances Innovation
&
\cellcolor{gray!15}
P3, P9, P11, P15, P16, P17, P19, P21
\\

\hhline{|>{\arrayrulecolor{gray!15}}->{\arrayrulecolor{black}}|-|-|}

\cellcolor{gray!15}
&
\cellcolor{gray!15}
1.2 Creativity \& Error Awareness Strengthen PS and Improve Innovation Quality
&
\cellcolor{gray!15}
P23, P24, P26, P27, P29, P111
\\

\hhline{|>{\arrayrulecolor{gray!15}}->{\arrayrulecolor{black}}|-|-|}

\cellcolor{gray!15}
&
\cellcolor{gray!15}
1.3 Exploration Behaviors Increase PS and Support Innovation Capability
&
\cellcolor{gray!15}
P31, P33, P37, P38, P41
\\

\hhline{|>{\arrayrulecolor{gray!15}}->{\arrayrulecolor{black}}|-|-|}

\cellcolor{gray!15}
&
\cellcolor{gray!15}
1.4 Learning \& Knowledge Sharing Boost PS and Innovation Outcomes
&
\cellcolor{gray!15}
P46, P60, P75, P77, P78, P83
\\

\hhline{|>{\arrayrulecolor{gray!15}}->{\arrayrulecolor{black}}|-|-|}

\cellcolor{gray!15}
\multirow[c]{-5}{0.30\textwidth}[10.5ex]{%
\raggedright
1. Innovation Processes Enabled by Psychological Safety
}
&
\cellcolor{gray!15}
1.5 Core PS Mechanisms Enable Innovation and High Performance
&
\cellcolor{gray!15}
P1, P9, P16, P111
\\

\arrayrulecolor{black}
\hline

% ---------------- THEME 2 — WHITE ----------------

\multirow[c]{2}{0.30\textwidth}[-1.5ex]{%
\raggedright
2. Social and Environmental Conditions Supporting Innovation
}
&
2.1 Inclusive Climate Enhances PS and
Innovation
&
P85, P88, P90, P91, P93
\\
\cline{2-3}

&
2.2 Positive Team Conditions Strengthen PS and Improve Creative Performance
&
P102, P103, P104, P106
\\
\hline

% ==================================================
% RQ 2.4 — SATISFACTION
% ==================================================

\rowcolor{yellow!25}
\multicolumn{3}{|l|}{%
\color{black}\bfseries
RQ 2.4: Identify that what are the aspects of Psychological Safety
on team performance (Satisfaction)
}
\\
\hline

\rowcolor{black}
\color{white}\bfseries Themes &
\color{white}\bfseries Sub-Themes &
\color{white}\bfseries Paper IDs
\\
\hline

% ---------------- THEME 1 — LIGHT GRAY ----------------

\cellcolor{gray!15}
&
\cellcolor{gray!15}
1.1 Core Psychological Safety effects Satisfaction Mechanisms
&
\cellcolor{gray!15}
P11, P13, P19, P22, P30, P33, P78, P112
\\

\hhline{|>{\arrayrulecolor{gray!15}}->{\arrayrulecolor{black}}|-|-|}

\cellcolor{gray!15}
&
\cellcolor{gray!15}
1.2 Satisfaction Through Trust, Safety Climate \& Reduced Conflict
&
\cellcolor{gray!15}
P21, P24, P26, P27, P28, P45, P46, P87
\\

\hhline{|>{\arrayrulecolor{gray!15}}->{\arrayrulecolor{black}}|-|-|}

\cellcolor{gray!15}
\multirow[c]{-3}{0.30\textwidth}[1.0ex]{%
\raggedright
1. PS and Human Factors Shaping Individual and Team Satisfaction
}
&
\cellcolor{gray!15}
1.3 Individual Well-Being, Motivation \& Developer Experience results Satisfaction
&
\cellcolor{gray!15}
P17, P3, P86, P60, P74, P75
\\

\arrayrulecolor{black}
\hline

% ---------------- THEME 2 — WHITE ----------------

\multirow[c]{2}{0.30\textwidth}[-0.5ex]{%
\raggedright
2. PS Enables Performance, Structure, and Organizational Conditions Influencing Satisfaction
}
&
2.1 Satisfaction Linked to Performance, Clarity \& Reflexivity
&
P7, P39, P40, P41, P42, P32, P112
\\
\cline{2-3}

&
2.2 Organizational Climate, Norms \& Systems Supporting
&
P76, P55, P62, P103, P102, P1
\\
\hline

\end{tabular}

}

\end{table*}

\begin{comment}
  \begin{figure*}[t]
    \centering
    \includegraphics[width=0.6\textwidth, keepaspectratio]{RQ2.4.png}
    \caption{Impact of Human Factors and PS on Team Satisfaction.}
    \label{fig:8}
\end{figure*}  
\end{comment}

A total of 22 studies (20.6\%) highlight the role of \textbf{\textit{Psychological Safety and Human Factors Shaping Individual and Team Satisfaction}}. The first subtheme, \textit{Core Psychological Safety Effects Satisfaction Mechanisms}, shows that psychologically safe environments directly enhance team satisfaction by enabling open communication and supportive interactions \cite{p11,p13,p19,p22,p30,p33,p78,p112a}. The second subtheme, \textit{Satisfaction Through Trust, Safety Climate, and Reduced Conflict}, indicates that trust, positive safety climates, and reduced interpersonal conflict contribute to higher satisfaction within teams \cite{p21,p24,p26,p27,p28,p45,p46,p87}. The third subtheme, \textit{Individual Well-Being, Motivation, and Developer Experience Results Satisfaction}, demonstrates that well-being, motivation, and positive developer experiences supported by psychological safety lead to increased individual and team satisfaction \cite{p17,p3,p86,p60,p74,p75}. Overall, the findings indicate that psychological safety enhances satisfaction by promoting trust, well-being, and supportive team interactions.\vspace{5pt}

A total of 13 studies (11.6\%) highlight the role of \textbf{\textit{Psychological Safety-Enabled Performance, Structure, and Organizational Conditions Influencing Satisfaction}}. The first subtheme, \textit{Satisfaction Linked to Performance, Clarity, and Reflexivity}, shows that clear roles, team reflexivity, and effective performance processes supported by psychological safety enhance satisfaction \cite{p7,p39,p40,p41,p42,p32,p112a}. The second subtheme, \textit{Organizational Climate, Norms, and Systems Supporting Satisfaction}, indicates that supportive organizational climates, norms, and structural systems strengthen psychological safety and contribute to higher satisfaction levels \cite{p76,p55,p62,p103,p102,p1}. Overall, the findings suggest that psychological safety with higher levels of team satisfaction through supportive interpersonal relationships, trust, open communication, inclusion, and positive team conditions.\vspace{5pt}

Overall, the findings for RQ2 show that psychological safety is associated with four major aspects of team performance: productivity, software quality, innovation, and team satisfaction. Across these aspects, the reported relationships are frequently connected to team-level mechanisms such as collaboration, communication, learning, knowledge sharing, trust, feedback, and supportive interpersonal conditions. The evidence therefore suggests that psychological safety is related to both technical performance outcomes and human-centered outcomes in software engineering teams.\vspace{5pt}

%------------------Takeaway RQ2
\begin{tcolorbox}[
    colback=gray!5,
    colframe=black,
    title=Takeaways (RQ2)
]

\begin{itemize}[
    label=\raisebox{-0.12\height}{\includegraphics[width=0.42cm]{tw1.png}},
    leftmargin=0.70cm,
    labelsep=0.12cm,
    itemsep=-1pt,
parsep=0pt,
topsep=0pt,
partopsep=0pt
]

\item Psychological safety improves team productivity by enabling collaboration, open communication, learning behaviors, and knowledge sharing within software development teams.

\item Psychological safety improves software quality by encouraging error reporting, open discussion of problems, trust-based collaboration, and effective communication among team members.

\item Psychological safety promotes innovation by supporting idea sharing, creativity, experimentation, and knowledge exchange in software teams.

\item Psychological safety improves team satisfaction by enhancing trust, well-being, motivation, and positive interpersonal relationships within the team.

\item Supportive leadership, inclusive team climate, and clear team structures strengthen psychological safety and improve overall team performance.

\end{itemize}

\end{tcolorbox}
%-------------------------END RQ2----------------------------------------

%-----------------RQ3------------------------
\subsection{\textbf{Influence of Psychological Safety on Team Processes (e.g.collaboration, communication, decision-making)(RQ3):}}
RQ3 examines the relationship between psychological safety and key team processes in software engineering teams. Of the 112 included studies, 58 (51.8\%) contributed evidence addressing this research question. The thematic analysis identified three major process dimensions: collaboration, communication, and decision-making. These dimensions capture how psychological safety is associated with team interaction, information exchange, knowledge sharing, participation, and collective decision processes.\vspace{5pt}

\subsubsection{\textbf{Psychological Safety and Team Collaboration}}
For the collaboration aspect of RQ3, 53 studies (47.3\%) discuss how human factors and psychological safety influence team collaboration, while 54 studies (48.2\%) do not explicitly address this aspect. The identified studies highlight that human factors and psychological safety create an environment where team members feel comfortable sharing ideas, coordinating tasks, and engaging in collaborative problem-solving. The relevant evidence is categorized into three themes and seven subthemes. The themes include \textit{Psychological Safety Enables Open and Constructive Collaboration}, \textit{PS Supports Collaborative Learning and Knowledge Sharing}, and \textit{Human and Contextual Factors Shape Psychological Safety for Collaboration}, each consisting of subthemes that explain how human factors and psychological safety facilitate open interaction, knowledge exchange, trust, and leadership support for effective collaboration. The relationships among these themes and their associated subthemes are illustrated in Table~\ref{tab:RQ3}.\vspace{5pt}

In the first theme, \textbf{Psychological Safety Enables Open and Constructive Collaboration}, findings from 13 studies (11.6\%) indicate that human factors and psychological safety enable open and constructive collaboration within teams. The subtheme \textit{PS Enables Open Interaction for Collaboration} shows that psychologically safe environments encourage team members to share ideas and participate actively in collaborative discussions \cite{p10,p11,p24,p26,p104,p40,p109a}. The subtheme \textit{PS Enables Constructive Conflict and Interpersonal Risk-Taking for Collaboration} explains that when teams experience psychological safety, members feel comfortable expressing differing opinions and engaging in constructive conflict that improves collaborative outcomes \cite{p38,p39,p31,p40,p56,p88}. These findings suggest that human factors and psychological safety promote open interaction and constructive dialogue, strengthening collaborative processes within teams.\vspace{5pt}

\begin{comment}
    \begin{figure*}[t]
    \centering
    \includegraphics[width=0.6\textwidth, keepaspectratio]{RQ3.1.png}
    \caption{Role of Human Factors and Psychological Safety in Team Collaboration.}
    \label{fig:11}
\end{figure*}
\end{comment}

Within the second theme, \textbf{Psychological Safety Supports Collaborative Learning and Knowledge Sharing}, 15 studies (13.3\%) show that human factors and psychological safety facilitate collaborative learning and knowledge sharing within teams. The subtheme \textit{PS Supports Collaborative Learning and Knowledge Exchange} highlights that psychologically safe environments enable team members to share experiences, learn from one another, and collaboratively solve problems \cite{p15,p19,p21,p41,p76,p79,p97}. The subtheme \textit{PS Facilitates Collaborative Learning and Knowledge Exchange} further shows that open communication and mutual support encourage knowledge sharing and collective learning among team members \cite{p3,p4,p22,p34,p42,p62,p108a}. These findings demonstrate that human factors and psychological safety play an important role in promoting knowledge exchange and collaborative learning that strengthen teamwork.\vspace{5pt}

In the third theme, \textbf{Human and Contextual Factors Shape Psychological Safety for Collaboration}, evidence from 29 studies (25.8\%) indicates that human factors and contextual conditions shape psychological safety for collaboration within teams. The subtheme \textit{Human Factors Enable Constructive Collaboration} shows that interpersonal behaviors, attitudes, and team member characteristics support constructive collaborative interactions \cite{p31,p36,p37,p45,p73,p56,p29,p23,p9,p17,p34,p112a}. The subtheme \textit{Human Relationships and Trust Strengthen PS and Collaborative Capacity} highlights that trust, positive relationships, and mutual respect reinforce psychological safety and improve collaborative engagement \cite{p5,p7,p25,p55,p57,p60,p65,p67,p70,p71,p112a}. The subtheme \textit{Leadership and Team Context Cultivate PS for Effective Collaboration} explains that leadership practices and supportive team environments play a key role in fostering psychological safety and collaborative teamwork \cite{p1,p13,p74,p102,p103}. These findings indicate that human factors, relationships, and leadership context collectively shape psychological safety and strengthen collaboration within teams.\vspace{5pt}

\subsubsection{\textbf{Psychological Safety and Team Communication} }
For the communication aspect of RQ4, 41 studies (36.6\%) examine how human factors and psychological safety influence team communication, while 66 studies (58.9\%) do not explicitly address this aspect. The relevant evidence is organized into three themes and nine subthemes, as illustrated in Table~\ref{tab:RQ3}. The first theme, \textit{Psychological Safety Enables Open, Honest, and Learning-Oriented Communication}, includes four subthemes: \textit{PS Enables Open and Honest Communication}, \textit{PS Enables Mistake-Sharing and Learning-Oriented Communication}, \textit{PS Enables Constructive Feedback and Conflict Resolution}, and \textit{PS Promotes Learning-Oriented and Exploratory Communication}. The second theme, \textit{Psychological Safety Promotes Information, Idea, and Knowledge Exchange}, consists of two subthemes: \textit{PS Promotes Information, Idea, and Knowledge Sharing} and \textit{PS Strengthens Knowledge Sharing and Information Exchange}. The third theme, \textit{Human and Contextual Factors Shape Communication Participation and Quality}, includes three subthemes: \textit{Trust and Relationships Enhance Communication Effectiveness}, \textit{Individual and Emotional Factors Shape Communication Participation}, and \textit{Context and Process Conditions Shape Communication Quality}.\vspace{5pt}

In the first theme, \textbf{Psychological Safety Enables Open, Honest, and Learning-Oriented Communication}, evidence from 23 studies (20.5\%) indicates that human factors and psychological safety enable open and learning-oriented communication within teams. In the first subtheme, \textit{PS Enables Open and Honest Communication}, psychologically safe environments encourage team members to express ideas and opinions freely, improving communication transparency \cite{p4,p13,p15,p21,p23,p28,p29,p73,p111a}. The subtheme \textit{PS Enables Mistake-Sharing and Learning-Oriented Communication} highlights that psychological safety enables team members to openly discuss mistakes and learn collectively from them \cite{p3,p26,p27,p61,p76}. Constructive dialogue is further emphasized in the subtheme \textit{PS Enables Constructive Feedback and Conflict Resolution}, where psychologically safe teams are more willing to provide feedback and address disagreements productively \cite{p12,p13,p15,p16,p24,p73,p102}. Finally, within the subtheme \textit{PS Promotes Learning-Oriented and Exploratory Communication}, human factors and psychological safety encourage exploratory discussions and knowledge-seeking communication among team members \cite{p21,p39,p76,p79,p80,p82,p87}. Overall, the findings suggest that human factors and psychological safety enable transparent dialogue, constructive feedback, and learning-focused communication within teams.\vspace{5pt}

In the second theme, \textbf{Psychological Safety Promotes Information, Idea, and Knowledge Exchange}, evidence from 19 studies (16.9\%) indicates that human factors and psychological safety promote information, ideas, and knowledge exchange among team members. In the first subtheme, \textit{PS Promotes Information, Idea, and Knowledge Sharing}, psychologically safe environments encourage team members to openly share information and contribute ideas that support collaborative communication \cite{p10,p30,p33,p37,p61,p76,p112a}. The subtheme \textit{PS Strengthens Knowledge Sharing and Information Exchange} further highlights that human factors and psychological safety enhance knowledge dissemination and improve the flow of information across teams \cite{p3,p10,p26,p27,p30,p33,p37,p51,p61,p72,p76,p87}. Overall, the findings indicate that human factors and psychological safety strengthen communication by facilitating idea sharing and knowledge exchange within teams.\vspace{5pt}

\begin{comment}
    \begin{figure*}[t]
    \centering
    \includegraphics[width=0.6\textwidth, keepaspectratio]{RQ3.2.png}
    \caption{Role of Human Factors and Psychological Safety in Team Communication.}
    \label{fig:12}
\end{figure*}
\end{comment}
Within the third theme, \textbf{Human and Contextual Factors Shape Communication Participation and Quality}, 21 studies (18.7\%) examine how human factors and psychological safety influence communication participation and the overall quality of team communication. In the first subtheme, \textit{Trust and Relationships Enhance Communication Effectiveness}, psychologically safe environments built on trust and positive interpersonal relationships strengthen openness and improve the effectiveness of team communication \cite{p1,p22,p57,p60,p70,p79,p80,p94,p107}. The next subtheme, \textit{Individual and Emotional Factors Shape Communication Participation}, indicates that personal characteristics and emotional conditions influence the extent to which team members engage in communication activities \cite{p6,p17,p23,p57,p94,p97}. Within the subtheme \textit{Context and Process Conditions Shape Communication Quality}, organizational context and team processes affect the clarity, structure, and efficiency of communication practices \cite{p9,p12,p16,p51,p62,p72,p97,p102}. These results indicate that human factors, interpersonal relationships, and contextual conditions play a significant role in the formation of effective communication within teams.\vspace{5pt}

\subsubsection{\textbf{Psychological Safety and Team Decision-Making}}
For the decision-making aspect of RQ4, 17 studies (15.1\%) examine how human factors and psychological safety influence team decision-making, while 90 studies (80.3\%) do not explicitly address this aspect. The relevant findings are organized into three themes and six subthemes, as illustrated in Table~\ref{tab:RQ3}. The first theme, \textit{Psychological Safety Enables Inclusive and High-Quality Decision-Making}, includes the subthemes \textit{PS Enables Inclusive and Participative Decision-Making} and \textit{PS Improves Decision Quality through Diversity and Cohesion}. The second theme, \textit{Psychological Safety Supports Learning-Oriented and Error-Sensitive Decisions}, consists of the subthemes \textit{PS Enables Error Discussion and Constructive Questioning in Decisions} and \textit{PS Reduces Dysfunction and Supports Collective Decision Alignment}. The third theme, \textit{Psychological Safety and Team Decision-Making}, includes the subthemes \textit{Leadership and Structure Condition PS in Decision-Making} and \textit{Communication Enables Coordinated and Informed Decision-Making}.\vspace{5pt}

\begin{table*}[t]
\centering
\caption{Influence of Psychological Safety on Team Processes}
\label{tab:RQ3}

{
\fontfamily{Carlito-TLF}\selectfont
\fontsize{7.5}{8.5}\selectfont
\renewcommand{\arraystretch}{1.05}
\setlength{\tabcolsep}{2pt}

% Clear table borders
\setlength{\arrayrulewidth}{0.7pt}
\arrayrulecolor{black}

\begin{tabular}{|
>{\raggedright\arraybackslash}m{0.30\textwidth}|
>{\raggedright\arraybackslash}m{0.38\textwidth}|
>{\raggedright\arraybackslash}m{0.24\textwidth}|}
\hline

% ==================================================
% RQ3.1 — COLLABORATION
% ==================================================

\rowcolor{yellow!25}
\multicolumn{3}{|l|}{%
\color{black}\bfseries
RQ3.1: How psychological safety influences team processes (collaboration)?
}
\\
\hline

\rowcolor{black}
\color{white}\bfseries Themes &
\color{white}\bfseries Sub-Themes &
\color{white}\bfseries Paper IDs
\\
\hline

% ---------------- THEME 1 ----------------

\cellcolor{gray!15}
&
\cellcolor{gray!15}
1.1 PS Enables Open Interaction for Collaboration
&
\cellcolor{gray!15}
P10, P11, P24, P26, P104, P40, P109
\\

\hhline{|>{\arrayrulecolor{gray!15}}->{\arrayrulecolor{black}}|-|-|}

\cellcolor{gray!15}
\multirow[c]{-2}{0.30\textwidth}[0.5ex]{%
\raggedright
1. Psychological Safety Enables Open and Constructive Collaboration
}
&
\cellcolor{gray!15}
1.2 PS enables constructive conflict and interpersonal risk-taking for collaboration
&
\cellcolor{gray!15}
P38, P39, P31, P40, P56, P88
\\

\arrayrulecolor{black}
\hline

% ---------------- THEME 2 ----------------

&
2.1 PS Supports Collaborative Learning and Knowledge Exchange
&
P15, P19, P21, P41, P76, P79, P97, P104
\\
\cline{2-3}

\multirow[c]{-2}{0.30\textwidth}[-0.2ex]{%
\raggedright
2. PS Supports Collaborative Learning and Knowledge Sharing
}
&
2.2 PS facilitates collaborative learning and knowledge exchange
&
P3, P4, P22, P34, P42, P62, P108
\\
\hline

% ---------------- THEME 3 ----------------

\cellcolor{gray!15}
&
\cellcolor{gray!15}
3.1 Human Factors Enable Constructive Collaboration
&
\cellcolor{gray!15}
P31, P36, P37, P45, P73, P56, P29, P23, P9, P17, P34, P112
\\

\hhline{|>{\arrayrulecolor{gray!15}}->{\arrayrulecolor{black}}|-|-|}

\cellcolor{gray!15}
&
\cellcolor{gray!15}
3.2 Human relationships and trust strengthen PS and collaborative capacity
&
\cellcolor{gray!15}
P5, P7, P25, P55, P57, P60, P65, P67, P70, P71, P112
\\

\hhline{|>{\arrayrulecolor{gray!15}}->{\arrayrulecolor{black}}|-|-|}

\cellcolor{gray!15}
\multirow[c]{-3}{0.30\textwidth}[6.5ex]{%
\raggedright
3. Human and Contextual Factors Shape Psychological Safety for Collaboration
}
&
\cellcolor{gray!15}
3.3 Leadership and team context cultivate PS for effective collaboration
&
\cellcolor{gray!15}
P1, P13, P74, P102, P103
\\

\arrayrulecolor{black}
\hline

% ==================================================
% RQ3.2 — COMMUNICATION
% ==================================================

\rowcolor{yellow!25}
\multicolumn{3}{|l|}{%
\color{black}\bfseries
RQ3.2: How psychological safety influences team processes (communication)?
}
\\
\hline

\rowcolor{black}
\color{white}\bfseries Themes &
\color{white}\bfseries Sub-Themes &
\color{white}\bfseries Paper IDs
\\
\hline

% ---------------- THEME 1 ----------------

\cellcolor{gray!15}
&
\cellcolor{gray!15}
1.1 PS Enables Open and Honest Communication
&
\cellcolor{gray!15}
P4, P13, P15, P21, P23, P28, P29, P73, P104
\\

\hhline{|>{\arrayrulecolor{gray!15}}->{\arrayrulecolor{black}}|-|-|}

\cellcolor{gray!15}
&
\cellcolor{gray!15}
1.2 PS Enables Mistake-Sharing and Learning-Oriented Communication
&
\cellcolor{gray!15}
P3, P26, P27, P61, P76
\\

\hhline{|>{\arrayrulecolor{gray!15}}->{\arrayrulecolor{black}}|-|-|}

\cellcolor{gray!15}
&
\cellcolor{gray!15}
1.3 PS enables constructive feedback and conflict resolution
&
\cellcolor{gray!15}
P12, P13, P15, P16, P24, P73, P111
\\

\hhline{|>{\arrayrulecolor{gray!15}}->{\arrayrulecolor{black}}|-|-|}

\cellcolor{gray!15}
\multirow[c]{-4}{0.30\textwidth}[4.5ex]{%
\raggedright
1. Psychological Safety Enables Open, Honest, and Learning-Oriented Communication
}
&
\cellcolor{gray!15}
1.4 PS promotes learning-oriented and exploratory communication
&
\cellcolor{gray!15}
P21, P39, P76, P79, P80, P82, P87
\\

\arrayrulecolor{black}
\hline

% ---------------- THEME 2 ----------------

&
2.1 PS Promotes Information, Idea, and Knowledge Sharing
&
P10, P30, P33, P37, P61, P76, P112
\\
\cline{2-3}

\multirow[c]{-2}{0.30\textwidth}[0.5ex]{%
\raggedright
2. Psychological Safety Promotes Information, Idea, and Knowledge Exchange
}
&
2.2 PS strengthens knowledge sharing and information exchange
&
P3, P10, P26, P27, P30, P33, P37, P51, P61, P72, P76, P87
\\
\hline

% ---------------- THEME 3 ----------------

\cellcolor{gray!15}
&
\cellcolor{gray!15}
3.1 Trust and relationships enhance communication effectiveness
&
\cellcolor{gray!15}
P1, P22, P57, P60, P70, P79, P80, P94, P107
\\

\hhline{|>{\arrayrulecolor{gray!15}}->{\arrayrulecolor{black}}|-|-|}

\cellcolor{gray!15}
&
\cellcolor{gray!15}
3.2 Individual and emotional factors shape communication
&
\cellcolor{gray!15}
P6, P17, P23, P57, P94, P97
\\

\hhline{|>{\arrayrulecolor{gray!15}}->{\arrayrulecolor{black}}|-|-|}

\cellcolor{gray!15}
\multirow[c]{-3}{0.30\textwidth}[3.5ex]{%
\raggedright
3. Human and Contextual Factors Shape Communication Participation and Quality
}
&
\cellcolor{gray!15}
3.3 Context and process conditions shape communication quality
&
\cellcolor{gray!15}
P9, P12, P16, P51, P62, P72, P97, P102
\\

\arrayrulecolor{black}
\hline

% ==================================================
% RQ3.3 — DECISION-MAKING
% ==================================================

\rowcolor{yellow!25}
\multicolumn{3}{|l|}{%
\color{black}\bfseries
RQ3.3: How psychological safety influences team processes (decision-making)?
}
\\
\hline

\rowcolor{black}
\color{white}\bfseries Themes &
\color{white}\bfseries Sub-Themes &
\color{white}\bfseries Paper IDs
\\
\hline

% ---------------- THEME 1 ----------------

\cellcolor{gray!15}
&
\cellcolor{gray!15}
1.1 PS enables inclusive and participative decision-making
&
\cellcolor{gray!15}
P9, P11, P27, P13, P24, P26, P27, P87, P104
\\

\hhline{|>{\arrayrulecolor{gray!15}}->{\arrayrulecolor{black}}|-|-|}

\cellcolor{gray!15}
\multirow[c]{-2}{0.30\textwidth}[-0.3ex]{%
\raggedright
1. Psychological Safety Enables Inclusive and High-Quality Decision-Making
}
&
\cellcolor{gray!15}
1.2 PS improves decision quality through diversity and cohesion
&
\cellcolor{gray!15}
P19, P24, P27, P103, P104
\\

\arrayrulecolor{black}
\hline

% ---------------- THEME 2 ----------------

&
2.1 PS enables error discussion and constructive questioning in decisions
&
P10, P26, P87, P87, P103
\\
\cline{2-3}

\multirow[c]{-2}{0.30\textwidth}[1.5ex]{%
\raggedright
2. Psychological Safety Supports Learning-Oriented and Error-Sensitive Decisions
}
&
2.2 PS reduces dysfunction and supports collective decision
&
P87 P103
\\
\hline

% ---------------- THEME 3 ----------------

\cellcolor{gray!15}
&
\cellcolor{gray!15}
3.1 Leadership and structure condition PS in decision-making
&
\cellcolor{gray!15}
P1, P4, P5, P9, P31, P102
\\

\hhline{|>{\arrayrulecolor{gray!15}}->{\arrayrulecolor{black}}|-|-|}

\cellcolor{gray!15}
\multirow[c]{-2}{0.30\textwidth}[0.5ex]{%
\raggedright
3. Human and Structural Factors Shape Psychological Safety in Decision-Making
}
&
\cellcolor{gray!15}
3.2 Communication enables coordinated and informed decision-making
&
\cellcolor{gray!15}
P2, P24, P27
\\

\arrayrulecolor{black}
\hline

\end{tabular}

}

\end{table*}

\begin{comment}
\begin{figure*}[t]
    \centering
    \includegraphics[width=0.6\textwidth, keepaspectratio]{RQ3.3.png}
    \caption{Role of Human Factors and Psychological Safety in Team Decision-Making.}
    \label{fig:13}
\end{figure*}
\end{comment}
In the first theme, \textbf{Psychological Safety Enables Inclusive and High-Quality Decision-Making}, findings reported in 11 studies (9.8\%) indicate that human factors and psychological safety contribute to inclusive and high-quality decision-making within teams. In the first subtheme, \textit{PS Enables Inclusive and Participative Decision-Making}, psychologically safe environments encourage team members to share opinions and participate actively in collective decision processes \cite{p1,p11,p13,p24,p26,p27,p87,p104}. The second subtheme, \textit{PS Improves Decision Quality through Diversity and Cohesion}, highlights that psychological safety enables teams to integrate diverse perspectives while maintaining cohesion, which leads to more balanced and effective decisions \cite{p19,p24,p27,p103,p104}. These findings indicate that human factors and psychological safety support inclusive participation and enhance the quality of team decision-making.\vspace{5pt}

In the second theme, \textbf{Psychological Safety Supports Learning-Oriented and Error-Sensitive Decisions}, a set of 4 studies (3.5\%) examines how human factors and psychological safety encourage learning-oriented decision processes within teams. In the first subtheme, \textit{PS Enables Error Discussion and Constructive Questioning in Decisions}, psychologically safe environments allow team members to openly discuss mistakes and raise critical questions during decision-making processes \cite{p10,p26,p87,p103}. The next subtheme, \textit{PS Reduces Dysfunction and Supports Collective Decision Alignment}, indicates that psychological safety helps reduce dysfunctional behaviors and enables teams to reach more aligned and coordinated decisions \cite{p87,p103}. These findings suggest that human factors and psychological safety support reflective decision-making by enabling error discussion and collective alignment within teams.\vspace{5pt}

Overall, the findings for RQ3 show that psychological safety is associated with team processes through open and constructive collaboration, communication and information exchange, knowledge sharing, participation, and collective decision-making. Collaboration and communication receive comparatively greater attention in the literature, whereas decision-making is less extensively investigated. The findings further indicate that human and contextual conditions, including trust, leadership, and interpersonal relationships, shape how psychological safety is expressed through these team processes.\vspace{5pt}

\begin{tcolorbox}[
    colback=gray!5,
    colframe=black,
    title=Takeaways (RQ3)
]

\begin{itemize}[
    label=\raisebox{-0.12\height}{\includegraphics[width=0.42cm]{tw1.png}},
    leftmargin=0.70cm,
    labelsep=0.12cm,
    itemsep=-1pt,
parsep=0pt,
topsep=0pt,
partopsep=0pt
]
\item Psychological safety is strongly influenced by team culture, leadership behavior, and communication practices within software teams.
\item Supportive, inclusive, and open environments encourage speaking up, knowledge sharing, and collaboration.
\item Leadership style and organizational culture play a critical role in shaping team dynamics and psychological safety levels.
\item Negative factors such as hierarchy, blame culture, and poor communication can significantly weaken psychological safety.
\item Psychological safety directly contributes to improved team performance, innovation, and overall team effectiveness.
\end{itemize}
\end{tcolorbox}

%---------------RQ4------------------

\subsection{\textbf{Factors (e.g.leadership, organizational culture, work environment, trust and nationality and regional differences)
Influencing the Relationship Between Psychological Safety and Team Performance (RQ4)}}

RQ4 examines the contextual factors that influence the relationship between psychological safety and team performance in software engineering teams. Of the 112 included studies, 74 (66.1\%) contributed evidence addressing this research question. The thematic analysis identified several recurring contextual factors, including leadership, organizational culture, work environment, trust, nationality and regional differences, personality characteristics, and gender- and power-related dynamics. These factors illustrate the organizational, interpersonal, and individual conditions under which psychological safety is associated with team functioning and performance.\vspace{5pt}

Although leadership, culture, work environment, and trust were specified as initial areas of interest, additional contextual factors emerged inductively from the thematic analysis, including nationality and regional differences, personality characteristics, and gender- and power-related dynamics.\vspace{5pt}

\subsubsection{\textbf{Leadership Factors Associated with Psychological Safety and Team Performance}}

Regarding the role of leadership style in shaping the relationship between human factors, psychological safety, and team performance, 56 studies (5.0\%) provide relevant evidence, while 51 studies (45.5\%) do not explicitly discuss this factor. The findings are categorized into three themes and ten subthemes, summarized in Lable~\ref{tab:RQ4}. The first theme, \textit{Positive Leadership Styles Improve Psychological Safety and Team Performance}, includes the subthemes \textit{Supportive Leadership and Trust}, \textit{Inclusive Leadership Effects}, \textit{Transformational Leadership Effects}, and \textit{Empowering Leadership Effects}. The second theme, \textit{Leadership Communication Quality, Emotional Signaling, and Safety Climate Improve Psychological Safety Results and Strong Team Outcomes}, consists of the subthemes \textit{Leadership Communication Effects}, \textit{Leader’s Emotions Effects}, and \textit{Leadership and Safety Climate}. The third theme, \textit{Leadership Power and Negative Behaviors Weaken the Performance Benefits of Psychological Safety}, includes the subthemes \textit{Leadership Power Structures}, \textit{Negative Leadership Effects}, and \textit{Contextual Leadership Support}. These themes illustrate how leadership behaviors can either strengthen or weaken psychologically safe environments and their influence on team performance.\vspace{5pt}

\begin{comment}
    \begin{figure*}[t]
    \centering
    \includegraphics[width=0.6\textwidth, keepaspectratio]{RQ4.1.png}
    \caption{Role of Leadership Style in Team Performance.}
    \label{fig:14}
\end{figure*}
\end{comment}

Within the first theme, \textbf{Positive Leadership Styles Improve Psychological Safety and Team Performance}, 24 studies (21.4\%) show that leadership behaviors significantly strengthen the relationship between human factors, psychological safety, and team performance. The subtheme \textit{Supportive Leadership and Trust} indicates that leaders who provide guidance and build trust create psychologically safe environments that encourage openness and team engagement \cite{p9,p10,p11,p15,p30,p44,p45,p67,p76,p78,p104}. Another subtheme, \textit{Inclusive Leadership Effects}, demonstrates that inclusive leadership encourages participation and respect for diverse viewpoints, strengthening psychological safety and collaboration \cite{p19,p21,p60,p61,p73,p97,p102}. The subtheme \textit{Transformational Leadership Effects} shows that transformational leaders motivate and inspire teams, reinforcing psychological safety and improving performance outcomes \cite{p5,p87,p91,p110a}. Finally, the subtheme \textit{Empowering Leadership Effects} highlights that empowering leaders who promote autonomy and shared responsibility further enhance team engagement and performance \cite{p23,p24,p26,p31,p39,p88,p90}. Overall, the reviewed studies associate supportive, inclusive, and empowering leadership behaviors with stronger psychological safety, whereas hierarchical, unsupportive, or controlling leadership conditions are associated with reduced openness and participation.\vspace{5pt}

The theme \textbf{Leadership Communication Quality, Emotional Signaling, and Safety Climate Improve Psychological Safety Results and Strong Team Outcomes} is reflected in 12 studies (11.2\%), indicating that leadership communication behaviors and emotional signals influence how psychological safety develops and affects team performance. The subtheme \textit{Leadership Communication Effects} shows that clear and supportive communication from leaders promotes openness, coordination, and collaborative interaction among team members \cite{p1,p34,p36,p27,p86}. The subtheme \textit{Leader’s Emotions Effects} shows that leaders’ emotional expressions shape team climate and influence employees’ willingness to share ideas or concerns \cite{p6,p14,p82}. In addition, the subtheme \textit{Leadership and Safety Climate} demonstrates that leadership behaviors help establish a supportive safety climate that strengthens trust and participation in team processes \cite{p13,p40,p56,p110a}. These findings indicate that leadership communication quality and emotional signaling play a key role in strengthening psychological safety and improving team outcomes.\vspace{5pt}

The theme \textbf{Leadership Power and Negative Behaviors Weaken the Performance Benefits of Psychological Safety} appears in 12 studies (11.2\%), indicating that certain leadership structures and negative behaviors can reduce the positive influence of psychological safety on team performance. The subtheme \textit{Leadership Power Structures} shows that hierarchical or rigid leadership power dynamics may limit open participation and reduce psychologically safe interactions within teams \cite{p4,p47,p77,p33}. The subtheme \textit{Negative Leadership Effects} highlights that harmful leadership behaviors, such as abusive supervision or unsupportive leadership practices, can weaken psychological safety and negatively affect team outcomes \cite{p59,p80}. In addition, the subtheme \textit{Contextual Leadership Support} demonstrates that leadership support within the organizational context plays a crucial role in sustaining psychological safety and enabling teams to perform effectively \cite{p16,p46,p96,p38,p74,p103,p108a}. These findings suggest that leadership power dynamics and negative leadership behaviors may undermine psychologically safe environments and weaken team performance benefits.\vspace{5pt}

\subsubsection{\textbf{Cultural Influences on Psychological Safety and Team Performance}}

The influence of organizational and team culture on the relationship between human factors, psychological safety, and team performance is examined in 35 studies (32.7\%), while 72 studies (67.3\%) do not explicitly discuss cultural influences. The reviewed literature indicates that cultural norms, shared values, and organizational practices shape how psychological safety develops within teams and how members communicate, collaborate, and share knowledge. The findings are organized into four themes and seven subthemes, as illustrated in Table~\ref{tab:RQ4}. The first theme, \textit{Supportive, Learning, and Feedback Culture Effects Strengthen Psychological Safety and Team Performance}, includes the subthemes \textit{Supportive and Open Culture Effects} and \textit{Learning and Feedback Climate}. The second theme, \textit{Agile, Innovative, and Practice-Oriented Culture Effects Psychological Safety and High Performance}, consists of the subthemes \textit{Learning, Agile, and Innovation Culture Effects} and \textit{Culturally Shaped Practices}. The third theme, \textit{Shared, National, and Normative Culture Effects Psychological Safety and Shape Performance Consistency}, includes the subthemes \textit{Shared and Balanced Culture Effects} and \textit{National Culture Effects}. The fourth theme, \textit{Toxic and Unsupportive Culture Weakens Psychological Safety and Team Performance}, highlights the subtheme \textit{Toxic Cultural Effects}.\vspace{5pt}

The theme \textbf{Supportive, Learning, and Feedback Cultures Strengthen Psychological Safety and Team Performance} appears in 12 studies (11.2\%), showing that organizational cultures emphasizing openness, learning, and constructive feedback strengthen psychological safety and improve team outcomes. The subtheme \textit{Supportive and Open Culture Effects} indicates that cultures promoting openness, mutual respect, and supportive interaction encourage team members to share ideas and participate actively in collaborative processes \cite{p10,p11,p21,p33,p36,p40,p26,p110a}. The subtheme \textit{Learning and Feedback Climate} highlights that environments encouraging continuous learning and constructive feedback help teams identify mistakes, improve practices, and strengthen psychological safety \cite{p23,p24,p25,p59}. These findings suggest that supportive and learning-oriented cultural environments play an important role in fostering psychological safety and enhancing team performance.\vspace{5pt}

The theme \textbf{Agile and Innovative Cultures Strengthen Psychological Safety and Team Performance} is reflected in 7 studies (6.5\%), suggesting that agile and innovation-oriented cultural environments help strengthen psychological safety and improve team performance. The subtheme \textit{Learning, Agile, and Innovation Culture Effects} indicates that cultures emphasizing agility, experimentation, and continuous learning encourage team members to share ideas, adapt to change, and collaborate effectively \cite{p1,p13,p16,p73,p91,p92}. The subtheme \textit{Culturally Shaped Practices} highlights that organizational practices shaped by cultural values influence how teams coordinate work, share knowledge, and implement improvements \cite{p98,p108a}. These findings suggest that agile and innovation-driven cultural contexts support psychologically safe environments that enable teams to perform more effectively.\vspace{5pt}

The theme \textbf{Shared and National Cultures Shape Psychological Safety and Performance Consistency} appears in 14 studies (13.1\%), indicating that broader cultural norms and shared values influence how psychological safety develops and how consistently teams perform. The subtheme \textit{Shared and Balanced Culture Effects} shows that teams operating in cultures emphasizing balance, mutual respect, and shared norms tend to maintain stable collaboration and psychologically safe interactions \cite{p27,p49,p55,p62}. The subtheme \textit{National Culture Effects} highlights that national cultural differences, including communication styles, power distance, and collective values, can shape psychological safety and affect team collaboration and decision-making \cite{p2,p4,p19,p20,p31,p53,p71,p74,p102,p104}. These findings suggest that both shared organizational norms and broader national cultural contexts influence the development of psychological safety and the consistency of team performance.\vspace{5pt}

\begin{comment}
    
\begin{figure*}[t]
    \centering
    \includegraphics[width=0.6\textwidth, keepaspectratio]{RQ4.2.png}
    \caption{Organizational Cultural Effects.}
    \label{fig:15}
\end{figure*}
\end{comment}

The theme \textbf{Toxic and Unsupportive Cultures Weaken Psychological Safety and Team Performance} is reported in 2 studies (1.9\%), indicating that negative cultural environments can undermine psychological safety and reduce team effectiveness. The subtheme \textit{Toxic Cultural Effects} highlights that cultures characterized by blame, lack of support, or unhealthy interpersonal dynamics discourage open communication and reduce employees’ willingness to share ideas or concerns \cite{p60,p112a}. Overall, the evidence associates supportive, inclusive, and learning-oriented organizational cultures with psychological safety, while rigid, blame-oriented, or unsupportive cultures are associated with reduced openness and participation.\vspace{5pt}

\subsubsection{\textbf{“Work Environment” affecting the relationship between  psychological safety and team performance}}

The role of the work environment in shaping the relationship between human factors, psychological safety, and team performance is examined in 62 studies (55.3\%), while 45 studies (40.1\%) do not explicitly address this factor. The findings are organized into five themes and eight subthemes, as illustrated in Table~\ref{tab:RQ4}
The first theme, \textit{Supportive and Trust-Based Work Environment Strengthens Psychological Safety and Team Performance}, includes the subthemes \textit{Supportive, No-Blame Climate Strengthening PS} and \textit{Trust and Relational Factors Increasing PS}. The second theme, \textit{Leadership, Structure, and Autonomy in the Work Environment Enable Psychological Safety to Drive High Team Performance}, consists of the subthemes \textit{Leadership and Management Practices Shaping PS} and \textit{Organizational Structures and Autonomy Supporting PS}. The third theme, \textit{Inclusive and Value-Aligned Work Environment Sustains Psychological Safety and Consistent Team Performance}, includes the subtheme \textit{Diversity, Inclusion, and Value Alignment Affecting PS}. The fourth theme, \textit{High Emotional and Contextual Work Pressures Weaken Psychological Safety and Reduce Team Performance}, highlights the subtheme \textit{Emotional and Contextual Pressures Weakening PS}. The fifth theme, \textit{Psychologically Safe Work Environments Directly Enhance Team Functioning, Innovation, and Performance Outcomes}, includes the subthemes \textit{PS Enabling Creativity, Innovation, and Performance} and \textit{Direct PS Outcomes on Team Functioning}.\vspace{5pt}

The theme \textbf{Supportive and Trust-Based Work Environments Strengthen Psychological Safety and Team Performance} is reflected in 15 studies (13.3\%), indicating that supportive and trust-oriented environments play a key role in enhancing psychological safety and improving team outcomes. The subtheme \textit{Supportive, No-Blame Climate Strengthening PS} shows that work environments characterized by openness, fairness, and a no-blame culture encourage team members to share ideas, report mistakes, and participate actively \cite{p4,p5,p11,p13,p15,p16,p19,p61}. The subtheme \textit{Trust and Relational Factors Increasing PS} highlights that strong interpersonal relationships, mutual trust, and positive team dynamics further strengthen psychological safety and collaboration within teams \cite{p8,p23,p24,p25,p49,p50,p62,p112a}. These findings indicate that supportive climates and trust-based relationships are essential for fostering psychological safety and enhancing team performance.\vspace{5pt}

The theme \textbf{Leadership, Structure, and Autonomy in the Work Environment Enable Psychological Safety to Drive High Team Performance} appears in 16 studies (14.2\%), indicating that organizational structures, leadership practices, and autonomy within the work environment shape psychological safety and team outcomes. The subtheme \textit{Leadership and Management Practices Shaping PS} shows that effective leadership practices, including guidance, support, and clear direction, help create psychologically safe environments that enhance team coordination and performance \cite{p30,p38,p39,p47,p52,p92,p96,p43}. The subtheme \textit{Organizational Structures and Autonomy Supporting PS} highlights that flexible structures and autonomy in decision-making empower team members, encourage participation, and strengthen psychological safety within teams \cite{p30,p38,p39,p47,p52,p92,p96,p43}. These findings suggest that leadership practices and organizational design play a crucial role in enabling psychological safety and improving team performance.\vspace{5pt}

The theme \textbf{Inclusive and Value-Aligned Work Environments Sustain Psychological Safety and Consistent Team Performance} is identified in 8 studies (7.5\%), indicating that inclusive and value-driven environments contribute to maintaining psychological safety and stable team performance. The subtheme \textit{Diversity, Inclusion, and Value Alignment Affecting PS} shows that environments promoting diversity, inclusion, and shared values encourage participation, mutual respect, and alignment among team members, which strengthens psychological safety and supports consistent performance outcomes \cite{p18,p20,p22,p56,p74,p89,p60,p112a}. These findings suggest that inclusive and value-aligned work environments play a key role in sustaining psychological safety and ensuring consistent team performance.\vspace{5pt}

The theme \textbf{High Emotional and Contextual Work Pressures Weaken Psychological Safety and Reduce Team Performance} is reported in 8 studies (7.1\%), indicating that stressful and demanding work conditions can negatively affect psychological safety and team outcomes. The subtheme \textit{Emotional and Contextual Pressures Weakening PS} shows that high workload, stress, and challenging work conditions reduce openness, limit communication, and discourage participation among team members \cite{p9,p12,p17,p27,p37,p63,p87,p97}. These findings suggest that high emotional and contextual pressures in the work environment can undermine psychological safety and weaken team performance.\vspace{5pt}

\begin{comment}
\begin{figure*}[t]
    \centering
    \includegraphics[width=0.7\textwidth, keepaspectratio]{RQ4.3.png}
    \caption{Work Environment on Team Performance.}
    \label{fig:16}
\end{figure*}
\end{comment}
The theme \textbf{Psychologically Safe Work Environments Directly Enhance Team Functioning, Innovation, and Performance Outcomes} is represented in 16 studies (14.2\%), demonstrating that psychologically safe environments have a direct and positive influence on team outcomes. The subtheme \textit{PS Enabling Creativity, Innovation, and Performance} shows that such environments promote creativity, innovation, and improved performance \cite{p31,p34,p35,p42,p104,p76,p90,p106}. The second subtheme \textit{Direct PS Outcomes on Team Functioning} indicates that psychological safety directly enhances team coordination, effectiveness, and overall functioning \cite{p86,p88,p90,p91,p99,p100,p63,p82}.\vspace{5pt}

Overall, the work-environment evidence highlights trust-based, inclusive, supportive, and well-structured working conditions as recurring contextual factors associated with psychological safety and team performance, whereas high-pressure and poorly supported environments are associated with less favorable team conditions.\vspace{5pt}

\subsubsection{\textbf{“Trust” affecting the relationship between  psychological safety and team performance}}

The influence of trust on the relationship between human factors, psychological safety, and team performance is examined in 30 studies (26.7\%), while 77 studies (68.7\%) do not explicitly address this factor. The findings are organized into three themes and six subthemes, as illustrated in Table~\ref{tab:RQ4}. 
The first theme, \textit{Trust as a Psychological Safety Enabler that Improves Team Performance}, includes the subthemes \textit{Trust as a Safety Enabler} and \textit{Trust, Voice, and Fear}. The second theme, \textit{Trust in Leadership and Team Relationships Strengthens Team Performance}, consists of the subthemes \textit{Leadership-Based Trust} and \textit{Trust in Teamwork}. The third theme, \textit{Trust Makes Team Performance Strong in Both Physical and Virtual Teams}, includes the subthemes \textit{Trust, Performance, and Team Outcomes} and \textit{Trust in Virtual Teams}. These themes explain how trust strengthens psychological safety and enhances team performance across different team contexts.\vspace{5pt}

The theme \textbf{Trust as a Psychological Safety Enabler that Improves Team Performance} is reflected in 10 studies (8.9\%), indicating that trust serves as a foundational mechanism for developing psychological safety and enhancing team outcomes. The subtheme \textit{Trust as a Safety Enabler} shows that trust among team members reduces interpersonal risk and encourages open communication, which strengthens psychological safety and improves collaboration and performance \cite{p10,p15,p21,p26,p33,p39,p87,p104}. Another dimension, captured in the subtheme \textit{Trust, Voice, and Fear}, highlights that trust reduces fear of negative consequences, enabling team members to express ideas, raise concerns, and actively participate in team processes \cite{p49,p83}. These findings indicate that trust plays a critical role in enabling psychological safety and improving team performance.\vspace{5pt}

The theme \textbf{Trust in Leadership and Team Relationships Strengthens Team Performance} is identified in 8 studies (7.1\%), showing that trust embedded in leadership and team relationships enhances psychological safety and team effectiveness. The subtheme \textit{Leadership-Based Trust} indicates that trust in leaders through fair behavior, support, and reliability encourages team members to engage openly and contributes to stronger psychological safety \cite{p1,p5,p13,p64}. The subtheme \textit{Trust in Teamwork} reflects that mutual trust among team members strengthens collaboration, coordination, and shared responsibility within teams \cite{p16,p32,p52,p77}. These observations suggest that trust at both leadership and team levels plays a crucial role in reinforcing psychological safety and improving team performance.\vspace{5pt}

\begin{comment}
\begin{figure*}[t]
    \centering
    \includegraphics[width=0.5\textwidth, keepaspectratio]{RQ4.4.png}
    \caption{Influence of Trust on PS and Team Performance.}
    \label{fig:17}
\end{figure*}
\end{comment}

The theme \textbf{Trust Makes Team Performance Strong in Both Physical and Virtual Teams} is observed in 13 studies (11.6\%), indicating that trust enhances team performance across both physical and virtual work environments. The subtheme \textit{Trust, Performance, and Team Outcomes} shows that trust improves coordination, collaboration, and overall team effectiveness, leading to stronger performance outcomes \cite{p11,p19,p27,p31,p41,p50,p58,p64,p70,p71,p92}. In addition, the subtheme \textit{Trust in Virtual Teams} highlights that trust is especially critical in virtual settings, where limited face-to-face interaction requires stronger reliance on communication and mutual confidence among team members \cite{p42,p96}.\vspace{5pt}

Across the reviewed studies, trust emerges as a recurring interpersonal condition associated with psychological safety, open communication, collaboration, and team functioning in both co-located and virtual environments.\vspace{5pt}

\subsubsection{\textbf{“Nationality/ Region/Personalities” affecting the relationship between Human Factors / psychological safety and team performance}}

The influence of nationality, regional context, and personality traits on the relationship between human factors, psychological safety, and team performance is examined in 18 studies (16.0\%), while 89 studies (89.4\%) do not explicitly address this dimension. The findings are organized into three themes and six subthemes, as illustrated in Table~\ref{tab:RQ4}. 
The first theme, \textit{Cultural, National, and Regional Differences Shape Psychological Safety and Team Performance}, includes the subthemes \textit{Cultural \& National Diversity Effect} and \textit{Contextual Team Effects}. The second theme, \textit{Personality Differences Affect Team Safety and How Well Teams Work}, consists of the subthemes \textit{Personality Dynamics Effects} and \textit{Individual Characteristics Effects}. The third theme, \textit{Gender and Power-Related Dynamics Influence Psychological Safety and Performance Outcomes}, includes the subthemes \textit{Gender Bias Effects} and \textit{Conflict and Silence Effect}. These themes explain how cultural, contextual, and individual differences influence psychological safety and team performance.\vspace{5pt}

The theme \textbf{Cultural, National, and Regional Differences Shape Psychological Safety and Team Performance} is reflected in 6 studies (5.3\%), indicating that differences in cultural background and regional context influence how psychological safety develops and affects team outcomes. The subtheme \textit{Cultural \& National Diversity Effect} shows that diversity in nationality and culture shapes communication styles, interaction patterns, and collaboration within teams  \cite{p2}, \cite{p4}, \cite{p50}, \cite{p71}, \cite{p109a} . The subtheme \textit{Contextual Team Effects} highlights that team-specific contextual factors, such as work setting and regional norms, further influence team dynamics and psychological safety \cite{p81}. These findings suggest that cultural and regional differences play an important role in shaping psychological safety and team performance.\vspace{5pt}

The theme \textbf{Personality Differences Affect Team Safety and How Well Teams Work} is identified in 7 studies (6.2\%), indicating that individual personality traits influence psychological safety and team functioning. The subtheme \textit{Personality Dynamics Effects} shows that variations in personality traits shape interaction patterns, participation, and collaboration within teams \cite{p6}, \cite{p28}, \cite{p102}, \cite{p103}. The subtheme \textit{Individual Characteristics Effects} highlights that individual attributes, such as confidence, behavior, and personal tendencies, affect how team members engage, communicate, and contribute to team processes \cite{p34}, \cite{p35}, \cite{p104}. These findings suggest that personality differences play a significant role in shaping psychological safety and overall team effectiveness.\vspace{5pt}

\begin{table*}[t]
\centering
\caption{Moderating and Influencing Factors Affecting the Relationship Between Psychological Safety and Team Performance}
\label{tab:RQ4}

{
\fontfamily{Carlito-TLF}\selectfont
\fontsize{7.5}{8.5}\selectfont
\renewcommand{\arraystretch}{1.05}
\setlength{\tabcolsep}{2pt}

% Clear table borders
\setlength{\arrayrulewidth}{0.7pt}
\arrayrulecolor{black}

\begin{tabular}{|
>{\raggedright\arraybackslash}m{0.30\textwidth}|
>{\raggedright\arraybackslash}m{0.38\textwidth}|
>{\raggedright\arraybackslash}m{0.24\textwidth}|}
\hline

% ==================================================
% RQ4.1 — LEADERSHIP
% ==================================================

\rowcolor{yellow!25}
\multicolumn{3}{|l|}{%
\color{black}\bfseries
\begin{minipage}[t]{0.91\textwidth}
RQ4.1: What moderating or influencing factors (leadership style) affecting the
relationship between psychological safety and team performance?
\end{minipage}
}
\\
\hline

\rowcolor{black}
\color{white}\bfseries Themes &
\color{white}\bfseries Sub-Themes &
\color{white}\bfseries Paper IDs
\\
\hline

% ---------------- THEME 1 — LIGHT GRAY ----------------

\cellcolor{gray!15}
&
\cellcolor{gray!15}
1.1 Supportive Leadership and Trust
&
\cellcolor{gray!15}
P9, P10, P11, P15, P30, P44, P45, P67, P76, P78, P104
\\

\hhline{|>{\arrayrulecolor{gray!15}}->{\arrayrulecolor{black}}|-|-|}

\cellcolor{gray!15}
&
\cellcolor{gray!15}
1.2 Inclusive Leadership Effects
&
\cellcolor{gray!15}
P19, P21, P60, P61, P73, P97, P102
\\

\hhline{|>{\arrayrulecolor{gray!15}}->{\arrayrulecolor{black}}|-|-|}

\cellcolor{gray!15}
&
\cellcolor{gray!15}
1.3 Transformational Leadership Effects
&
\cellcolor{gray!15}
P5, P87, P91, P110
\\

\hhline{|>{\arrayrulecolor{gray!15}}->{\arrayrulecolor{black}}|-|-|}

\cellcolor{gray!15}
\multirow[c]{-4}{0.30\textwidth}[4.0ex]{%
\raggedright
1. Positive Leadership Styles Improve PS and Team Performance
}
&
\cellcolor{gray!15}
1.4 Empowering Leadership Effects
&
\cellcolor{gray!15}
P23, P24, P26, P31, P39, P88, P90
\\

\arrayrulecolor{black}
\hline

% ---------------- THEME 2 — WHITE ----------------

&
2.1 Leadership Communication Effects
&
P1, P34, P36, P27, P86
\\
\cline{2-3}

&
2.2 Leader's Emotions Effects
&
P6, P14, P82
\\
\cline{2-3}

\multirow[c]{-3}{0.30\textwidth}[-0.2ex]{%
\raggedright
2. Leadership Communication Quality, Emotional Signaling, Safety Climate Improve PS Results Strong Team Outcomes
}
&
2.3 Leadership and Safety Climate
&
P13, P40, P56, P110
\\
\hline

% ---------------- THEME 3 — LIGHT GRAY ----------------

\cellcolor{gray!15}
&
\cellcolor{gray!15}
3.1 Leadership Power Structures
&
\cellcolor{gray!15}
P4, P47, P77, P33
\\

\hhline{|>{\arrayrulecolor{gray!15}}->{\arrayrulecolor{black}}|-|-|}

\cellcolor{gray!15}
&
\cellcolor{gray!15}
3.2 Negative Leadership Effects
&
\cellcolor{gray!15}
P59, P80
\\

\hhline{|>{\arrayrulecolor{gray!15}}->{\arrayrulecolor{black}}|-|-|}

\cellcolor{gray!15}
\multirow[c]{-3}{0.30\textwidth}[-0.5ex]{%
\raggedright
3. Leadership Power and Negative Behaviors Weaken the Performance Benefits of Psychological Safety
}
&
\cellcolor{gray!15}
3.3 Contextual Leadership Support
&
\cellcolor{gray!15}
P16, P46, P96, P38, P74, P103, P108
\\

\arrayrulecolor{black}
\hline

% ==================================================
% RQ4.2 — CULTURE
% ==================================================

\rowcolor{yellow!25}
\multicolumn{3}{|l|}{%
\color{black}\bfseries
\begin{minipage}[t]{0.91\textwidth}
RQ4.2: What moderating or influencing factors (culture) affecting the
relationship between psychological safety and team performance?
\end{minipage}
}
\\
\hline

\rowcolor{black}
\color{white}\bfseries Themes &
\color{white}\bfseries Sub-Themes &
\color{white}\bfseries Paper IDs
\\
\hline

% ---------------- THEME 1 — LIGHT GRAY ----------------

\cellcolor{gray!15}
&
\cellcolor{gray!15}
1.1 Supportive and Open Culture Effects
&
\cellcolor{gray!15}
P10, P11, P21, P33, P36, P40, P26, P110
\\

\hhline{|>{\arrayrulecolor{gray!15}}->{\arrayrulecolor{black}}|-|-|}

\cellcolor{gray!15}
\multirow[c]{-2}{0.30\textwidth}[-0.5ex]{%
\raggedright
1. Supportive, Learning, and Feedback Culture Effects Strength PS and Team Performance
}
&
\cellcolor{gray!15}
1.2 Learning and Feedback Climate
&
\cellcolor{gray!15}
P23, P24, P25, P59
\\

\arrayrulecolor{black}
\hline

% ---------------- THEME 2 — WHITE ----------------

&
2.1 Learning, Agile, and Innovation Culture Effects
&
P1, P13, P16, P73, P91, P92
\\
\cline{2-3}

\multirow[c]{-2}{0.30\textwidth}[-0.5ex]{%
\raggedright
2. Agile and Innovative Cultures Strengthen Psychological Safety and Team Performance
}
&
2.2 Culturally Shaped Practices
&
P98, P108
\\
\hline

% ---------------- THEME 3 — LIGHT GRAY ----------------

\cellcolor{gray!15}
&
\cellcolor{gray!15}
3.1 Shared and Balanced Culture Effects
&
\cellcolor{gray!15}
P27, P49, P55, P62
\\

\hhline{|>{\arrayrulecolor{gray!15}}->{\arrayrulecolor{black}}|-|-|}

\cellcolor{gray!15}
\multirow[c]{-2}{0.30\textwidth}[0.5ex]{%
\raggedright
3. Shared and National Cultures Shape Psychological Safety and Performance Consistency
}
&
\cellcolor{gray!15}
3.2 National Culture Effects
&
\cellcolor{gray!15}
P2, P4, P19, P20, P31, P53, P71, P74, P102, P104
\\

\arrayrulecolor{black}
\hline

% ---------------- THEME 4 — WHITE ----------------

\raggedright
4. Toxic and Unsupportive Culture weak PS and team Performance
&
4.1 Toxic Cultural Effects
&
P60, P112
\\
\hline

% ==================================================
% RQ4.3 — WORK ENVIRONMENT
% ==================================================

\rowcolor{yellow!25}
\multicolumn{3}{|l|}{%
\color{black}\bfseries
RQ4.3: What influencing factors (Work Environment) affecting the
relationship between psychological safety and team performance?
}
\\
\hline

\rowcolor{black}
\color{white}\bfseries Themes &
\color{white}\bfseries Sub-Themes &
\color{white}\bfseries Paper IDs
\\
\hline

% ---------------- THEME 1 — LIGHT GRAY ----------------

\cellcolor{gray!15}
&
\cellcolor{gray!15}
1.1 Supportive, No-Blame Climate Strengthening PS
&
\cellcolor{gray!15}
P4, P5, P11, P13, P15, P16, P19, P61
\\

\hhline{|>{\arrayrulecolor{gray!15}}->{\arrayrulecolor{black}}|-|-|}

\cellcolor{gray!15}
\multirow[c]{-2}{0.30\textwidth}[-0.4ex]{%
\raggedright
1. Supportive and Trust-Based Work Environment Strengthens PS and Team Performance
}
&
\cellcolor{gray!15}
1.2 Trust \& Relational Factors Increasing PS
&
\cellcolor{gray!15}
P8, P23, P24, P25, P49, P50, P62, P112
\\

\arrayrulecolor{black}
\hline

% ---------------- THEME 2 — WHITE ----------------

\multirow[c]{2}{0.30\textwidth}{%
\raggedright
2. Leadership, Structure, and Autonomy in the Work Environment Enable PS to Drive High Team Performance
}
&
2.1 Leadership \& Management Practices Shaping PS
&
P30, P38, P39, P47, P52, P92, P96, P43
\\
\cline{2-3}

&
2.2 Organizational Structures \& Autonomy Supporting PS
&
P30, P38, P39, P47, P52, P92, P96, P43
\\
\hline

% ---------------- THEME 3 — LIGHT GRAY ----------------

\cellcolor{gray!15}
3. Inclusive and Value-Aligned Work Environment Sustains PS and Consistent Team Performance
&
\cellcolor{gray!15}
3.1 Diversity, Inclusion \& Value Alignment Affecting PS
&
\cellcolor{gray!15}
P18, P20, P22, P56, P74, P89, P60, P112
\\
\hline

% ---------------- THEME 4 — WHITE ----------------

4. High Emotional and Contextual Work Pressures Weaken PS and Reduce Team Performance
&
4.1 Emotional \& Contextual Pressures Weakening PS
&
P9, P12, P17, P27, P37, P63, P87, P97
\\
\hline

% ---------------- THEME 5 — LIGHT GRAY ----------------

\cellcolor{gray!15}
&
\cellcolor{gray!15}
\rule{0pt}{3.3ex}%
5.1 PS Enabling Creativity, Innovation \& Performance
&
\cellcolor{gray!15}
P31, P34, P35, P42, P104, P76, P90, P106
\\

\hhline{|>{\arrayrulecolor{gray!15}}->{\arrayrulecolor{black}}|-|-|}

\cellcolor{gray!15}
\multirow[c]{-2}{0.30\textwidth}[2.0ex]{%
\raggedright
5. Psychologically Safe Work Environments Directly Enhance Team Functioning,
Innovation, and Performance Outcomes
}
&
\cellcolor{gray!15}
\rule{0pt}{3.3ex}%
5.2 Direct PS Outcomes on Team Functioning
&
\cellcolor{gray!15}
P86, P88, P90, P91, P99, P100, P63, P82
\\

\arrayrulecolor{black}
\hline

% ==================================================
% RQ4.4 — TRUST
% ==================================================

\rowcolor{yellow!25}
\multicolumn{3}{|l|}{%
\color{black}\bfseries
RQ4.4: What influencing factor (Trust) affecting the relationship
between psychological safety and team performance?
}
\\
\hline

\rowcolor{black}
\color{white}\bfseries Themes &
\color{white}\bfseries Sub-Themes &
\color{white}\bfseries Paper IDs
\\[-1pt]
\hline

% ---------------- THEME 1 — LIGHT GRAY ----------------

\cellcolor{gray!15}
&
\cellcolor{gray!15}
1.1 Trust as Safety Enabler
&
\cellcolor{gray!15}
P10, P15, P21, P26, P33, P39, P87, P104
\\

\hhline{|>{\arrayrulecolor{gray!15}}->{\arrayrulecolor{black}}|-|-|}

\cellcolor{gray!15}
\multirow[c]{-2}{0.30\textwidth}[0ex]{%
\raggedright
1. Trust as a Psychological Safety Enabler that Improves Team Performance
}
&
\cellcolor{gray!15}
1.2 Trust, Voice and Fear
&
\cellcolor{gray!15}
P49, P83
\\

\arrayrulecolor{black}
\hline

% ---------------- THEME 2 — WHITE ----------------

&
2.1 Leadership-Based Trust
&
P1, P5, P13, P64
\\
\cline{2-3}

\multirow[c]{-2}{0.30\textwidth}[0ex]{%
\raggedright
2. Trust in Leadership and Team Relationships Strengthens Team Performance
}
&
2.2 Trust in Teamwork
&
P16, P32, P52, P77
\\
\hline

% ---------------- THEME 3 — LIGHT GRAY ----------------

\cellcolor{gray!15}
&
\cellcolor{gray!15}
3.1 Trust, Performance \& Team Outcomes
&
\cellcolor{gray!15}
P11, P19, P27, P31, P41, P50, P58, P64, P70, P71, P92
\\

\hhline{|>{\arrayrulecolor{gray!15}}->{\arrayrulecolor{black}}|-|-|}

\cellcolor{gray!15}
\multirow[c]{-2}{0.30\textwidth}[0ex]{%
\raggedright
3. Trust Makes Team Performance Strong in Both Physical and Virtual Teams
}
&
\cellcolor{gray!15}
3.2 Trust in Virtual Teams
&
\cellcolor{gray!15}
P42, P96
\\

\arrayrulecolor{black}
\hline

% ==================================================
% RQ4.5 — NATIONALITY, REGION AND PERSONALITY
% ==================================================

\rowcolor{yellow!25}
\multicolumn{3}{|l|}{%
\color{black}\bfseries
RQ4.5: What influencing factors (Nationality, Region and Personality)
affecting the relationship between psychological safety and team performance?
}
\\
\hline

\rowcolor{black}
\color{white}\bfseries Themes &
\color{white}\bfseries Sub-Themes &
\color{white}\bfseries Paper IDs
\\
\hline

% ---------------- THEME 1 — LIGHT GRAY ----------------

\cellcolor{gray!15}
&
\cellcolor{gray!15}
1.1 Cultural \& National Diversity Effect
&
\cellcolor{gray!15}
P2, P4, P50, P71, P109
\\

\hhline{|>{\arrayrulecolor{gray!15}}->{\arrayrulecolor{black}}|-|-|}

\cellcolor{gray!15}
\multirow[c]{-2}{0.30\textwidth}[0ex]{%
\raggedright
1. Cultural, National, and Regional Differences Shape PS and Team Performance
}
&
\cellcolor{gray!15}
1.2 Contextual Team Effects
&
\cellcolor{gray!15}
P81
\\

\arrayrulecolor{black}
\hline

% ---------------- THEME 2 — WHITE ----------------

&
2.2 Personality Dynamics Effects
&
P6, P28, P102, P103
\\
\cline{2-3}

\multirow[c]{-2}{0.30\textwidth}[0ex]{%
\raggedright
2. Personality Differences Affect Team Safety and How Well Teams Work
}
&
2.3 Individual Characteristics Effects
&
P34, P35, P104
\\
\hline

% ---------------- THEME 3 — LIGHT GRAY ----------------

\cellcolor{gray!15}
&
\cellcolor{gray!15}
3.1 Gender Bias Effects
&
\cellcolor{gray!15}
P20, P56, P94, P99
\\

\hhline{|>{\arrayrulecolor{gray!15}}->{\arrayrulecolor{black}}|-|-|}

\cellcolor{gray!15}
\multirow[c]{-2}{0.30\textwidth}[0ex]{%
\raggedright
3. Gender and Power-Related Dynamics in Psychological Safety and Team Performance
}
&
\cellcolor{gray!15}
3.2 Conflict and Silence Effect
&
\cellcolor{gray!15}
P24
\\

\arrayrulecolor{black}
\hline

\end{tabular}

}

\end{table*}

\begin{comment}
\begin{figure*}[t]
    \centering
    \includegraphics[width=0.5\textwidth, keepaspectratio]{RQ4.5.png}
    \caption{Effect of Nationality, Region and Personality on PS and Team Performance.}
    \label{fig:18}
\end{figure*}
\end{comment}

The theme \textbf{Gender and Power-Related Dynamics in Psychological Safety and Team Performance} is represented in 5 studies (4.4\%), indicating that gender-related issues and power dynamics shape psychological safety and team outcomes. The subtheme \textit{Gender Bias Effects} shows that gender bias and unequal power relations can influence participation, communication, and inclusion within teams \cite{p20}, \cite{p56}, \cite{p94}, \cite{p99}. The subtheme \textit{Conflict and Silence Effect} highlights that power imbalances and social dynamics may lead to silence, reduced voice, and limited participation among team members \cite{p24}.\vspace{5pt}

Overall, the findings for RQ4 indicate that the relationship between psychological safety and team performance is shaped by multiple contextual conditions. Leadership, organizational culture, work environment, and trust receive the greatest attention, while nationality, regional differences, personality characteristics, and gender- and power-related dynamics represent additional but less extensively investigated influences. These findings demonstrate that psychological safety should be understood within the organizational, interpersonal, and sociocultural contexts in which software teams operate.\vspace{5pt}

\begin{tcolorbox}[
    colback=gray!5,
    colframe=black,
    title=Takeaways (RQ4)
]

\begin{itemize}[
    label=\raisebox{-0.12\height}{\includegraphics[width=0.42cm]{tw1.png}},
    leftmargin=0.70cm,
    labelsep=0.12cm,
    itemsep=-1pt,
parsep=0pt,
topsep=0pt,
partopsep=0pt
]
\item Leadership emerges as an important contextual factor, with supportive, inclusive, and empowering leadership frequently associated with stronger psychological safety.
\item Organizational culture and work environment shape the conditions in which psychological safety develops, particularly through openness, inclusion, learning orientation, workload, and organizational support.
\item Trust is a recurring interpersonal condition associated with open communication, collaboration, and effective team functioning.
\item Nationality, regional differences, personality characteristics, and gender- and power-related dynamics receive comparatively less attention but indicate that psychological safety may vary across individual and sociocultural contexts.
\item Overall, the evidence indicates that the relationship between psychological safety and team performance is context-dependent rather than uniform across software engineering teams
\end{itemize}
\end{tcolorbox}

%-----------RQ6--------------

\subsection{\textbf{Challenges or barriers to establishing psychological safety in software development teams (RQ 5)}:}

This research question examines the key barriers that hinder the development of psychological safety in software development teams. The analysis includes 89 studied papers (79.4\%), while 23 papers were not discussed, indicating strong coverage of challenges in the literature. By conducting thematic analysis, the literature was organized into six major themes and multiple sub-themes. The identified themes include Communication \& Information Flow Barriers, Cultural \& Interpersonal Trust Challenges, Structural \& Organizational Barriers, Leadership \& Management Issues, Diversity, Bias, and Inclusion Barriers, and Workload, Pressure \& Resource Constraints as illustrated in Table~\ref{tab:rq5_barriers}.\vspace{5pt}
\begin{table*}[t]
\centering
\caption{Challenges and Barriers to Establishing Psychological Safety in Software Development Teams}
\label{tab:rq5_barriers}

{
\fontfamily{Carlito-TLF}\selectfont
\fontsize{7.5}{8.5}\selectfont
\renewcommand{\arraystretch}{1.05}
\setlength{\tabcolsep}{2pt}

% Clear table borders
\setlength{\arrayrulewidth}{0.7pt}
\arrayrulecolor{black}

\begin{tabular}{|
>{\raggedright\arraybackslash}m{0.27\textwidth}|
>{\raggedright\arraybackslash}m{0.36\textwidth}|
>{\raggedright\arraybackslash}m{0.29\textwidth}|}
\hline

% ==================================================
% RQ HEADING
% ==================================================

\rowcolor{yellow!25}
\multicolumn{3}{|l|}{%
\color{black}\bfseries
RQ5: What challenges or barriers to establishing psychological safety
in software development teams?
}
\\
\hline

% ==================================================
% COLUMN HEADINGS
% ==================================================

\rowcolor{black}
\color{white}\bfseries Themes &
\color{white}\bfseries Sub-Themes &
\color{white}\bfseries Paper IDs
\\
\hline

% ==================================================
% THEME 1 — LIGHT GRAY
% ==================================================

\cellcolor{gray!15}
&
\cellcolor{gray!15}
1.1 Silence, Withholding, and Lack of Open Dialogue
&
\cellcolor{gray!15}
P3, P15, P24, P28, P30, P34, P41, P57, P79, P87, P91, P108
\\

\hhline{|>{\arrayrulecolor{gray!15}}->{\arrayrulecolor{black}}|-|-|}

\cellcolor{gray!15}
&
\cellcolor{gray!15}
1.2 Communication \& Coordination Breakdowns
&
\cellcolor{gray!15}
P2, P4, P8, P12, P31, P51, P55, P112
\\

\hhline{|>{\arrayrulecolor{gray!15}}->{\arrayrulecolor{black}}|-|-|}

\cellcolor{gray!15}
&
\cellcolor{gray!15}
1.3 Silence, Conflict \& Interpersonal Barriers Undermining Psychological Safety
&
\cellcolor{gray!15}
P18, P19, P20, P24, P27, P28, P29, P34, P59, P62
\\

\hhline{|>{\arrayrulecolor{gray!15}}->{\arrayrulecolor{black}}|-|-|}

\cellcolor{gray!15}
&
\cellcolor{gray!15}
1.4 Diversity, Cultural \& Multi-Group Communication Barriers
&
\cellcolor{gray!15}
P45, P48, P102
\\

\hhline{|>{\arrayrulecolor{gray!15}}->{\arrayrulecolor{black}}|-|-|}

\cellcolor{gray!15}
&
\cellcolor{gray!15}
1.5 Distributed, Virtual, and Technology-Driven Barriers
&
\cellcolor{gray!15}
P2, P8, P33, P46, P51, P95, P96
\\

\hhline{|>{\arrayrulecolor{gray!15}}->{\arrayrulecolor{black}}|-|-|}

\cellcolor{gray!15}
\multirow[c]{-6}{0.27\textwidth}[8.5ex]{%
\raggedright
1. Communication \& Information Flow Barriers
}
&
\cellcolor{gray!15}
1.6 Technology, Tools, and Contextual Interference
&
\cellcolor{gray!15}
P33, P57, P62, P88, P100
\\

\arrayrulecolor{black}
\hline

% ==================================================
% THEME 2 — WHITE
% ==================================================

&
2.1 Trust, Familiarity \& Relationship Breakdown
&
P4, P7, P22, P32, P40, P42, P46, P51, P53, P65, P67, P71, P95
\\
\cline{2-3}

&
2.2 Communication, Conflict \& Coordination Barriers
&
P17, P24, P29, P38, P62, P72, P64, P82, P86
\\
\cline{2-3}

&
2.3 Structural, Cultural, Leadership \& Inclusion Barriers
&
P9, P13, P16, P45, P60, P78, P80, P89, P96
\\
\cline{2-3}

&
2.4 Fear-Driven Barriers to Openness, Learning \& Contribution
&
P10, P11, P30, P31, P41, P42, P76, P79, P83, P87, P98
\\
\cline{2-3}

&
2.5 Hierarchy, Blame \& Structural Power Dynamics Undermining Safety
&
P15, P21, P26, P73, P91, P108
\\
\cline{2-3}

&
2.6 Trust, Belonging \& Relational Climate
&
P65, P67, P71, P82, P98
\\
\cline{2-3}

\multirow[c]{-7}{0.27\textwidth}[12.5ex]{%
\raggedright
2. Cultural \& Interpersonal Trust Challenges
}
&
2.7 Feedback and Learning Culture Challenges
&
P23, P24, P59, P62, P64, P77, P82, P100, P104
\\
\hline

% ==================================================
% THEME 3 — LIGHT GRAY
% ==================================================

\cellcolor{gray!15}
&
\cellcolor{gray!15}
3.1 Organizational and Structural Constraints
&
\cellcolor{gray!15}
P27, P39, P75, P81, P90, P93, P94, P103
\\

\hhline{|>{\arrayrulecolor{gray!15}}->{\arrayrulecolor{black}}|-|-|}

\cellcolor{gray!15}
\multirow[c]{-2}{0.27\textwidth}[0.5ex]{%
\raggedright
3. Structural \& Organizational Barriers
}
&
\cellcolor{gray!15}
3.2 Role, Structure, and Goal Ambiguity
&
\cellcolor{gray!15}
P5, P20, P29, P38, P53, P56, P74, P85, P100, P103, P112
\\

\arrayrulecolor{black}
\hline

% ==================================================
% THEME 4 — WHITE
% ==================================================

&
4.1 Leadership Gaps and Poor Management Practices
&
P13, P47, P55, P60, P67, P73, P80, P94, P103
\\
\cline{2-3}

&
4.2 Fear-Driven Barriers to Openness, Learning \& Contribution
&
P10, P11, P30, P31, P41, P42, P76, P79, P83, P87, P98
\\
\cline{2-3}

&
4.3 Hierarchy, Blame Culture \& Structural Power Imbalance
&
P15, P21, P26, P73, P91, P92, P94
\\
\cline{2-3}

\multirow[c]{-4}{0.27\textwidth}[4.5ex]{%
\raggedright
4. Leadership \& Management Issues
}
&
4.4 Trust, Belonging, Relationship Climate \& Support
&
P28, P65, P67, P71, P82
\\
\hline

% ==================================================
% THEME 5 — LIGHT GRAY
% ==================================================

\cellcolor{gray!15}
&
\cellcolor{gray!15}
5.1 Gender Bias, Gendered Barriers \& Women's Participation
&
\cellcolor{gray!15}
P1, P99
\\

\hhline{|>{\arrayrulecolor{gray!15}}->{\arrayrulecolor{black}}|-|-|}

\cellcolor{gray!15}
&
\cellcolor{gray!15}
5.2 Bias, Prejudice \& Unsafe Norms Undermining Psychological Safety
&
\cellcolor{gray!15}
P61, P97, P60, P94
\\

\hhline{|>{\arrayrulecolor{gray!15}}->{\arrayrulecolor{black}}|-|-|}

\cellcolor{gray!15}
\multirow[c]{-3}{0.27\textwidth}[3.5ex]{%
\raggedright
5. Diversity, Bias, and Inclusion Barriers
}
&
\cellcolor{gray!15}
5.3 Cultural Resistance, Trust Issues \& Collaboration Barriers
&
\cellcolor{gray!15}
P16, P89, P110
\\

\arrayrulecolor{black}
\hline

% ==================================================
% THEME 6 — WHITE
% ==================================================

&
6.1 High Pressure, Stress \& Weak Support Undermining Safety
&
P9, P13, P28, P101
\\
\cline{2-3}

&
6.2 Communication, Coordination \& Collaboration Barriers
&
P18, P19, P31
\\
\cline{2-3}

&
6.3 Resource, Skill \& Constraint-Based Limitations
&
P74, P90
\\
\cline{2-3}

&
6.4 Demographic Barriers \& Participation Limitations
&
P99
\\
\cline{2-3}

\multirow[c]{-5}{0.27\textwidth}[4.5ex]{%
\raggedright
6. Workload, Pressure \& Resource Constraints
}
&
6.5 Contextual Factors Affecting Findings
&
P39
\\
\hline

\end{tabular}

}

\end{table*}

\begin{comment}
\begin{figure*}[ht]
    \centering
    \includegraphics[width=1.05\linewidth]{RQ5.png}
    \caption{Barriers to PS in Software Development teams}
    \label{fig:19}
\end{figure*}
\end{comment}
In the first theme, \textbf{Communication \& Information Flow Barriers}, a total of 45 studies (37.5\%) highlight that breakdowns in communication significantly hinder the development of psychological safety in teams. The subtheme \textit{Silence, Withholding, and Lack of Open Dialogue}, shows that team members often hesitate to speak up due to fear or discomfort \cite{p3,p15,p24,p28,p30,p34,p41,p57,p79,p87,p91,p108a}. The subtheme \textit{Communication \& Coordination Breakdowns} indicates that ineffective coordination and unclear communication reduce team alignment \cite{p2,p4,p8,p12,p31,p51,p55,p112a}. Additionally, \textit{Silence, Conflict \& Interpersonal Barriers Undermining Psychological Safety} emphasizes how interpersonal conflicts and communication avoidance weaken team trust and collaboration \cite{p18,p19,p20,p24,p27,p28,p29,p34,p59,p62}. Further, \textit{Diversity, Cultural \& Multi-Group Communication Barriers} reveal challenges arising from diverse team compositions \cite{p45,p48,p102}, while \textit{Distributed, Virtual, and Technology-Driven Barriers} highlight limitations in remote and technology-mediated communication \cite{p2,p8,p33,p46,p51,p95,p96}. The subtheme \textit{Technology, Tools, and Contextual Interference} reflects how tools and contextual constraints disrupt effective information flow \cite{p33,p57,p62,p88,p100}. These findings indicate that communication-related barriers are a critical obstacle to psychological safety, as they limit openness, reduce participation, and negatively impact team coordination and performance.\vspace{5pt}

Overall, communication and information-flow barriers are characterized by silence, information withholding, limited opportunities for open dialogue, and coordination difficulties, which the reviewed studies associate with reduced participation, openness, and interpersonal trust within software teams.\vspace{5pt}

In the second theme, \textbf{Cultural \& Interpersonal Trust Challenges}, a total of 46 studies (41.0\%) demonstrate that breakdowns in trust, relationships, and team dynamics act as major barriers to psychological safety. Within RQ5, trust is considered from a barrier perspective, particularly where distrust, limited familiarity, or fear restrict open participation and interpersonal interaction. The subtheme \textit{Trust, Familiarity \& Relationship Breakdown} indicates that weak interpersonal relationships and lack of familiarity reduce openness \cite{p4,p7,p22,p32,p40,p42,p46,p51,p53,p65,p67,p71,p95}. The subtheme \textit{Communication, Conflict \& Coordination Barriers} highlights how unresolved conflicts and poor coordination disrupt team interaction \cite{p17,p24,p29,p38,p62,p72,p64,p82,p109a}. Further, the subtheme \textit{Structural, Cultural, Leadership \& Inclusion Barriers} reflects how organizational and cultural conditions limit inclusivity \cite{p9,p13,p16,p45,p60,p78,p80,p89,p96}, while the subtheme \textit{Fear-Driven Barriers to Openness, Learning \& Contribution} shows that fear of evaluation or failure suppresses participation \cite{p10,p11,p30,p31,p41,p42,p76,p79,p83,p87,p98}. The subtheme \textit{Hierarchy, Blame \& Structural Power Dynamics} further emphasizes how power imbalances discourage voice and risk-taking \cite{p15,p21,p26,p73,p91,p108a}. The subtheme, \textit{Trust, Belonging \& Relational Climate} highlights the importance of inclusion and belonging \cite{p65,p67,p71,p82,p98}, and \textit{Feedback and Learning Culture Challenges} point to difficulties in establishing open feedback environments \cite{p23,p24,p59,p62,p64,p77,p82,p100,p104}. Overall, the findings identify distrust, limited familiarity, fear-driven team conditions, and unsupportive cultural norms as recurring barriers associated with reduced openness, participation, and collaboration in software teams.\vspace{5pt}

In the third theme, \textbf{Structural \& Organizational Barriers}, a total of 18 studies (16.0\%) indicate that organizational structures and unclear roles act as key obstacles to establishing psychological safety. The subtheme \textit{Organizational and Structural Constraints} highlights how rigid structures, policies, and organizational limitations restrict open communication and participation \cite{p27,p39,p75,p81,p90,p93,p94,p103}. The subtheme \textit{Role, Structure, and Goal Ambiguity} shows that unclear roles, responsibilities, and goals create confusion and reduce team confidence in speaking up \cite{p5,p20,p29,p38,p53,p56,p74,p85,p100,p103,p112a}. Overall, the findings identify rigid organizational arrangements, unclear roles and responsibilities, restrictive processes, and structural conditions that limit autonomy or participation as recurring organizational barriers to psychological safety in software teams.\vspace{5pt}

In the fourth theme, \textbf{Leadership \& Management Issues}, a total of 28 studies (25.0\%) indicate that ineffective leadership and management practices significantly hinder the establishment of psychological safety. The subtheme \textit{Leadership Gaps and Poor Management Practices} highlight how inadequate leadership behaviors and weak management structures limit support and openness \cite{p13,p47,p55,p60,p67,p73,p80,p94,p103}. The subtheme \textit{Fear-Driven Barriers to Openness, Learning \& Contribution} shows that fear of negative consequences discourages participation and knowledge sharing \cite{p10,p11,p30,p31,p41,p42,p76,p79,p83,p87,p98}. The subtheme, \textit{Hierarchy, Blame Culture \& Structural Power Imbalance} demonstrates how hierarchical systems and blame-oriented cultures suppress voice and innovation \cite{p15,p21,p26,p73,p91,p92,p94}, while \textit{Trust, Belonging, Relationship Climate \& Support} emphasizes the lack of supportive relational environments \cite{p28,p65,p67,p71,p82}. Overall, the findings identify insufficient managerial support, restrictive or controlling leadership practices, and ineffective responses to employee voice, feedback, and mistakes as recurring leadership-related barriers to psychological safety in software teams.\vspace{5pt}

In the fifth theme, \textbf{Diversity, Bias, and Inclusion Barriers}, a total of 9 studies (8.0\%) indicate that diversity-related challenges and bias significantly hinder psychological safety in teams. The subtheme \textit{Gender Bias, Gendered Barriers \& Women’s Participation} highlights how gender-related challenges limit participation and voice \cite{p1,p99}. The reviewed studies identify gender bias and gender-related barriers as factors associated with unequal participation and reduced opportunities for open contribution within software teams. The subtheme \textit{Bias, Prejudice \& Unsafe Norms Undermining Psychological Safety} shows that discriminatory behaviors and unsafe norms reduce openness and inclusion \cite{p61,p97,p60,p94}. Bias, prejudice, and unsafe team norms were reported as barriers associated with reduced inclusion, openness, and willingness to participate. Additionally, \textit{Cultural Resistance, Trust Issues \& Collaboration Barriers} reflects how cultural differences and resistance to inclusion weaken collaboration and trust \cite{p16,p89,p110a}. These results show that bias and inclusion-related barriers remain underexplored yet critical factors that negatively affect psychological safety and team performance. Overall, the findings identify gender bias, prejudice, exclusionary or unsafe team norms, and cultural resistance as diversity- and inclusion-related barriers to psychological safety. Although these issues receive comparatively limited attention in the reviewed literature, the available evidence associates them with reduced inclusion, openness, and participation within software teams.\vspace{5pt}

In the sixth theme, \textbf{Workload, Pressure \& Resource Constraints}, a total of 11 studies (9.8\%) indicate that workload-related and contextual pressures act as significant barriers to psychological safety.  The subtheme \textit{High Pressure, Stress \& Weak Support Undermining Safety} shows that excessive workload and lack of support reduce team members’ willingness to speak up \cite{p9,p13,p28,p101}. The subtheme \textit{Communication, Coordination \& Collaboration Barriers} highlights how pressure affects effective interaction and teamwork \cite{p18,p19,p31}. The subtheme, \textit{Resource, Skill \& Constraint-Based Limitations} reflects how lack of resources and skills restrict team functioning \cite{p74,p90}, while \textit{Demographic Barriers \& Participation Limitations} and \textit{Contextual Factors Affecting Findings} indicate that participation constraints and situational factors further influence psychological safety \cite{p99,p39}. Overall, the findings identify excessive workload, time pressure, work-related stress, and resource constraints as recurring work-environment barriers associated with less favorable conditions for psychological safety. These pressures may limit opportunities for open communication, reflection, participation, and supportive interpersonal interaction within software teams.\vspace{5pt}

Overall, the findings for RQ5 show that barriers to psychological safety in software engineering teams arise across multiple levels. Communication and information-flow problems restrict openness and information exchange; cultural and interpersonal trust challenges affect team relationships; structural and organizational conditions constrain participation and autonomy; leadership and management issues influence employee voice; diversity-, bias-, and inclusion-related barriers affect equitable participation; and workload, pressure, and resource constraints create additional challenges for maintaining supportive team interactions. These findings indicate that barriers to psychological safety are not limited to individual behavior but are embedded in interpersonal, organizational, and work-environment conditions.\vspace{5pt}

\begin{tcolorbox}[
    colback=gray!5,
    colframe=black,
    title=Takeaways (RQ5)
]

\begin{itemize}[
    label=\raisebox{-0.12\height}{\includegraphics[width=0.42cm]{tw1.png}},
    leftmargin=0.70cm,
    labelsep=0.12cm,
    itemsep=-1pt,
parsep=0pt,
topsep=0pt,
partopsep=0pt
]

\item 	Communication breakdowns and lack of open dialogue are the most dominant barriers to psychological safety.
\item Trust issues and fear-driven environments significantly limit participation and knowledge sharing.
\item Leadership gaps, hierarchy, and blame culture suppress openness and team voice.
\item Organizational constraints and unclear roles reduce coordination and team confidence.
\item Diversity and workload-related barriers are less studied but represent important research gaps.
\end{itemize}
\end{tcolorbox}
%---------RQ 6---------------

\subsection{\textbf{Strategies, Practices, and Interventions for Enhancing Psychological Safety in Software Teams (RQ6):}}

This research question examines the strategies, practices, and interventions used to enhance psychological safety in software development teams. The analysis includes 88 studied papers (78.5\%), while 19 papers were not discussed, highlighting strong research attention on improvement approaches. Through thematic analysis, the findings are organized into five major themes, including \textit{Leadership and Organizational Practices That Strengthen Psychological Safety in Software Teams}, \textit{Agile, Role, and Process Practices That Embed Psychological Safety into Daily Team Work}, \textit{Learning, Feedback, and Capability-Building Practices That Increase Psychological Safety}, \textit{How Teams Build Trust and Belonging Through Everyday Practices}, and \textit{Well-Being, Stress, and Technology Practices That Sustain Psychological Safety Over Time}. These are further detailed through thirteen subthemes, as illustrated in the corresponding Table~\ref{tab:rq6_strategies}.\vspace{5pt}

\textbf{Leadership and Organizational Practices That Strengthen Psychological Safety in Software Teams} is supported by 22 studies, highlighting how leadership and organizational approaches contribute to safer team environments. The subtheme \textit{Leadership Practices That Foster Psychological Safety} includes studies \cite{p1,p10,p11,p13,p23,p26,p31,p73,p87,p92,p103,p104}, showing that supportive, open, and inclusive leadership behaviors encourage speaking up and participation. The role of policies is reflected in subtheme \textit{Organizational Policies and HR Practices That Support Psychological Safety} \cite{p51,p95,p75,p81,p91,p103}, where structured support systems and HR initiatives help institutionalize psychological safety. Additionally, \textit{Motivation and Reward Practices That Support Team Safety} \cite{p38,p102,p74,p111a} demonstrates how recognition, incentives, and fair evaluation systems reinforce safe behaviors and team engagement. These insights indicate that leadership behaviors and formal organizational systems jointly create conditions that enable openness, trust, and sustained team performance.\vspace{5pt}

\begin{table*}[t]
\centering
\caption{Strategies, Practices, and Interventions for Enhancing Psychological Safety in Software Teams}
\label{tab:rq6_strategies}

{
\fontfamily{Carlito-TLF}\selectfont
\fontsize{7.5}{8.5}\selectfont
\renewcommand{\arraystretch}{1.05}
\setlength{\tabcolsep}{2pt}

% Clear table borders
\setlength{\arrayrulewidth}{0.7pt}
\arrayrulecolor{black}

\begin{tabular}{|
>{\raggedright\arraybackslash}m{0.30\textwidth}|
>{\raggedright\arraybackslash}m{0.38\textwidth}|
>{\raggedright\arraybackslash}m{0.24\textwidth}|}
\hline

% ==================================================
% RQ HEADING
% ==================================================

\rowcolor{yellow!25}
\multicolumn{3}{|l|}{%
\color{black}\bfseries
RQ 6: What strategies, practices, or interventions does the paper report for enhancing psychological safety in software teams?
}
\\
\hline

% ==================================================
% COLUMN HEADINGS
% ==================================================

\rowcolor{black}
\color{white}\bfseries Themes &
\color{white}\bfseries Sub-Themes &
\color{white}\bfseries Paper IDs
\\
\hline

% ==================================================
% THEME 1 — LIGHT GRAY
% ==================================================

\cellcolor{gray!15}
&
\cellcolor{gray!15}
1.1 Leadership Practices That Foster PS
&
\cellcolor{gray!15}
P1, P10, P11, P13, P23, P26, P31, P73, P87, P92, P103, P104
\\

\hhline{|>{\arrayrulecolor{gray!15}}->{\arrayrulecolor{black}}|-|-|}

\cellcolor{gray!15}
&
\cellcolor{gray!15}
1.2 Organizational Policies and HR Practices That Support PS
&
\cellcolor{gray!15}
P51, P95, P75, P81, P91, P103
\\

\hhline{|>{\arrayrulecolor{gray!15}}->{\arrayrulecolor{black}}|-|-|}

\cellcolor{gray!15}
\multirow[c]{-3}{0.30\textwidth}[2.0ex]{%
\raggedright
1: Leadership and Organizational Practices That Strengthen PS in Software Teams
}
&
\cellcolor{gray!15}
1.3 Motivation and Reward Practices That Support Team Safety
&
\cellcolor{gray!15}
P38, P102, P74, P111
\\

\arrayrulecolor{black}
\hline

% ==================================================
% THEME 2 — WHITE
% ==================================================

&
2.1 Agile Practices That Enhance PS
&
P1, P4, P27, P73, P85, P108
\\
\cline{2-3}

&
2.2 Role Clarity, Task Rotation, and Collective Ownership
&
P1, P5, P7, P20, P29, P56, P64, P85
\\
\cline{2-3}

\multirow[c]{-3}{0.30\textwidth}[0.5ex]{%
\raggedright
2: Agile, Role, and Process Practices That Embed PS into Daily Team Work
}
&
2.3 Tools and Processes That Support PS
&
P62, P12, P59, P70, P72, P92
\\
\hline

% ==================================================
% THEME 3 — LIGHT GRAY
% ==================================================

\cellcolor{gray!15}
&
\cellcolor{gray!15}
3.1 Feedback and Learning Practices That Promote PS
&
\cellcolor{gray!15}
P1, P12, P23, P27, P59, P64, P100, P101
\\

\hhline{|>{\arrayrulecolor{gray!15}}->{\arrayrulecolor{black}}|-|-|}

\cellcolor{gray!15}
\multirow[c]{-2}{0.30\textwidth}[-0.2ex]{%
\raggedright
3: Learning, Feedback, and Capability-Building Practices That Increase PS
}
&
\cellcolor{gray!15}
3.2 Training and Coaching Interventions for PS
&
\cellcolor{gray!15}
P5, P16, P60, P97, P47, P80, P92, P85
\\

\arrayrulecolor{black}
\hline

% ==================================================
% THEME 4 — WHITE
% ==================================================

&
4.1 Team-Building and Relationship-Building Practices for PS
&
P2, P22, P32, P40, P46, P96, P99, P112
\\
\cline{2-3}

&
4.2 Inclusion and Diversity Practices for Team Safety
&
P1, P60, P61, P45, P89, P99, P97
\\
\cline{2-3}

\multirow[c]{-3}{0.30\textwidth}[0.8ex]{%
\raggedright
4: How Teams Build Trust and Belonging Through Everyday Practices
}
&
4.3 Team Norms and Practices That Build PS
&
P11, P26, P27, P31, P42, P87
\\
\hline

% ==================================================
% THEME 5 — LIGHT GRAY
% ==================================================

\cellcolor{gray!15}
&
\cellcolor{gray!15}
5.1 Stress Management and Resilience Practices for Team Safety
&
\cellcolor{gray!15}
P18, P9, P74, P78, P90, P101, P108
\\

\hhline{|>{\arrayrulecolor{gray!15}}->{\arrayrulecolor{black}}|-|-|}

\cellcolor{gray!15}
\multirow[c]{-2}{0.30\textwidth}[-0.2ex]{%
\raggedright
5: Well-Being, Stress, and Technology Practices That Sustain PS Over Time
}
&
\cellcolor{gray!15}
5.2 AI Interaction Practices That Support PS
&
\cellcolor{gray!15}
P57, P62
\\

\arrayrulecolor{black}
\hline

\end{tabular}

}

\end{table*}
\begin{comment}
\begin{figure*}[t]
    \centering
    \includegraphics[width=0.8\textwidth, keepaspectratio]{RQ6.png}
    \caption{Strategies and Practices for Enhancing Psychological Safety in Software Teams.}
    \label{fig:20}
\end{figure*}
\end{comment}
\textbf{Agile, Role, and Process Practices That Embed Psychological Safety into Daily Teamwork} is supported by 18 studies, showing how everyday practices and structures integrate psychological safety into team operations. The subtheme \textit{Agile Practices That Enhance Psychological Safety} includes studies \cite{p1,p4,p27,p73,p85,p108a}, indicating that agile methods promote openness, continuous feedback, and team interaction. The subtheme \textit{Role Clarity, Task Rotation, and Collective Ownership Strategies} is reflected in \cite{p1,p5,p7,p20,p29,p56,p64,p85}, where clear roles and shared responsibilities help reduce ambiguity and encourage participation. Furthermore, \textit{Tools and Processes That Support Psychological Safety} \cite{p62,p12,p59,p70,p72,p92}, highlights how collaboration tools and structured workflows facilitate communication, coordination, and knowledge sharing within teams. These findings indicate that embedding psychological safety into daily routines, roles, and tools enables consistent team interaction, reduces uncertainty, and strengthens collaborative performance.\vspace{5pt}

\textbf{Learning, Feedback, and Capability-Building Practices That Increase Psychological Safety} is supported by 16 studies, showing how continuous learning and development practices strengthen safe team environments. The subtheme \textit{Feedback and Learning Practices That Promote Psychological Safety} includes studies \cite{p1,p12,p23,p27,p59,p64,p100,p101}, emphasizing that open feedback, reflection, and shared learning processes encourage voice and improvement. The subtheme \textit{Training and Coaching Interventions for Psychological Safety} is reflected in \cite{p5,p16,p60,p97,p47,p80,p92,p85}, where structured training and coaching support skill development, confidence, and interpersonal understanding within teams. These findings suggest that continuous learning and feedback mechanisms build competence and confidence, enabling teams to engage more openly and collaboratively.\vspace{5pt}

\textbf{How Teams Build Trust and Belonging Through Everyday Practices} draws on 19 studies, illustrating how routine team interactions and social practices contribute to psychological safety. The subtheme \textit{Team building and Relationship-Building Practices for Psychological Safety} includes studies \cite{p2,p22,p32,p40,p46,p96,p99,p112a}, highlighting the role of relationship-building activities in strengthening trust among team members. The subtheme \textit{Inclusion and Diversity Practices for Team Safety} appears in \cite{p1,p60,p61,p45,p89,p99,p97}, where inclusive approaches foster a sense of belonging and encourage broader participation. In addition, the theme, \textit{Team Norms and Practices That Build Psychological Safety} \cite{p11,p26,p27,p31,p42,p87} emphasizes how shared norms and agreed team behaviors promote respectful communication and collaboration. Together, these insights show that trust, inclusion, and everyday team norms are central to maintaining psychological safety and strengthening team cohesion.\vspace{5pt}

\textbf{Well-Being, Stress, and Technology Practices That Sustain Psychological Safety Over Time} draws on 9 studies, showing how long-term conditions and emerging practices influence psychological safety in teams. The subtheme \textit{Stress Management and Resilience Practices for Team Safety} includes studies \cite{p18,p9,p74,p78,p90,p101,p108a}, highlighting how managing workload, reducing stress, and building resilience help maintain a safe and supportive team climate. The subtheme \textit{AI Interaction Practices That Support Psychological Safety} is reflected in \cite{p57,p62}, where the use of AI tools and technologies can assist communication, collaboration, and team support mechanisms. Taken together, these insights suggest that sustaining psychological safety requires ongoing attention to well-being, stress reduction, and the effective integration of supportive technologies.\vspace{5pt}

\begin{tcolorbox}[
    colback=gray!5,
    colframe=black,
    title=Takeaways (RQ6)
]

\begin{itemize}[
    label=\raisebox{-0.12\height}{\includegraphics[width=0.42cm]{tw1.png}},
    leftmargin=0.70cm,
    labelsep=0.12cm,
    itemsep=-1pt,
parsep=0pt,
topsep=0pt,
partopsep=0pt
]
\item 	Leadership behaviors and organizational support systems are central to creating and sustaining psychological safety.
\item Embedding PS into daily practices (agile methods, roles, tools) ensures consistent team interaction and collaboration.
\item Continuous feedback, learning, and coaching strengthen confidence and encourage open participation.
\item Trust, inclusion, and shared team norms play a key role in fostering belonging and team cohesion.
\item Long-term sustainability of PS depends on managing stress, supporting well-being, and leveraging enabling technologies.
\end{itemize}
\end{tcolorbox}

\section{Discussion}
This section discusses the study findings in relation to research questions and the existing literature on psychological safety and human factors in software development teams. Although the results section presented the thematic findings, this section interprets these findings and explains their implications for team processes, team performance, and organizational practices. The discussion is organized according to the research questions, focusing on how psychological safety is conceptualized, how it influences team performance and team processes, what factors influence it, what barriers exist, and what strategies and practices can be used to improve psychological safety in software development teams.
\subsection{\textbf{Discussion of RQ1:}}
The findings for RQ1 show that psychological safety is conceptualized across the literature through seven interconnected dimensions involving collaboration, communication, team dynamics, social environment, learning and knowledge sharing, interpersonal openness, and individual security and well-being.

\noindent
\raisebox{-0.22\height}{%
\includegraphics[height=1.8em]{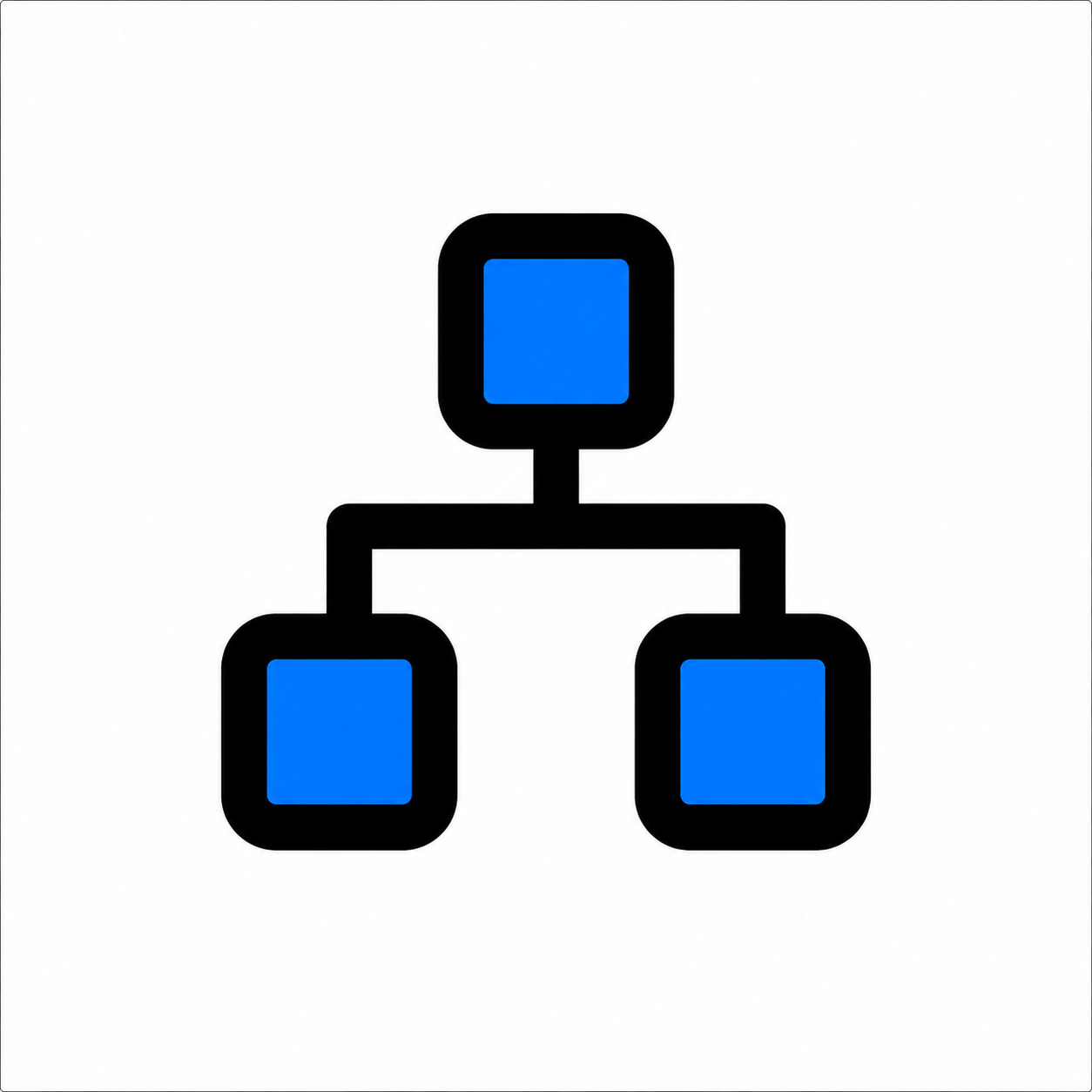}%
}%
\hspace{0.15em}%
{\textbf{Theme 1:}}
Previous studies \cite{p3} associate collaboration with psychological safety through knowledge sharing and collective problem-solving. The findings of this SMS reinforce this relationship, showing that collaborative practices and trust-based interactions frequently co-occur with psychologically safe team environments. However, the literature largely focuses on the positive aspects of collaboration, while challenges such as conflict, unequal participation, and power imbalance receive limited attention. This suggests that existing research may present an overly optimistic perspective, lacking critical examination of real-world team complexities.

\noindent
\raisebox{-0.22\height}{%
\includegraphics[height=1.8em]{themeP.png}%
}%
\hspace{0.15em}%
{\textbf{Theme 2:}} Communication emerges as the most prominent dimension in the conceptualization of psychological safety across the reviewed studies. As highlighted in prior work, supportive leadership and open dialogue enable individuals to express ideas and take interpersonal risks without fear \cite{p8}. The findings therefore suggest that communication-related behaviors, including speaking up, open dialogue, and freedom of expression, are central characteristics through which psychological safety is represented in software teams. Nevertheless, the strong emphasis on communication may reflect an imbalance in the literature, with structural and organizational dimensions receiving comparatively less attention. This raises concerns about whether communication alone is sufficient to ensure psychological safety in all contexts.

\noindent
\raisebox{-0.22\height}{%
\includegraphics[height=1.8em]{themeP.png}%
}%
\hspace{0.15em}%
{\textbf{Theme 3:}}
The role of team dynamics and human factors is consistently recognized in shaping psychological safety. Research has shown that leadership behavior and interpersonal relationships significantly influence team participation and engagement \cite{p2}. The findings further show that leadership behaviors, interpersonal relationships, and individual characteristics form part of the broader team context through which psychological safety is interpreted in software engineering research  \cite{p25}. However, limited attention is given to individual differences, such as personality traits, cultural diversity, or experience levels. This suggests that the current literature tends to generalize human factors rather than addressing their context-specific impacts.

\noindent
\raisebox{-0.22\height}{%
\includegraphics[height=1.8em]{themeP.png}%
}%
\hspace{0.15em}%
{\textbf{Theme 4:}}
The social and organizational environment emerges as an important contextual dimension of psychological safety, particularly through trust-based and supportive climates. Studies such as \cite{p21}demonstrate that trust enhances open communication and collaboration, while supportive environments further strengthen team cohesion \cite{p79}. The findings align with these perspectives; however, many studies focus on ideal organizational conditions and overlook constraints such as hierarchical structures, workload pressures, and organizational politics. As a result, the applicability of these findings in complex real-world settings remains uncertain.

\noindent
\raisebox{-0.22\height}{%
\includegraphics[height=1.8em]{themeP.png}%
}%
\hspace{0.15em}%
{\textbf{Theme 5:}}
Psychological safety is strongly associated with learning and knowledge sharing within teams. Prior research \cite{p4} indicates that safe environments encourage individuals to share information and learn from mistakes, which is further supported by findings in \cite{p10}. The results reinforce the association between psychological safety, continuous learning, and knowledge-sharing behaviors. However, literature primarily focuses on outcomes rather than explaining the mechanisms through which psychological safety enables learning. This highlights a gap between theoretical understanding and practical implementation.

\noindent
\raisebox{-0.22\height}{%
\includegraphics[height=1.8em]{themeP.png}%
}%
\hspace{0.15em}%
{\textbf{Theme 6:}}
The findings associate psychological safety with interpersonal openness, including feedback exchange, idea contribution, and shared team dialogue. Evidence from existing studies \cite{p28} supports the role of psychological safety in facilitating open communication, while \cite{p32} highlights its importance in achieving team consensus. Despite its relevance, this theme is comparatively underexplored in literature. This indicates that aspects such as constructive disagreement and honest feedback require further investigation to better understand their role in team performance.

\noindent
\raisebox{-0.22\height}{%
\includegraphics[height=1.8em]{themeP.png}%
}%
\hspace{0.15em}%
{\textbf{Theme 7:}}
Psychological safety is also associated with individual security and well-being by reducing concerns about negative interpersonal consequences and threats to professional identity. Studies such as \cite{p95} show that psychologically safe environments enhance role comfort, while \cite{p43} links psychological safety to employee well-being and organizational outcomes.The findings support these insights by associating psychological safety with greater role comfort, confidence, well-being, and engagement. However, the literature tends to prioritize team-level outcomes over individual experiences, suggesting a need for more fine-grained analysis at the individual level.\vspace{5pt}

Overall, the findings show that psychological safety is conceptualized in software engineering as a multidimensional team-level construct centered on interpersonal risk-taking and expressed through open communication, collaboration, learning and knowledge sharing, supportive social environments, interpersonal openness, and individual security and well-being. Communication receives the greatest attention across the reviewed literature, while individual and contextual dimensions are comparatively less developed. Although these dimensions are strongly interconnected, the literature often examines them separately, indicating a need for more integrated conceptual models of psychological safety in software engineering.\vspace{5pt}

\noindent
\raisebox{-0.22\height}{%
\includegraphics[height=1.4em]{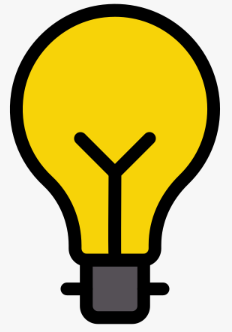}%
}%
\hspace{0.15em}%
\textbf{Practical implications:} Organizations seeking to foster psychological safety should pay attention to communication quality, collaborative norms, trust-based interactions, and supportive leadership, as these dimensions consistently characterize psychologically safe software teams. Clear norms for speaking up, feedback, and inclusion may further support these conditions.\vspace{5pt}

\noindent
\raisebox{-0.22\height}{%
\includegraphics[height=1.4em]{TA1.png}%
}%
\hspace{0.15em}%
\textbf{Theoretical implications:}Future research should develop more integrative and theory-driven models that explain how communication, collaboration, social context, learning, interpersonal openness, and individual-level experiences jointly shape psychological safety across different software engineering contexts.\vspace{5pt}

\subsection{\textbf{Discussion of RQ2:}}
The findings for RQ2 show that psychological safety is associated with four major aspects of team performance: productivity, software quality, innovation, and team satisfaction. The following discussion interprets these relationships and highlights the mechanisms, contextual conditions, and research gaps identified across the literature.\vspace{5pt}

\subsubsection{\textbf{Psychological Safety and Team Productivity (RQ 2.1)}}
The findings for team productivity show that psychological safety is associated with productivity-related outcomes through multiple team and organizational mechanisms. The evidence is organized into four major themes: Team Dynamics and Psychological Safety, Leadership and Management Practices, Organizational and Structural Factors, and Team Performance and Outcomes. Together, these themes indicate that the relationship between psychological safety and team productivity is primarily reflected through team interaction, supportive leadership, organizational conditions, collaboration, communication, and performance-related processes.

\noindent
\raisebox{-0.22\height}{%
\includegraphics[height=1.8em]{themeP.png}%
}%
\hspace{0.15em}%
{\textbf{Theme 1: }}
Collaboration and communication emerge as key mechanisms associated with the relationship between psychological safety and team productivity. Prior studies [1] highlight that collaborative environments encourage collective problem-solving, while effective communication improves coordination and information sharing \cite{p12}. The findings indicate that psychologically safe teams are frequently associated with more open interaction, coordination, and information sharing, which the reviewed literature links to productivity-related outcomes. However, literature often assumes that collaboration and communication automatically lead to better outcomes, without considering challenges such as miscommunication, coordination overhead, or conflict. This suggests a need for more nuanced research on how these factors operate under different team conditions.

\noindent
\raisebox{-0.22\height}{%
\includegraphics[height=1.8em]{themeP.png}%
}%
\hspace{0.15em}%
{\textbf{Theme 2:}}
Learning-oriented behaviors and feedback mechanisms emerge as the most dominant contributors to productivity. Research shows that psychological safety supports learning and capability development \cite{p9}, while reflective practices enhance innovation and R\&D productivity \cite{p35}. Additionally, feedback and problem-solving behaviors strengthen knowledge-driven productivity \cite{p46}. The findings reinforce that learning and knowledge-sharing processes are central pathways linking psychological safety to performance. Nevertheless, most studies focus on the presence of these processes rather than their effectiveness or quality. The reviewed evidence suggests a gap in understanding how learning practices translate into measurable productivity gains.

\noindent
\raisebox{-0.22\height}{%
\includegraphics[height=1.8em]{themeP.png}%
}%
\hspace{0.15em}%
{\textbf{Theme 3:}}
Human factors such as trust and emotional climate are also frequently associated with team productivity in the reviewed literature. Studies such as \cite{p13}demonstrate that trust strengthens team functioning, while psychologically safe environments directly support performance outcomes \cite{p22}. The findings suggest that emotional and relational aspects are critical for enabling effective teamwork. However, literature tends to generalize these factors without considering individual differences or cultural influences. This limits the understanding of how emotional climate varies across teams and affects productivity in diverse contexts.

\noindent
\raisebox{-0.22\height}{%
\includegraphics[height=1.8em]{themeP.png}%
}%
\hspace{0.15em}%
{\textbf{Theme 4:}}
Creativity, motivation, and proactive behaviors are identified as additional pathways through which psychological safety influences productivity. Prior work \cite{p19} indicates that psychologically safe environments encourage creativity and innovation, while motivation and initiative further enhance productivity \cite{p33}. The findings support that psychological safety enables individuals to contribute beyond routine tasks. However, these aspects are comparatively underrepresented in literature. This suggests that current research places greater emphasis on collaboration and learning, while the role of motivation and creativity in driving productivity remains insufficiently explored.

Overall, the productivity-related evidence suggests that psychological safety is associated with productivity primarily through collaboration, communication, learning, feedback, knowledge sharing, trust, and proactive behavior. However, the literature provides stronger evidence for the presence of these mechanisms than for their direct and measurable effects on productivity, highlighting a need for more outcome-oriented and context-sensitive empirical research.

\subsubsection{\textbf{Psychological Safety and Software Quality (RQ 2.2)}}
The findings for software quality show that psychological safety is associated with quality-related outcomes through both interpersonal and learning-oriented mechanisms. The evidence is organized into two major themes: Human Interaction Factors and Software Quality, and Learning, Error Management, and Innovation. Together, these themes indicate that the relationship between psychological safety and software quality is primarily reflected through team interaction, communication, trust, learning, knowledge sharing, and error-management processes.

\noindent
\raisebox{-0.22\height}{%
\includegraphics[height=1.8em]{themeP.png}%
}%
\hspace{0.15em}%
{\textbf{Theme 1:}}
Human interaction factors, particularly collaboration, communication, and trust, emerge as prominent mechanisms associated with software quality. Prior studies \cite{p3}  indicate that collaborative environments supported by psychological safety encourage open dialogue and early error reporting, which are associated with software quality outcomes. Similarly, emotional and interpersonal factors are linked to communication effectiveness and team coordination \cite{p7}. The findings therefore suggest that psychologically safe environments may support software quality through improved interaction and coordination within development teams. However, the literature often assumes that improved interaction translates directly into better software quality without sufficiently examining potential challenges such as miscommunication, coordination complexity, or conflicting team priorities. This suggests that the relationship between psychological safety, team interaction, and software quality may be more complex and context-dependent than currently portrayed.

\noindent
\raisebox{-0.22\height}{%
\includegraphics[height=1.8em]{themeP.png}%
}%
\hspace{0.15em}%
{\textbf{Theme 2:}}
Learning behaviors, error management, and innovation emerge as additional mechanisms through which the literature connects psychological safety with software quality. Research indicates that psychologically safe teams are more likely to speak up, share knowledge, and identify errors early in the development process \cite{p4}. Psychological safety is also associated with creativity and innovation, which may support the development of higher-quality solutions \cite{p15}. The findings therefore indicate that learning-oriented and innovation-related practices represent another pathway linking psychological safety with software quality. However, these mechanisms receive comparatively less attention than interpersonal factors. This imbalance suggests that existing research emphasizes collaboration, communication, and trust while providing less evidence on how learning, error management, and innovation translate into measurable software-quality outcomes. 

Taken together, these findings suggest that the relationship between psychological safety and software quality is predominantly indirect, operating through team interaction, learning, error-management, and knowledge-sharing processes rather than through a direct technical mechanism.\vspace{5pt}

\subsubsection{\textbf{Psychological Safety and Innovation (RQ 2.3)}}
The innovation-related findings show that psychological safety is associated with innovation through both team-level innovation processes and broader social and environmental conditions. The evidence is organized into two major themes: Innovation Processes Enabled by Psychological Safety and Social and Environmental Conditions Supporting Innovation. Together, these themes highlight the roles of idea sharing, creativity, experimentation, learning, knowledge sharing, inclusion, and supportive team environments in innovation-related outcomes.

\noindent
\raisebox{-0.22\height}{%
\includegraphics[height=1.8em]{themeP.png}%
}%
\hspace{0.15em}%
{\textbf{Theme 1:}}
Psychological safety is strongly associated with innovation processes such as idea sharing, creativity, experimentation, and learning. Prior studies \cite{p7} show that psychologically safe environments encourage open idea sharing, while creativity and awareness of errors further enhance innovation quality \cite{p19}.The findings suggest that psychological safety is associated with exploration, experimentation, and knowledge-sharing behaviors that the reviewed literature links to innovative outcomes. However, literature often assumes that increased idea sharing, and experimentation automatically lead to successful innovation. This indicates that the relationship between psychological safety and innovation may depend not only on generating and sharing ideas but also on teams' capacity to evaluate, implement, and sustain those ideas.

\noindent
\raisebox{-0.22\height}{%
\includegraphics[height=1.8em]{themeP.png}%
}%
\hspace{0.15em}%
{\textbf{Theme 2:}}
Supportive social and organizational environments emerge as important contextual conditions associated with innovation. Research indicates that inclusive climates promote idea expression and participation \cite{p36}, while positive team conditions strengthen creative performance \cite{p41}. The findings indicate that innovation is associated not only with individual and team behaviors but also with broader social and organizational conditions. Nevertheless, these conditions are relatively underexplored compared to process-related factors. This suggests that the role of organizational context, including culture, structure, and leadership support, requires deeper investigation to fully understand how innovation emerges in psychologically safe teams.\vspace{5pt}

Overall, the innovation-related evidence suggests that psychological safety is associated with innovation through idea sharing, creativity, experimentation, learning, and knowledge sharing, while supportive and inclusive environments provide important contextual conditions for these processes. However, the literature places considerably greater emphasis on innovation-related team behaviors than on the organizational conditions under which these behaviors translate into successful innovation. Future research should therefore examine not only whether psychological safety encourages innovative behaviors, but also when and under what contextual conditions these behaviors result in implemented and measurable innovation outcomes.\vspace{5pt}

\subsubsection{\textbf{Psychological Safety and Team Satisfaction (RQ 2.4)}}
The satisfaction-related findings show that psychological safety is associated with both individual and team satisfaction through interpersonal, structural, and organizational conditions. The evidence is organized into two major themes: Psychological Safety and Human Factors Shaping Satisfaction, and Performance, Structure, and Organizational Conditions. Together, these themes highlight the roles of communication, trust, well-being, team processes, role clarity, and supportive organizational environments in satisfaction-related outcomes.

\noindent
\raisebox{-0.22\height}{%
\includegraphics[height=1.8em]{themeP.png}%
}%
\hspace{0.15em}%
{\textbf{Theme 1:}}
Psychological safety and related human factors are frequently associated with both individual and team satisfaction. Prior studies \cite{p12} link psychologically safe environments with satisfaction through open communication and supportive interactions, while trust and positive safety climates are also associated with favorable team experiences \cite{p15}. The findings further identify reduced conflict, trust, and well-being as recurring mechanisms associated with satisfaction. However, the literature often assumes a relatively direct relationship between psychological safety and satisfaction without sufficiently considering potential moderating factors such as workload, organizational pressure, or individual expectations. This suggests that satisfaction outcomes may vary across different team and organizational contexts.

\noindent
\raisebox{-0.22\height}{%
\includegraphics[height=1.8em]{themeP.png}%
}%
\hspace{0.15em}%
{\textbf{Theme 2:}}
Structural and organizational conditions represent another dimension associated with satisfaction. Prior research \cite{p3} links clear roles and effective team processes with satisfaction, while supportive organizational climates and team norms are also associated with more positive experiences \cite{p1}. These findings suggest that psychological safety operates alongside structural and organizational conditions rather than independently in shaping satisfaction-related outcomes. Nevertheless, these contextual conditions receive comparatively limited attention in the reviewed literature. Further research is therefore needed to examine how organizational structures, team arrangements, and contextual conditions interact with psychological safety across different software engineering environments.

Overall, the findings for RQ2 show that psychological safety is associated with multiple dimensions of team performance, including productivity, software quality, innovation, and team satisfaction. Across these outcomes, collaboration, communication, learning, knowledge sharing, trust, and supportive team conditions emerge as recurring mechanisms through which psychological safety is related to performance. However, the literature is unevenly distributed across these dimensions, with greater attention to productivity and learning-related mechanisms than to innovation, satisfaction, and contextual influences. This imbalance highlights the need for more integrated and outcome-specific research on how psychological safety relates to different dimensions of software team performance.\vspace{5pt}

\noindent
\raisebox{-0.22\height}{%
\includegraphics[height=1.4em]{TA1.png}%
}%
\hspace{0.15em}%
\textbf{Practical implications:} Practical implications: Organizations should consider psychological safety when designing collaborative practices, feedback and learning processes, and mechanisms for knowledge sharing and error reporting. Attention to experimentation, trust, supportive interaction, and employee well-being may further help create conditions associated with both technical and human-centered performance outcomes.\vspace{5pt}

\noindent
\raisebox{-0.22\height}{%
\includegraphics[height=1.4em]{TA1.png}%
}%
\hspace{0.15em}%
\textbf{Theoretical implications:}Theoretical implications: Future research should develop more comprehensive models that integrate multiple performance dimensions and explicitly examine mediating and moderating mechanisms, including task complexity, team structure, and organizational context, to explain when and how psychological safety is associated with different software-team outcomes.

\subsection{\textbf{Discussion of RQ3}}
The findings for RQ3 show that psychological safety is associated with three key team processes in software engineering: collaboration, communication, and decision-making. Across these processes, the literature highlights open participation, knowledge exchange, interpersonal risk-taking, trust, and supportive contextual conditions as recurring mechanisms through which psychological safety is related to team interaction. The following subsections discuss these three team processes and the gaps identified in the existing literature.\vspace{5pt}

\subsubsection{\textbf{Psychological Safety and Team Collaboration (RQ 3.1)}}
The collaboration-related findings are organized into three themes: Open and Constructive Collaboration, Collaborative Learning and Knowledge Sharing, and Human and Contextual Factors Shaping Collaboration. Together, these themes indicate that the relationship between psychological safety and collaboration involves both team interaction processes and the interpersonal and contextual conditions in which collaboration occurs.

\noindent
\raisebox{-0.22\height}{%
\includegraphics[height=1.8em]{themeP.png}%
}%
\hspace{0.15em}%
{\textbf{Theme 1:}}
Psychological safety is strongly associated with open and constructive collaboration within software teams. Prior studies \cite{p9} indicate that psychologically safe environments encourage active participation and idea sharing, while the ability to engage in constructive conflict further enhances collaborative outcomes \cite{p29}. The findings indicate that psychological safety is associated with open dialogue, active participation, and interpersonal risk-taking, which are recurring characteristics of effective collaboration in the reviewed literature. However, literature often assumes that constructive conflict naturally leads to positive outcomes, without considering situations where conflict may escalate or remain unresolved. This suggests that the relationship between psychological safety and collaboration may depend on how conflict is managed within teams.

\noindent
\raisebox{-0.22\height}{%
\includegraphics[height=1.8em]{themeP.png}%
}%
\hspace{0.15em}%
{\textbf{Theme 2:}}
Collaborative learning and knowledge sharing emerge as important mechanisms associated with the relationship between psychological safety and team collaboration. Research shows that psychologically safe teams are more likely to exchange knowledge and learn collectively \cite{p10}, while open communication and mutual support further strengthen these processes \cite{p2}. The findings suggest that psychological safety is associated with shared understanding, knowledge exchange, and collective problem-solving within collaborative team environments. Future research should therefore examine not only whether knowledge sharing occurs in psychologically safe teams, but also the quality, relevance, and contribution of that knowledge exchange to meaningful collaboration outcomes. This indicates a need for further research on how knowledge exchange contributes to meaningful collaboration outcomes.

\noindent
\raisebox{-0.22\height}{%
\includegraphics[height=1.8em]{themeP.png}%
}%
\hspace{0.15em}%
{\textbf{Theme 3:}}
Human and contextual factors, including interpersonal relationships, trust, leadership, and supportive team environments, emerge as important conditions associated with psychological safety and collaboration. Studies such as \cite{p7} highlight that individual behaviors and attitudes influence collaborative interactions, while trust-based relationships enhance team engagement \cite{p3}. Additionally, leadership and supportive team environments are critical in fostering psychological safety and enabling collaboration [1]. Taken together, the findings indicate that the relationship between psychological safety and collaboration is shaped not only by interpersonal interactions but also by broader team and organizational conditions. However, literature often generalizes these factors without considering variations across team types, cultures, or organizational settings, suggesting a need for more context-specific analysis.\vspace{5pt}

Overall, the collaboration-related evidence suggests that psychological safety is associated with open participation, constructive dialogue, collaborative learning, knowledge sharing, and collective problem-solving. However, these relationships are not uniform and appear to depend on how teams manage conflict, the quality of knowledge exchange, interpersonal trust, leadership, and broader organizational conditions. Future research should therefore move beyond examining whether psychological safety is associated with collaboration and investigate the conditions under which psychologically safe interactions translate into effective collaborative outcomes.\vspace{5pt}

\subsubsection{\textbf{Psychological Safety and Team Communication (RQ 3.2)}}
The communication-related findings are organized into three themes: Open, Honest, and Learning-Oriented Communication; Information, Idea, and Knowledge Exchange; and Human and Contextual Factors Shaping Communication Participation and Quality. Together, these themes indicate that psychological safety is associated not only with the willingness of team members to communicate openly, but also with the exchange of information and knowledge and the contextual conditions that shape communication within software teams.\vspace{5pt}

\noindent
\raisebox{-0.22\height}{%
\includegraphics[height=1.8em]{themeP.png}%
}%
\hspace{0.15em}%
{\textbf{Theme 1:}}
Psychological safety is strongly associated with enabling open, honest, and learning-oriented communication within teams. Prior studies \cite{p2} indicate that psychologically safe environments encourage individuals to express ideas freely, while mistake-sharing and feedback practices further support collective learning \cite{p8}. The findings confirm that psychological safety promotes transparent dialogue, constructive feedback, and effective conflict resolution. However, literature often assumes that openness automatically leads to better communication outcomes, without considering potential challenges such as information overload, misinterpretation, or unproductive conflict. This suggests that the quality of communication depends not only on openness but also on how communication is structured and managed. However, greater willingness to communicate does not necessarily imply higher communication quality. The effectiveness of psychologically safe communication may also depend on whether dialogue is constructive, relevant, respectful, and appropriately integrated into team processes.

\noindent
\raisebox{-0.22\height}{%
\includegraphics[height=1.8em]{themeP.png}%
}%
\hspace{0.15em}%
{\textbf{Theme 2:}}
Psychological safety also plays a key role in facilitating the exchange of information, ideas, and knowledge within teams. Research shows that psychologically safe environments encourage knowledge sharing and idea contribution \cite{p9}, while improved information flow enhances coordination and collaboration \cite{p23}. The findings reinforce that psychological safety strengthens communication by enabling more effective knowledge dissemination. Nevertheless, many studies focus on the presence of knowledge-sharing behaviors rather than their depth, relevance, or impact on decision-making. This indicates a need for more detailed examination of how knowledge exchange contributes to communication effectiveness. Nevertheless, the existing literature gives comparatively less attention to the quality, accuracy, relevance, and practical use of exchanged information. Increased information sharing alone may therefore be insufficient to improve team processes if the exchanged knowledge is incomplete, poorly coordinated, or not effectively incorporated into development activities.

\noindent
\raisebox{-0.22\height}{%
\includegraphics[height=1.8em]{themeP.png}%
}%
\hspace{0.15em}%
{\textbf{Theme 3:}}
Human and contextual factors, including trust, relationships, individual characteristics, and organizational conditions, significantly influence communication participation and quality. Studies such as \cite{p1} highlight that trust-based relationships enhance openness and communication effectiveness, while individual emotional factors shape participation levels \cite{p11}. Additionally, team processes and organizational context affect communication clarity and efficiency \cite{p4}. The findings confirm that communication is not solely determined by psychological safety but is also shaped by broader interpersonal and contextual dynamics. However, literature often generalizes these factors without considering variations across cultural, organizational, or distributed team settings, suggesting a need for more context-aware research. The findings indicate that communication within psychologically safe teams is also shaped by human and contextual conditions. Interpersonal relationships, trust, leadership, team norms, and individual willingness to participate can influence how psychological safety is expressed through communication. This suggests that psychological safety alone may not be sufficient to produce high-quality communication when contextual conditions restrict participation or information exchange. Future research should therefore examine how these contextual conditions interact with psychological safety across different software-team settings rather than treating communication as a uniform outcome of psychological safety.\vspace{5pt}

Overall, the communication-related evidence suggests that psychological safety is associated with open and learning-oriented dialogue, information and knowledge exchange, and greater participation in team communication. While RQ1 identifies communication as a central dimension through which psychological safety is conceptualized, RQ3 shows how it operates as a team process involving speaking up, idea exchange, feedback, and information sharing. However, the effectiveness of these communication processes appears to depend on communication quality and broader human and contextual conditions. Future research should therefore move beyond examining whether psychological safety increases communication and investigate when psychologically safe communication results in accurate, constructive, and useful information exchange within software teams.\vspace{5pt}

\subsubsection{\textbf{Psychological Safety and Team Decision-Making (RQ 3.3)}}
The decision-making findings are organized into three themes: Inclusive and High-Quality Decision-Making, Learning-Oriented and Error-Sensitive Decision Processes, and Human and Structural Factors in Decision-Making. Together, these themes indicate that psychological safety is associated with participation, diversity of perspectives, error discussion, constructive questioning, and contextual conditions that shape collective decision processes. Compared with collaboration and communication, however, decision-making receives considerably less attention in the reviewed literature.

\noindent
\raisebox{-0.22\height}{%
\includegraphics[height=1.8em]{themeP.png}%
}%
\hspace{0.15em}%
{\textbf{Theme 1:}}
Psychological safety is associated with more inclusive and higher-quality decision-making processes within teams. Prior studies \cite{p1} indicate that psychologically safe environments encourage active participation and allow team members to share diverse perspectives, while cohesion supports balanced decision outcomes \cite{p22}.The findings suggest that psychological safety is associated with more inclusive participation and the integration of diverse perspectives in collective decision-making. However, the available evidence does not necessarily establish that greater participation directly results in higher decision quality. However, literature often assumes that increased participation and diversity automatically lead to better decisions, without considering potential challenges such as decision delays, conflict, or difficulty in reaching consensus. This suggests that the value of inclusive participation may depend on how teams integrate diverse perspectives, manage disagreement, and reach decisions efficiently.

\noindent
\raisebox{-0.22\height}{%
\includegraphics[height=1.8em]{themeP.png}%
}%
\hspace{0.15em}%
{\textbf{Theme 2:}}
Psychological safety is also associated with learning-oriented and reflective decision processes characterized by error discussion and constructive questioning. Research indicates that psychologically safe teams are more willing to discuss mistakes and critically evaluate decisions \cite{p9}, while reduced dysfunctional behavior and greater collective alignment are also reported in this context \cite{p48}. The findings therefore suggest a relationship between psychological safety and more reflective decision processes. However, this aspect remains comparatively underexplored, with relatively few studies examining it. Further research is needed to determine how error discussion, constructive questioning, and reflective decision practices relate to decision quality and longer-term team outcomes.

\noindent
\raisebox{-0.22\height}{%
\includegraphics[height=1.8em]{themeP.png}%
}%
\hspace{0.15em}%
{\textbf{Theme 3:}}
Human and structural factors, including leadership, communication, and team processes, emerge as important contextual conditions associated with psychological safety in decision-making. Prior studies indicate that leadership and communication practices are related to participation and coordination \cite{p12}, while structured processes can support informed decision-making \cite{p4}. These findings suggest that decision-making is shaped not only by psychological safety but also by broader organizational and process-related conditions. However, the literature provides limited evidence on how these relationships vary across different team settings, including distributed and cross-functional teams. More context-specific research is therefore needed to understand how psychological safety interacts with structural and organizational conditions during team decision-making.\vspace{5pt}

Overall, the findings for RQ3 show that psychological safety is consistently associated with key team processes, including collaboration, communication, and decision-making. Across these processes, open participation, knowledge sharing, constructive dialogue, learning, trust, and inclusive interaction emerge as recurring mechanisms associated with psychologically safe team environments. However, the literature is unevenly distributed, with substantially greater attention to collaboration and communication than to decision-making. This imbalance limits current understanding of how psychological safety relates to collective decision processes and highlights decision-making as an important area for further research.\vspace{5pt}

\noindent
\raisebox{-0.22\height}{%
\includegraphics[height=1.4em]{TA1.png}%
}%
\hspace{0.15em}%
\textbf{Practical implications:} Practical implications: Software teams should consider practices that support open interaction, structured communication, inclusive participation, and constructive discussion during collaborative and decision-making processes. Creating opportunities for team members to raise concerns, exchange knowledge, discuss mistakes, and engage in constructive disagreement may help support effective team interaction and coordination.\vspace{5pt}

\noindent
\raisebox{-0.22\height}{%
\includegraphics[height=1.4em]{TA1.png}%
}%
\hspace{0.15em}%
\textbf{Theoritical implications:} Theoretical implications: Future research should investigate the process-level mechanisms through which psychological safety is associated with collaboration, communication, and decision-making, with particular attention to the comparatively underexplored decision-making dimension. Research should also examine how conflict management, communication quality, team diversity, leadership, and organizational context shape these relationships across different software engineering environments.

\subsection{\textbf{Discussion of RQ4:}}
The findings for RQ4 show that the relationship between psychological safety and team performance is shaped by several contextual and interpersonal factors. Four factors receive particular attention in the reviewed literature: leadership style, organizational culture, work environment, and trust. These factors may strengthen, constrain, or otherwise shape how psychological safety is expressed and associated with team performance across different software engineering contexts.\vspace{5pt}

\subsubsection{\textbf{Leadership Style, Psychological Safety, and Team Performance (RQ 4.1)}}
The leadership-related findings are organized into three themes: positive leadership styles, leadership communication and emotional signaling, and leadership power structures and negative behaviors. Together, these themes show that leadership is represented in the literature through both supportive and restrictive behaviors, as well as through communication and emotional dynamics that shape the conditions surrounding psychological safety and team performance.

\noindent
\raisebox{-0.22\height}{%
\includegraphics[height=1.8em]{themeP.png}%
}%
\hspace{0.15em}%
{\textbf{Theme 1:}}
Positive leadership styles are consistently associated with psychological safety and favorable team outcomes across the reviewed literature. Prior studies \cite{p7} link supportive leadership with trust and openness, while inclusive leadership is associated with participation and respect for diverse perspectives \cite{p15}. Transformational leadership is also linked to motivation and engagement \cite{p6}, whereas empowering leadership emphasizes autonomy and shared responsibility \cite{p19}. Taken together, the findings indicate that supportive, inclusive, transformational, and empowering leadership styles frequently co-occur with conditions favorable to psychological safety and team functioning. However, the literature often treats these leadership approaches as broadly applicable without sufficiently examining contextual conditions such as organizational culture, team maturity, or situational constraints. Their relationship with psychological safety and performance may therefore vary across different software-team environments.

\noindent
\raisebox{-0.22\height}{%
\includegraphics[height=1.8em]{themeP.png}%
}%
\hspace{0.15em}%
{\textbf{Theme 2:}}
Leadership communication quality, emotional signaling, and safety climate emerge as additional leadership-related conditions associated with psychological safety and team performance. Prior research \cite{p1} links clear and supportive leadership communication with coordination and openness, while leaders' emotional expressions are reported to shape team climate and participation \cite{p11}. Leadership behaviors are also associated with the development of safety climates characterized by trust and collaboration \cite{p17}. These findings broaden the role of leadership beyond formal authority by highlighting communication patterns and emotional dynamics. However, these dimensions receive comparatively less attention than traditional leadership styles, indicating a need for further research on how leadership communication and emotional dynamics interact with psychological safety across different software-team contexts.

\noindent
\raisebox{-0.22\height}{%
\includegraphics[height=1.8em]{themeP.png}%
}%
\hspace{0.15em}%
{\textbf{Theme 3:}}
Leadership power structures and negative leadership behaviors represent a contrasting set of conditions associated with psychological safety and team performance. Prior studies \cite{p2} report that rigid or highly hierarchical leadership structures can restrict participation and openness, while negative leadership behaviors are associated with lower trust and less favorable team outcomes \cite{p52}. These findings indicate that leadership should not be examined only through supportive or positive behaviors; power relations and ineffective leadership practices also form part of the context in which psychological safety develops. However, the reviewed literature places considerably greater emphasis on positive leadership approaches than on negative or ineffective leadership, leaving the consequences of restrictive leadership practices comparatively underexplored.

Overall, the leadership-related evidence indicates that psychological safety and team performance are associated with multiple dimensions of leadership, including leadership style, communication, emotional signaling, safety climate, and power relations. Supportive, inclusive, transformational, and empowering leadership receive considerably greater attention than negative or restrictive leadership behaviors. At the same time, the evidence suggests that leadership should be understood as a context-dependent influence rather than a universally effective mechanism, as its relationship with psychological safety may vary according to organizational culture, team characteristics, and situational conditions.\vspace{5pt}

\subsubsection{\textbf{Organizational and Team Culture, Psychological Safety, and Team Performance (RQ 4.2)}}
The culture-related findings are organized into four themes: Supportive, Learning, and Feedback-Oriented Cultures; Agile and Innovation-Oriented Cultures; Shared, Normative, and National Cultural Influences; and Toxic and Unsupportive Cultures. Together, these themes show that psychological safety is associated with both organizational and broader cultural conditions, ranging from supportive and learning-oriented environments to restrictive or blame-oriented cultures.

\noindent
\raisebox{-0.22\height}{%
\includegraphics[height=1.8em]{themeP.png}%
}%
\hspace{0.15em}%
{\textbf{Theme 1:}}
Supportive, learning-oriented, and feedback-oriented cultures are frequently associated with psychological safety and favorable team outcomes. Prior studies \cite{p9} link open and respectful cultural environments with participation and idea sharing, while learning and feedback climates are associated with continuous improvement and error identification \cite{p19}. The findings therefore indicate that openness, trust, learning, and reflective practices are recurring characteristics of organizational cultures associated with psychological safety. However, the literature often assumes that supportive cultures naturally emerge or can be readily established, without sufficiently considering organizational resistance, legacy practices, or misalignment between stated values and actual behaviors. This highlights an important gap between cultural ideals and their implementation in practice.

\noindent
\raisebox{-0.22\height}{%
\includegraphics[height=1.8em]{themeP.png}%
}%
\hspace{0.15em}%
{\textbf{Theme 2:}}
Agile and innovation-oriented cultures emerge as another contextual condition associated with psychological safety and team performance. Prior research \cite{p1} links cultures emphasizing adaptability, experimentation, and continuous learning with collaboration and idea sharing, while culturally shaped practices are also associated with coordination and knowledge exchange \cite{p77}. These findings suggest that agile and innovation-oriented cultural characteristics may provide conditions favorable to psychologically safe interaction. However, the literature often presents agile culture as broadly beneficial without sufficiently examining situations in which agile values are superficially adopted, inconsistently implemented, or constrained by existing organizational practices. This raises questions about how the relationship between agile culture and psychological safety varies across organizational contexts.

\noindent
\raisebox{-0.22\height}{%
\includegraphics[height=1.8em]{themeP.png}%
}%
\hspace{0.15em}%
{\textbf{Theme 3:}}
Shared organizational norms and broader national-cultural contexts represent additional conditions associated with psychological safety and team performance. Prior studies \cite{p23} link balanced and respectful cultural norms with stable collaboration, while national-cultural differences are associated with variations in communication styles and participation \cite{p2}. These findings suggest that psychological safety should not be understood solely as a team-level phenomenon, as it may also be shaped by wider organizational and sociocultural contexts. However, the literature often treats culture as relatively static, giving limited attention to how cultural dynamics evolve over time or interact with organizational change. This indicates a need for more longitudinal and context-sensitive research on cultural influences in software teams.

\noindent
\raisebox{-0.22\height}{%
\includegraphics[height=1.8em]{themeP.png}%
}%
\hspace{0.15em}%
{\textbf{Theme 4:}}
Toxic and unsupportive cultural conditions represent a contrasting set of factors associated with lower psychological safety and less favorable team outcomes. Prior evidence \cite{p47} links blame-oriented cultures and insufficient support with reduced communication and participation. These findings highlight the importance of examining negative cultural conditions alongside supportive cultural characteristics. However, this area remains comparatively underexplored, with limited research examining how toxic cultures emerge, persist, or change over time. Greater attention is therefore needed to understand both the development of harmful cultural conditions and the organizational processes through which they may be transformed.\vspace{5pt}

Overall, the culture-related evidence indicates that psychological safety is associated with multiple levels and forms of culture, including supportive and learning-oriented organizational climates, agile and innovation-oriented cultures, shared organizational norms, broader national-cultural contexts, and toxic or blame-oriented environments. The findings also reveal an imbalance in the literature, with greater emphasis on desirable cultural characteristics than on how supportive cultures are implemented, how cultural conditions evolve, or how harmful cultures can be transformed. Within RQ4, culture is considered as a contextual condition shaping the relationship between psychological safety and team performance, whereas RQ5 specifically examines problematic cultural conditions as barriers to establishing psychological safety.\vspace{5pt}

\subsubsection{\textbf{Work Environment, Psychological Safety, and Team Performance (RQ 4.3)}}
The work-environment findings are organized into five themes: Supportive and Trust-Based Work Environments; Leadership, Structure, and Autonomy in the Work Environment; Inclusive and Value-Aligned Work Environments; Emotional and Contextual Work Pressures; and the Direct Impact of Psychologically Safe Work Environments. Together, these themes indicate that work environment represents a multidimensional context in which interpersonal relationships, organizational arrangements, inclusion, work pressures, and psychologically safe conditions are associated with team functioning and performance.

\noindent
\raisebox{-0.22\height}{%
\includegraphics[height=1.8em]{themeP.png}%
}%
\hspace{0.15em}%
{\textbf{Theme 1:}}
Supportive and trust-based work environments are frequently associated with psychological safety and favorable team outcomes. Prior studies \cite{p2} indicate that no-blame climates are associated with openness and active participation, while trust-based relationships are linked to collaboration and positive team dynamics \cite{p16}. The findings therefore suggest that supportive work environments provide conditions in which team members may feel more comfortable sharing ideas, reporting mistakes, and participating in team processes. However, the literature often assumes that such environments can be readily established without sufficiently considering organizational constraints such as hierarchical structures or competing performance pressures. This suggests that creating and maintaining supportive and trust-based environments may be more complex and context-dependent in practice.

\noindent
\raisebox{-0.22\height}{%
\includegraphics[height=1.8em]{themeP.png}%
}%
\hspace{0.15em}%
{\textbf{Theme 2:}}
Leadership practices, organizational structures, and autonomy are key elements shaping psychological safety within the work environment. Research shows that effective leadership and management practices provide direction and support, enhancing coordination and performance \cite{p46}, while flexible structures and autonomy empower team members and encourage participation \cite{p70}. The findings indicate that both structural design and leadership influence the extent to which psychological safety translates into performance outcomes. However, literature often overlooks potential trade-offs, such as the challenges of balancing autonomy with accountability or maintaining coordination in highly flexible environments.

\noindent
\raisebox{-0.22\height}{%
\includegraphics[height=1.8em]{themeP.png}%
}%
\hspace{0.15em}%
{\textbf{Theme 3:}}
Inclusive and value-aligned work environments represent another contextual dimension associated with psychological safety and team performance. Prior studies \cite{p14} link diversity and inclusion practices with participation and mutual respect, while shared values are associated with alignment and collaboration \cite{p21}. The findings suggest that inclusion and value alignment may provide important conditions for psychologically safe team interaction. However, these factors receive comparatively limited attention in the reviewed literature, indicating a need for further research on how inclusion, diversity, and value alignment interact with psychological safety across different software-team contexts.

\noindent
\raisebox{-0.22\height}{%
\includegraphics[height=1.8em]{themeP.png}%
}%
\hspace{0.15em}%
{\textbf{Theme 4:}}
Emotional and contextual work pressures, particularly workload and stress, emerge as adverse conditions associated with psychological safety and team performance. Prior research \cite{p4} links stressful work conditions with reduced openness and communication, while excessive workload is associated with lower participation and collaboration \cite{p7}. These findings suggest that psychological safety should be considered in relation to the demands and pressures under which software teams operate. However, the literature often treats workload and stress as external contextual variables rather than integrating them into explanations of how psychological safety develops and functions. Future research should therefore examine environmental stressors as part of the broader socio-organizational context surrounding psychological safety and team dynamics.

\noindent
\raisebox{-0.22\height}{%
\includegraphics[height=1.8em]{themeP.png}%
}%
\hspace{0.15em}%
{\textbf{Theme 5:}}
Psychologically safe work environments are also associated with team functioning, innovation, and performance outcomes. Unlike the preceding themes, which describe specific work-environment conditions surrounding psychological safety, this theme reflects studies that conceptualize the psychologically safe environment itself as directly associated with team outcomes. Prior studies \cite{p29}  link psychologically safe environments with creativity and innovation, while other evidence associates these environments with coordination and team effectiveness \cite{p36}. These findings indicate that some studies conceptualize psychological safety as being closely related to team outcomes rather than solely as a contextual condition. However, the literature often presents this relationship in relatively direct terms without sufficiently examining potential mediating or moderating mechanisms, such as task complexity, team composition, or organizational context. Future research should therefore test these mechanisms explicitly rather than assuming a uniform direct relationship between psychological safety and team performance.\vspace{5pt}

Overall, the work-environment evidence indicates that the relationship between psychological safety and team performance is embedded within multiple contextual conditions, including supportive and trust-based climates, leadership and organizational structures, autonomy, inclusion and value alignment, and workload or emotional pressures. The findings also show that some studies conceptualize psychologically safe work environments as being directly associated with team functioning and innovation. However, the literature provides limited attention to the interactions and trade offs among these conditions, particularly how autonomy, organizational pressures, team composition, and other contextual factors jointly shape psychological safety and performance. This highlights the need for more integrated and context-sensitive models of the software-team work environment.\vspace{5pt}

\subsubsection{\textbf{Trust, Psychological Safety, and Team Performance (RQ 4.4)}}
The trust-related findings are organized into three themes: Trust as a Foundation for Psychological Safety and Performance, Trust in Leadership and Team Relationships, and Trust Across Physical and Virtual Team Environments. Together, these themes indicate that trust is associated with psychological safety at interpersonal, leadership, and team levels and that its role may vary across co-located and virtual software-team contexts.

\noindent
\raisebox{-0.22\height}{%
\includegraphics[height=1.8em]{themeP.png}%
}%
\hspace{0.15em}%
{\textbf{Theme 1:}}
Trust emerges as a recurring interpersonal condition associated with psychological safety and team performance. Prior studies \cite{p9} link trust with lower perceptions of interpersonal risk and greater openness in communication, while reduced fear is associated with greater willingness to express ideas and participate in team processes \cite{p84}. The findings therefore indicate a close relationship between trust and psychological safety, particularly in relation to voice, openness, and interpersonal risk-taking. However, the literature often assumes that trust develops naturally within teams without sufficiently considering factors such as prior negative experiences, organizational culture, or team diversity. This suggests that trust formation is a complex and context-dependent process that requires further investigation.

\noindent
\raisebox{-0.22\height}{%
\includegraphics[height=1.8em]{themeP.png}%
}%
\hspace{0.15em}%
{\textbf{Theme 2:}}
Trust in leadership and trust among team members represent distinct but related relational conditions associated with psychological safety and team functioning. Prior research \cite{p1} links trust in leadership, particularly perceptions of fairness and reliability, with openness and engagement, while mutual trust among team members is associated with collaboration and shared responsibility \cite{p20}. These findings indicate that trust operates at multiple levels within software teams. However, the literature often examines leadership trust and team-level trust separately, providing limited understanding of how these forms of trust interact or reinforce one another. More integrative research is therefore needed to examine multi-level trust dynamics in relation to psychological safety and team performance.

\noindent
\raisebox{-0.22\height}{%
\includegraphics[height=1.8em]{themeP.png}%
}%
\hspace{0.15em}%
{\textbf{Theme 3:}}
Trust is also associated with team functioning across both co-located and virtual team environments. Prior studies \cite{p12} link trust with coordination and collaboration, while research on virtual teams \cite{p51} highlights its particular relevance when opportunities for face-to-face interaction are limited. These findings suggest that the role and development of trust may differ across physical and distributed team contexts. However, the literature provides limited understanding of how trust is established and maintained in virtual software teams, particularly in the presence of communication barriers, cultural differences, and reliance on digital collaboration tools. Further research is therefore needed to examine trust dynamics and their relationship with psychological safety across different forms of distributed teamwork.\vspace{5pt}

\subsubsection{\textbf{Sociocultural and Individual Factors, Psychological Safety, and Team Performance (RQ 4.5)}}
The findings further identify sociocultural and individual factors associated with psychological safety and team performance. These findings are organized into three themes: Cultural, National, and Regional Influences; Personality and Individual Differences; and Gender and Power-Related Dynamics. Compared with leadership, work environment, and trust, these dimensions receive relatively limited attention, indicating that individual and sociocultural variation remains an underexplored area of psychological safety research in software engineering.

\noindent
\raisebox{-0.22\height}{%
\includegraphics[height=1.8em]{themeP.png}%
}%
\hspace{0.15em}%
{\textbf{Theme 1:}}
Cultural, national, and regional differences emerge as contextual factors associated with how psychological safety is experienced and expressed within software teams. Prior studies \cite{p5} indicate that cultural diversity is associated with variations in communication styles and interaction patterns, while regional and contextual conditions are also linked to differences in team dynamics \cite{p43}. These findings suggest that psychological safety may not be experienced uniformly across cultural and regional settings. However, the literature often describes cultural differences without sufficiently examining the mechanisms through which they interact with psychological safety or how teams manage these differences in practice. More comparative and context-sensitive research is therefore needed to understand how cultural and regional conditions shape psychological safety across software engineering environments.

\noindent
\raisebox{-0.22\height}{%
\includegraphics[height=1.8em]{themeP.png}%
}%
\hspace{0.15em}%
{\textbf{Theme 2:}}
Personality traits and individual characteristics emerge as additional factors associated with psychological safety and team functioning. Prior research \cite{p11} links personality dynamics with participation and collaboration, while individual characteristics such as confidence and behavioral tendencies are associated with communication and engagement \cite{p57}. These findings suggest that team members may perceive and respond to psychologically safe conditions differently even within the same team. However, the literature predominantly emphasizes team-level outcomes and provides comparatively limited attention to individual-level variation. Future research should therefore examine how individual characteristics interact with shared team perceptions of psychological safety and whether different personality profiles influence participation, communication, and team outcomes in distinct ways.

\noindent
\raisebox{-0.22\height}{%
\includegraphics[height=1.8em]{themeP.png}%
}%
\hspace{0.15em}%
{\textbf{Theme 3:}}
Gender-related issues and power dynamics represent additional sociostructural conditions associated with psychological safety and team participation. Prior studies \cite{p21} identify gender bias and unequal power relations as barriers to participation and inclusion, while power imbalances are associated with silence and reduced voice \cite{p22}. These findings indicate that psychological safety may be constrained by social and structural inequalities within teams. However, gender and power-related dynamics receive comparatively limited attention in the reviewed literature, particularly in relation to diverse and global software teams. This limited evidence base highlights the need for more focused research on how gender, power relations, and inclusion interact with psychological safety and team functioning.\vspace{5pt}

Overall, the findings for RQ4 show that the relationship between psychological safety and team performance is associated with multiple contextual, interpersonal, and individual factors, including leadership, organizational culture, work environment, trust, cultural and regional context, personality, and gender- and power-related dynamics. Leadership, work environment, and trust receive comparatively greater attention, whereas sociocultural and individual factors remain less extensively investigated. The evidence therefore suggests that psychological safety should be understood as context-dependent rather than as operating uniformly across software teams. However, existing studies often examine these factors separately, highlighting a need for more integrated research on how multiple contextual conditions interact in shaping psychological safety and team outcomes.\vspace{5pt}

\noindent
\raisebox{-0.22\height}{%
\includegraphics[height=1.4em]{TA1.png}%
}%
\hspace{0.15em}%
\textbf{Practical implications:} Practical implications: Organizations should consider psychological safety within the broader context of leadership practices, organizational culture, work-environment conditions, trust, and team diversity. Rather than applying uniform approaches across teams, managers should recognize that cultural, individual, and power-related differences may shape how team members experience and participate in psychologically safe environments. Attention to inclusion and equitable participation is therefore important when designing practices intended to support psychological safety.\vspace{5pt}

\noindent
\raisebox{-0.22\height}{%
\includegraphics[height=1.4em]{TA1.png}%
}%
\hspace{0.15em}%
\textbf{Theoretical implications:}  Theoretical implications: Future research should adopt multi-level and context-sensitive models that examine how leadership, culture, work environment, trust, and individual and sociocultural factors interact in relation to psychological safety and team performance. Longitudinal research is particularly needed to understand how these contextual relationships evolve over time rather than treating influencing factors as static and independent conditions.

\subsection{\textbf{Discussion of RQ5:}}
The findings for RQ5 identify six major categories of barriers to establishing psychological safety in software development teams: Communication and Information Flow Barriers; Cultural and Interpersonal Trust Challenges; Structural and Organizational Barriers; Leadership and Management Issues; Diversity, Bias, and Inclusion Barriers; and Workload, Pressure, and Resource Constraints. Together, these themes show that barriers to psychological safety arise not from a single source but from interacting interpersonal, organizational, structural, and work-related conditions.

\noindent
\raisebox{-0.22\height}{%
\includegraphics[height=1.8em]{themeP.png}%
}%
\hspace{0.15em}%
{\textbf{Theme 1:}}
Communication and information-flow barriers emerge as one of the most frequently reported challenges to establishing psychological safety. Prior studies \cite{p8} indicate that lack of open dialogue and hesitation to speak up can restrict team interaction, while coordination breakdowns further disrupt information exchange \cite{p2}. The broader findings show that these barriers extend beyond silence and unclear communication to include interpersonal conflict, communication challenges in culturally diverse teams, difficulties associated with distributed and virtual collaboration, and interference arising from technologies, tools, and contextual constraints. This indicates that communication barriers are not simply individual speaking-up problems but can arise from interpersonal, technological, cultural, and organizational conditions. However, the literature often examines these communication problems separately, providing limited understanding of how they interact. Communication barriers should therefore be considered as part of the broader socio-technical environment in which psychological safety develops. This finding also complements RQ3, where communication appears as a team process associated with psychological safety, suggesting that communication and psychological safety may be mutually reinforcing rather than operating in only one direction.

\noindent
\raisebox{-0.22\height}{%
\includegraphics[height=1.8em]{themeP.png}%
}%
\hspace{0.15em}%
{\textbf{Theme 2:}}
Cultural and interpersonal trust challenges emerge as one of the most frequently reported categories of barriers to psychological safety. Prior research indicates that weak trust, limited familiarity, and strained interpersonal relationships are associated with reduced openness and collaboration \cite{p3}, while fear-driven environments can discourage participation and contribution \cite{p9}. The broader findings show that these challenges also involve unresolved conflict and coordination problems, restrictive cultural and leadership conditions, hierarchy and blame, limited belonging, and difficulties in establishing open feedback and learning climates. This suggests that trust-related barriers are not solely interpersonal but are closely connected with organizational culture, power structures, inclusion, and team-learning conditions. However, the literature often treats trust as something that can be readily developed without sufficiently examining how previous team experiences, cultural differences, organizational conditions, and power relations shape trust formation. More context-sensitive and longitudinal research is therefore needed to understand how trust and relational climates develop, deteriorate, and recover within software teams. This finding complements RQ4, where trust is identified as a contextual factor associated with psychological safety and team performance; RQ5 further shows that breakdowns in trust, belonging, and relational conditions can constitute barriers to establishing psychological safety.

\noindent
\raisebox{-0.22\height}{%
\includegraphics[height=1.8em]{themeP.png}%
}%
\hspace{0.15em}%
{\textbf{Theme 3:}}
Structural and organizational barriers emerge through both restrictive organizational arrangements and ambiguity concerning roles, responsibilities, and team goals. Prior studies \cite{p23} associate rigid organizational structures and policies with restricted communication and participation, while unclear roles and responsibilities are linked to reduced confidence in speaking up \cite{p6}. The findings therefore suggest that structural clarity and supportive organizational arrangements represent important conditions associated with psychological safety. However, the literature gives limited attention to the possibility that highly flexible or agile organizational arrangements may also introduce challenges, including role ambiguity, unclear responsibilities, or coordination complexity. This suggests that the relationship between organizational structure and psychological safety should not be understood as a simple contrast between rigid and flexible structures; rather, the degree of role clarity, coordination, autonomy, and structural support may determine whether particular arrangements support or constrain psychologically safe participation.
This finding complements RQ4, where organizational structure and autonomy emerge as contextual factors, by showing that structural rigidity and ambiguity can specifically operate as barriers to psychological safety.

\noindent
\raisebox{-0.22\height}{%
\includegraphics[height=1.8em]{themeP.png}%
}%
\hspace{0.15em}%
{\textbf{Theme 4:}}
Leadership and management issues emerge as frequently reported barriers to establishing psychological safety. Prior work \cite{p17}  associates inadequate leadership and weak management practices with limited support and openness, while hierarchical and blame-oriented environments are linked to reduced voice and participation \cite{p10}. The broader findings further show that leadership-related barriers operate through fear of negative consequences, weak relational support, limited trust and belonging, and structural power imbalances that can discourage learning and contribution. This indicates that leadership barriers extend beyond individual management styles to include the wider relational and organizational conditions created or reinforced through leadership practices. However, much of the literature emphasizes ideal or supportive leadership approaches while giving comparatively less attention to the practical constraints leaders face, such as organizational pressure, conflicting priorities, or restrictive structures. This highlights a gap between theoretical leadership models and the conditions under which leadership practices are implemented in real software organizations.
This finding complements RQ4, where leadership is examined as a contextual factor, by showing that ineffective leadership, fear-driven management, hierarchy, and insufficient relational support can specifically operate as barriers to establishing psychological safety.

\noindent
\raisebox{-0.22\height}{%
\includegraphics[height=1.8em]{themeP.png}%
}%
\hspace{0.15em}%
{\textbf{Theme 5:}}
Bias, exclusion, and diversity-related challenges emerge as important but comparatively underexplored barriers to psychological safety. Prior studies \cite{p1} identify gender-related issues as potential constraints on participation, while discriminatory or exclusionary norms are associated with reduced inclusion and openness \cite{p47}. The findings indicate that inclusive conditions and equitable participation are closely associated with psychological safety, while bias and exclusion can restrict team members' willingness to participate and speak openly. However, these issues receive comparatively limited attention in the reviewed literature. This imbalance indicates that the evidence base on bias, exclusion, and inclusion remains limited, particularly for diverse, global, and distributed software-team contexts. Future research should therefore examine more explicitly how bias, exclusion, power relations, and inclusion practices interact with psychological safety across different software-team settings.

\noindent
\raisebox{-0.22\height}{%
\includegraphics[height=1.8em]{themeP.png}%
}%
\hspace{0.15em}%
{\textbf{Theme 6:}}
Workload, pressure, and resource constraints emerge as additional barriers associated with psychological safety in software teams. Prior studies \cite{p7} indicate that high workload, pressure, and stress can reduce team members’ willingness to speak up, while limited resources may constrain effective collaboration \cite{p36}. The broader findings show that these barriers also involve weak organizational support, communication and coordination difficulties, limitations in available skills and resources, participation constraints, and other contextual conditions. This suggests that workload and resource-related barriers can affect psychological safety not only through individual stress but also by shaping opportunities for participation, communication, and collaborative work. However, these factors receive comparatively limited attention in the reviewed literature and are often treated as secondary contextual issues rather than being integrated into explanations of psychological safety. Future research should therefore examine more explicitly how workload, resource availability, organizational support, and contextual pressures interact with psychological safety across different software development environments.\vspace{5pt}

Overall, the findings for RQ5 show that barriers to psychological safety are multidimensional and interconnected, spanning communication and information flow, cultural and interpersonal trust, organizational structures, leadership and management, bias and inclusion, and workload and resource conditions. Communication and trust-related barriers receive the greatest research attention, whereas diversity-, bias-, inclusion-, and workload-related barriers are comparatively less investigated. The findings further suggest that these barriers should not be viewed in isolation, as communication problems, trust breakdowns, leadership practices, structural conditions, and work pressures may interact in shaping psychologically safe team environments.\vspace{5pt}

\noindent
\raisebox{-0.22\height}{%
\includegraphics[height=1.4em]{TA1.png}%
}%
\hspace{0.15em}%
\textbf{Practical implications:} Organizations should address psychological safety barriers through a coordinated approach that improves communication and information flow, builds interpersonal trust, reduces unnecessary hierarchical and blame-oriented practices, clarifies roles and responsibilities, supports inclusive participation, and manages workload and resource pressures. Because these barriers may interact, isolated interventions may be insufficient to create and sustain psychologically safe software-team environments.\vspace{5pt}

\noindent
\raisebox{-0.22\height}{%
\includegraphics[height=1.4em]{TA1.png}%
}%
\hspace{0.15em}%
\textbf{Theoretical implications:}  Future research should move beyond examining barriers independently and develop more integrated models that explain how communication, trust, leadership, organizational structures, inclusion, and workload-related conditions interact in shaping psychological safety. Greater attention should also be given to comparatively underexplored barriers, particularly bias and inclusion, workload and resource constraints, and the contextual conditions under which different barriers become more or less influential.

\subsection{\textbf{Discussion of RQ6:}}
The findings for RQ6 identify five categories of strategies, practices, and interventions reported for enhancing psychological safety in software teams: Leadership and Organizational Practices; Agile, Role, and Process Practices; Learning, Feedback, and Capability-Building Practices; Trust, Belonging, and Everyday Team Practices; and Well-Being, Stress, and Technology Practices. Together, these themes show that improvement approaches operate across leadership, organizational, team, process, learning, well-being, and technological dimensions.

\noindent
\raisebox{-0.22\height}{%
\includegraphics[height=1.8em]{themeP.png}%
}%
\hspace{0.15em}%
{\textbf{Theme 1:}}
Leadership and organizational practices emerge as one of the main categories of strategies reported for supporting psychological safety. Prior studies \cite{p1} associate supportive and inclusive leadership with greater willingness to speak up and participate, while organizational policies and formal support mechanisms are reported as approaches for reinforcing psychologically safe environments \cite{p43}. The mapped evidence also identifies motivation, recognition, and reward-related practices as approaches for encouraging participation and reinforcing behaviors associated with psychologically safe teamwork. Taken together, the findings suggest that leadership behaviors, formal organizational support, and motivation- or recognition-related practices represent complementary approaches reported for supporting psychological safety. However, much of the literature assumes that such practices can be readily implemented without sufficiently considering organizational constraints such as resistance to change, limited resources, or conflicting priorities. This reveals a gap between identifying or recommending practices and demonstrating their feasibility and effectiveness in real-world software organizations.

\noindent
\raisebox{-0.22\height}{%
\includegraphics[height=1.8em]{themeP.png}%
}%
\hspace{0.15em}%
{\textbf{Theme 2:}}
Agile, role-related, and process practices emerge as another category of approaches reported for integrating psychological safety into everyday team activities. Prior research \cite{p2} associates agile practices with continuous feedback, open communication, and team interaction, while role clarity and shared ownership are reported as approaches for reducing uncertainty and supporting participation \cite{p3}. The mapped evidence further identifies task rotation, collective ownership, tools, and structured team processes as practices used to support psychological safety within everyday workflows. Taken together, the findings suggest that psychological safety is addressed not only through interpersonal behaviors but also through the design of roles, processes, and routine team practices. However, the literature often presents these approaches as broadly beneficial without sufficiently examining implementation quality or the organizational conditions under which they are applied. In particular, agile practices may be superficially or inconsistently implemented, raising questions about whether the presence of an agile practice alone is sufficient to support psychological safety. Future research should therefore evaluate how implementation quality, team context, and organizational conditions influence the effectiveness of these practices in supporting psychological safety.

\noindent
\raisebox{-0.22\height}{%
\includegraphics[height=1.8em]{themeP.png}%
}%
\hspace{0.15em}%
{\textbf{Theme 3:}}
Learning, feedback, training, and coaching emerge as another category of practices reported for supporting psychological safety and capability development within software teams. Prior studies \cite{p4} associate open feedback, reflection, and shared learning with greater voice and opportunities for improvement, while training and coaching interventions are reported as approaches for supporting skill development, confidence, and interpersonal understanding \cite{p6}. Taken together, the findings suggest that learning-oriented practices address psychological safety through both everyday feedback and reflection processes and more structured capability-building interventions. However, the reviewed literature focuses more strongly on identifying these practices than on evaluating their implementation quality, effectiveness, or longer-term impact. Future research should therefore evaluate how feedback, reflection, training, and coaching interventions are implemented, which forms are most effective under different team conditions, and whether their effects on psychological safety are sustained over time.

\noindent
\raisebox{-0.22\height}{%
\includegraphics[height=1.8em]{themeP.png}%
}%
\hspace{0.15em}%
{\textbf{Theme 4:}}
Trust, belonging, and everyday team practices emerge as another category of approaches reported for supporting psychological safety. Prior studies \cite{p5} indicate that team- and relationship-building practices are associated with stronger interpersonal trust and supportive team interaction, while inclusion and diversity-oriented practices are reported as approaches for encouraging equitable participation and a sense of belonging. The mapped evidence further identifies team norms and everyday practices \cite{p47}, indicating that psychological safety can be supported through routine interactions and shared behavioral expectations rather than only through formal organizational interventions. Taken together, these findings suggest that relational and everyday team practices complement leadership, organizational, and process-level approaches to psychological safety. However, the literature provides comparatively limited evidence on how these practices are implemented consistently, how their effects develop over time, and whether they produce sustained changes in psychological safety across different software-team contexts. Future research should therefore evaluate the implementation and longer-term effectiveness of trust-building, inclusion-oriented, and everyday team practices rather than focusing primarily on their recommendation or presence.

\noindent
\raisebox{-0.22\height}{%
\includegraphics[height=1.8em]{themeP.png}%
}%
\hspace{0.15em}%
{\textbf{Theme 5:}}
Well-being, stress-management, and technology-related practices emerge as additional approaches reported for supporting psychological safety. Prior research \cite{p7} indicates that reducing workload and managing stress can contribute to supportive team conditions, while emerging practices involving AI-supported tools are associated with communication and coordination \cite{p54}. Taken together, these findings suggest that supporting psychological safety may involve attention not only to interpersonal and organizational practices but also to employee well-being and the technological environment in which teams operate. However, these dimensions remain comparatively underexplored, particularly regarding the long-term effectiveness of stress-management practices and the role of emerging technologies in team interaction. Future research should therefore examine how well-being practices and technology-supported collaboration contribute to psychological safety across different software-team contexts.

Overall, the findings for RQ6 identify a broad range of strategies, practices, and interventions reported for supporting psychological safety in software teams, including leadership and organizational practices, agile and process-related practices, learning and feedback mechanisms, trust- and belonging-oriented team practices, and well-being and technology-related approaches. These findings indicate that psychological safety is addressed through multiple organizational, team, process, interpersonal, and contextual dimensions rather than through a single type of intervention. However, the literature focuses more strongly on identifying and recommending practices than on evaluating their implementation quality, effectiveness, scalability, or long-term impact. This limits the current evidence regarding which strategies are most effective, under what conditions they work, and whether their effects can be sustained over time.\vspace{5pt}

\noindent
\raisebox{-0.22\height}{%
\includegraphics[height=1.4em]{TA1.png}%
}%
\hspace{0.15em}%
\textbf{Practical implications:}Organizations should consider integrated approaches that combine supportive leadership and organizational practices, structured team and agile processes, continuous learning and feedback, trust- and belonging-oriented practices, and support for employee well-being. Rather than adopting individual practices in isolation, organizations should adapt these approaches to their team and organizational contexts and consider how they can be incorporated into everyday software-development activities. Emerging technology-supported practices should also be approached carefully, given the currently limited evidence regarding their role in psychological safety.\vspace{5pt}

\noindent
\raisebox{-0.22\height}{%
\includegraphics[height=1.4em]{TA1.png}%
}%
\hspace{0.15em}%
\textbf{Theoretical implications:}Future research should move beyond identifying and recommending psychological-safety practices toward systematically evaluating their effectiveness, scalability, implementation quality, and sustainability over time. Greater attention is also needed to how leadership, organizational, process, learning, relational, well-being, and technology-related strategies interact rather than operating independently. Longitudinal and context-sensitive studies would help determine which approaches are effective, for whom, and under what software-engineering conditions.

Table ~\ref{tab:rq_summary} summarizes the main findings across the six research questions and highlights their broader implications. Overall, the findings show that psychological safety in software engineering is a multidimensional and context-dependent phenomenon associated with team performance and team processes, while also being shaped by organizational, interpersonal, and individual conditions. The synthesis further identifies recurring barriers and a broad range of reported practices for supporting psychological safety, although evidence regarding the effectiveness and sustainability of these practices remains limited.

%-------------Table 6--------------------
\renewcommand{\arraystretch}{1.3}

\begin{table*}[t]
\centering
\caption{Summary of Findings and Implications Across the Six Research Questions}
\label{tab:rq_summary}

\footnotesize
\setlength{\tabcolsep}{5pt}

\begin{tabularx}{\textwidth}{c c X X}
\toprule

\rowcolor{gray!40}
\textbf{RQ} & \textbf{Key Focus} & \textbf{Main Finding} & \textbf{Implication} \\

\midrule

\textbf{RQ1} & Conceptualization of PS 
& PS is conceptualized through interpersonal, communication, collaboration, learning, social, and well-being dimensions. 
& PS is represented as a multidimensional construct in software-team research. \\

\textbf{RQ2} & PS and Team Performance 
& PS is predominantly associated with productivity, software quality, innovation, and team satisfaction. 
& PS is associated with multiple dimensions of team performance. \\

\textbf{RQ3} & Team Processes 
& PS is associated with collaboration, communication, and decision-making processes.
& PS is closely related to key team-interaction and coordination processes. \\

\textbf{RQ4} & Contextual and Influencing Factors 
& Leadership, culture, work environment, trust, and sociocultural and individual factors are associated with the PS performance relationship.
& The relationship between PS and team performance is context-dependent. \\

\textbf{RQ5} & Barriers 
& Communication, trust, structural, leadership, inclusion, and workload-related barriers are reported across the literature.
& Barriers to PS are multidimensional and interconnected. \\

\textbf{RQ6} & Strategies 
& Leadership, organizational, agile/process, learning, relational, well-being, and technology related practices are reported for supporting PS.
& A broad range of strategies is reported, but evidence on their effectiveness and sustainability remains limited \\

%\textbf{RQ7} & Recommendations 
%& Leadership, training, trust-building, and HR practices improve PS 
%& PS should be embedded at the organizational level \\

\bottomrule
\end{tabularx}

\end{table*}

\section{Comparison with Existing Studies}
To position this study within the existing body of research, this section compares the present systematic mapping study with representative prior studies on psychological safety and related human factors in software engineering. The selected studies vary in research context, focus, treatment of psychological safety, consideration of team processes and performance outcomes, and methodological approach. Table ~\ref{tab:comparison} summarizes these differences and provides a basis for identifying how the present SMS extends and integrates the existing evidence. The studies included in the comparison were selected to represent different research contexts, methodological approaches, and levels of emphasis on psychological safety, team processes, and performance outcomes within the mapped literature.\vspace{5pt}

Existing research on psychological safety has roots in organizational and management contexts, where it has been examined in relation to team learning, communication, and performance outcomes \cite{p37,p40}. Within software engineering, prior studies have examined psychological safety and related human factors in connection with collaboration, leadership, communication, and team dynamics \cite{p3,p4,p26}. However, these studies vary in their emphasis on psychological safety and in the specific team processes, contextual conditions, and outcomes they examine. Consequently, the evidence remains distributed across different dimensions of software-team interaction, making it difficult to obtain an integrated view of psychological safety, team processes, and performance outcomes \cite{p4,p15}.\vspace{5pt}

Moreover, many empirical studies in software engineering examine psychological safety within specific contexts, such as agile teams or distributed development environments \cite{p1,p2,p42,p46}. While these studies provide important context-specific insights, they typically examine a narrower set of variables and do not systematically capture the broader relationships among psychological safety, team processes, and performance outcomes \cite{p3,p26}. Similarly, existing reviews and surveys address important behavioral and organizational aspects but provide comparatively limited integration of psychological safety with team processes, performance outcomes, contextual factors, barriers, and improvement strategies within software development teams \cite{p4,p6,p15}. In contrast, the present SMS synthesizes these dimensions across multiple software-engineering contexts, providing a broader mapping of the existing evidence and highlighting areas that remain fragmented or underexplored.\vspace{5pt}

Another challenge in the existing literature concerns variation in how psychological safety is conceptualized and measured. Prior studies use different definitions, theoretical perspectives, and measurement approaches \cite{p15,p33}, which can make direct comparison and synthesis across studies difficult. Psychological safety is also examined alongside related constructs such as trust, communication, and team climate \cite{p26,p37}, creating potential conceptual overlap when construct boundaries are not clearly specified. In addition, differences in research design and operationalization further contribute to variation in how psychological safety and its relationships with team outcomes are reported \cite{p45}. The present SMS addresses this fragmented evidence base by systematically mapping how psychological safety is conceptualized across the literature and by organizing the reported relationships, contextual factors, barriers, and improvement practices into a structured thematic framework.\vspace{5pt}

In contrast to the narrower or context-specific perspectives identified in prior studies, the present SMS provides a structured synthesis of multiple dimensions of psychological safety in software engineering. Specifically, it maps how psychological safety is conceptualized, how it is associated with team-performance outcomes, how it relates to team processes such as communication, collaboration, and decision-making, and how contextual and influencing factors are represented in the literature. The study further integrates reported barriers to establishing psychological safety and strategies, practices, and interventions proposed for supporting it. By organizing these dimensions within a unified analytical perspective, the SMS provides a broader view of the existing evidence and helps clarify areas where the literature remains fragmented or comparatively underexplored.\vspace{4pt}
\renewcommand{\arraystretch}{1.12}

\begin{table*}[t]
\centering
\caption{Comparison with Representative Existing Studies}
\label{tab:comparison}

{
\fontsize{7.5}{8.5}\selectfont
\setlength{\tabcolsep}{2.5pt}

\begin{tabularx}{\textwidth}{
>{\raggedright\arraybackslash}p{0.065\textwidth}
>{\raggedright\arraybackslash}p{0.115\textwidth}
>{\raggedright\arraybackslash}X
>{\centering\arraybackslash}p{0.115\textwidth}
>{\centering\arraybackslash}p{0.105\textwidth}
>{\centering\arraybackslash}p{0.135\textwidth}
>{\centering\arraybackslash}p{0.115\textwidth}
}

\toprule

\rowcolor{gray!40}
\textbf{Study} &
\textbf{Context} &
\textbf{Focus} &
\mbox{\textbf{PS as Main Focus}} &
\mbox{\textbf{Team Processes}} &
\mbox{\textbf{Performance Outcomes}} &
\textbf{Method}
\\

\midrule

P3 &
Agile SE &
Human factors + PS &
Partial &
Limited &
Yes &
Survey
\\

P4 &
SE (General) &
Human aspects &
No &
Yes &
\mbox{Not Explicitly Studied} &
SLR
\\

P8 &
SE (General) &
Personality / human aspects &
No &
Yes &
\mbox{Not Explicitly Studied} &
SLR
\\

P15 &
SE (General) &
PS Antecedents + Consequences &
Yes &
Yes &
Yes &
SLR
\\

P26 &
SE (General) &
Psychological Safety (review study) &
Yes &
Yes &
Yes &
SLR
\\

P42 &
\mbox{Large-Scale Agile} &
Psychological Safety in Agile Context &
Yes &
Yes &
Yes &
Interviews
\\

P46 &
Virtual Teams &
Trust, Knowledge Sharing + PS &
Partial &
Yes &
Yes &
Survey
\\

P102 &
SE (General) &
Social + Psychological Factors &
No &
Yes &
Yes &
\mbox{Survey + Interviews}
\\

P108 &
Agile SE &
Psychological Safety + Quality &
Yes &
Yes &
Yes &
\mbox{Survey + Interviews}
\\
\rowcolor{gray!30}
\mbox{\textbf{Our Study}} &
\textbf{SE Domain} &
\textbf{PS + Team Performance (Integrated)} &
\textbf{Yes} &
\textbf{Integrated} &
\textbf{Comprehensive} &
\textbf{SMS}
\\

\bottomrule

\end{tabularx}

}
\end{table*}

Overall, the comparison indicates that existing studies provide important evidence on psychological safety and related human factors in software engineering, but differ in their scope, context, and dimensions examined. The present SMS extends this body of work by integrating conceptualizations of psychological safety, team-performance outcomes, team processes, contextual and influencing factors, reported barriers, and improvement strategies within a single mapping framework. Rather than replacing context-specific or focused studies, this broader synthesis complements them by identifying recurring patterns, areas of convergence, and dimensions that remain comparatively underexplored across the software engineering literature.
%--------------Gap--------------
\section{Research Gaps}
The systematic mapping of the literature reveals several research gaps concerning psychological safety and related human factors in software development teams. Although the existing literature provides valuable insights into team processes, performance outcomes, contextual factors, barriers, and improvement strategies, the mapping identifies imbalances in research attention, underexplored areas, and methodological limitations that warrant further investigation. These gaps are organized into conceptual, contextual, and methodological dimensions.
\subsection{Conceptual Gaps}
\textit{Limited Attention to Decision-Making Compared with Communication and Collaboration:}
The findings across RQ1 and RQ3 indicate comparatively greater research attention to communication and collaboration in relation to psychological safety, while decision-making processes receive less attention. Although some studies address inclusive and participative decision-making, the mechanisms linking psychological safety with decision quality, conflict resolution, and strategic decision outcomes remain comparatively underexplored. Future research should therefore examine decision-making more explicitly, including how psychologically safe conditions are associated with participation, disagreement, conflict management, and decision processes within software development teams.\vspace{5pt}

\begin{comment}

\subsubsection{Limited Exploration of Negative and Boundary Effects of Psychological Safety}
Although RQ3 shows that psychological safety is predominantly associated with positive outcomes, relatively few studies examine its potential negative or unintended effects. Issues such as overconfidence, reduced constructive conflict, or complacency in highly safe environments are not sufficiently addressed. This highlights a conceptual gap in understanding the boundary conditions under which psychological safety may have diminishing or adverse effects.\vspace{5pt}
\end{comment}

\textit{Insufficient Integration of Human Factors and Psychological Safety:}
The literature often examines individual-level human factors, such as personality and individual characteristics, alongside psychological safety without fully integrating these dimensions conceptually. Findings from RQ1 and RQ4 indicate connections between individual characteristics, team-level psychological safety, and team functioning; however, limited theoretical integration explains how individual-level differences interact with shared perceptions of psychological safety in relation to team performance. Future research should therefore develop and evaluate multi-level conceptual frameworks that integrate individual human factors with team-level psychological safety and examine how these relationships vary across software-engineering contexts.

\subsection{Contextual Gaps}
\textit{Under representation of Diversity, Inclusion, and Power Dynamics:}
Findings from RQ4 and RQ5 indicate that diversity-, inclusion-, gender-, and power-related conditions receive comparatively limited research attention relative to leadership, communication, trust, and other organizational factors. Existing studies identify issues related to cultural and regional differences, gender, bias, exclusion, and power imbalances, but the evidence remains limited regarding how these conditions are associated with psychological safety across different software-team contexts. This gap is particularly relevant for diverse, global, and distributed software teams, where team members may differ in cultural backgrounds, communication practices, and participation conditions. Future research should therefore examine more systematically how diversity, inclusion, bias, and power relations interact with psychological safety and equitable participation in software development teams.\vspace{5pt}

\textit{Limited Focus on Workload, Stress, and Environmental Constraints:}
Findings from RQ4 and RQ5 indicate that diversity-, inclusion-, gender-, and power-related conditions receive comparatively limited research attention relative to leadership, communication, trust, and other organizational factors. Existing studies identify issues related to cultural and regional differences, gender, bias, exclusion, and power imbalances, but the evidence remains limited regarding how these conditions are associated with psychological safety across different software-team contexts. This gap is particularly relevant for diverse, global, and distributed software teams, where team members may differ in cultural backgrounds, communication practices, and participation conditions. Future research should therefore examine more systematically how diversity, inclusion, bias, and power relations interact with psychological safety and equitable participation in software development teams.\vspace{5pt}

\textit{Insufficient Attention to Virtual, Distributed, and Hybrid Teams:}
The findings indicate that virtual and distributed working conditions appear across several dimensions of psychological safety, particularly in relation to communication, coordination, technology-mediated interaction, and contextual differences. However, these conditions are often examined as part of broader team or organizational factors rather than as a sustained focus of psychological safety research. Consequently, there remains limited understanding of how psychological safety develops and is maintained when software-team interaction occurs across geographical, cultural, and technological boundaries. Future research should therefore examine psychological safety more explicitly in virtual, distributed, and hybrid software teams, including how communication practices, collaboration technologies, team relationships, and working arrangements are associated with psychologically safe participation over time.\vspace{5pt}

\textit{Limited Exploration of Emerging Technologies (e.g. AI-Supported Collaboration):}
Findings from RQ6 indicate that AI-supported interaction and other technology-related practices receive comparatively limited attention within the literature on psychological safety in software teams. Although emerging evidence considers the use of AI-supported tools in relation to team interaction, the current evidence base remains insufficient to establish how such technologies are associated with psychological safety across different software-development contexts. Important questions remain regarding how AI-supported collaboration may affect communication, participation, trust, team relationships, and willingness to speak up. Future research should therefore examine both the opportunities and challenges associated with AI-supported collaboration and investigate how the design and use of emerging technologies relate to psychological safety in software development teams.

\subsection{Methodological Gaps}
\textit{Dominance of Qualitative and Conceptual Studies:}
The mapping indicates that a substantial part of the existing literature relies on qualitative methods, conceptual discussions, and case-based research. While these approaches provide valuable contextual and explanatory insights, comparatively limited evidence is available from large-scale empirical studies that systematically examine relationships between psychological safety and team-performance outcomes across different software-engineering settings. This methodological imbalance makes it difficult to compare findings across larger populations and organizational contexts. Future research should therefore complement existing qualitative evidence with larger-scale empirical and mixed-method studies to strengthen the breadth and comparability of the evidence base.\vspace{5pt}

\textit{Limited Longitudinal and Causal Research Designs:}
The mapped literature contains comparatively limited longitudinal evidence examining how psychological safety develops and changes throughout project and team life cycles. Cross-sectional research provides valuable evidence on associations but offers limited ability to establish temporal ordering or determine the direction of relationships between psychological safety, team processes, contextual factors, and performance outcomes. Future research should therefore employ longitudinal and, where appropriate, stronger causal research designs to examine how psychological safety develops, fluctuates, and is associated with changes in team functioning and performance over time. Such studies could also help clarify whether relationships between psychological safety and factors such as communication, trust, and team performance are unidirectional or reciprocal.\vspace{5pt}

\textit{Lack of Standardized Measurement and Evaluation Frameworks:}
The mapped literature shows variation in how psychological safety and related constructs are defined, operationalized, and measured. Differences in definitions, metrics, measurement instruments, and evaluation approaches make direct comparison and synthesis across studies more difficult. Future research should therefore place greater emphasis on the use and validation of clearly defined and comparable measurement approaches, while also examining whether existing instruments require adaptation for specific software-engineering contexts. Greater consistency in construct definitions, measurement procedures, and reporting would strengthen comparability and support the accumulation of evidence across studies.\vspace{5pt}

Overall, the identified research gaps show that psychological safety has received substantial attention in relation to software-team performance and team processes, but the existing evidence remains uneven across research dimensions and contexts. The literature gives comparatively greater attention to communication, collaboration, and positive team outcomes, while decision-making, individual-level human factors, diversity and inclusion, power dynamics, workload and well-being, and virtual or distributed work environments remain comparatively less explored. Emerging areas such as AI-supported collaboration also have a limited evidence base, indicating opportunities for further investigation.\vspace{5pt}

Methodologically, the evidence base would benefit from greater methodological diversity, larger-scale empirical studies, longitudinal and stronger causal designs, and greater consistency in measurement and evaluation approaches. Addressing these gaps requires a more balanced and context-sensitive research agenda that integrates conceptual development, diverse software-engineering contexts, and rigorous empirical evaluation. Such research would strengthen cumulative understanding of psychological safety and clarify how it is associated with team processes and performance across different software-development environments.

%--------------Validity--------------
\section{Threats to Validity}
This systematic mapping study is subject to several threats that may affect the reliability and validity of its findings, as summarized in Table ~\ref{tab:validity}. To mitigate these threats, the study followed established guidelines for conducting systematic studies and thematic analysis in software engineering research \cite{p110}, \cite{p111}. The study design, including the search strategy, study selection process, data extraction, and classification procedures, was systematically defined to improve consistency and transparency. Following established validity classifications \cite{p112}, the identified threats are discussed in terms of internal validity, external validity, construct validity, and conclusion validity.

\subsection{Internal Validity}
Internal validity refers to potential issues that may affect the correctness and consistency of the data analysis performed on the selected studies. In this SMS, such threats may occur at four following stages of the study process.\vspace{5pt}

\textbf{\textit{Stage 1:}}A potential threat is the omission of relevant studies during the search process. To reduce this risk, primary database searches were complemented by snowballing techniques \cite{p112}, following established systematic-review guidelines \cite{p110}, as described in Table~\ref{string} and Figure~\ref{fig:2}. Pilot searches were also conducted to refine the search strings before the final search was executed. These measures were intended to increase search coverage and reduce the likelihood of missing relevant studies.\vspace{5pt}

\textbf{\textit{Stage 2:}} A potential threat arises from subjective judgments during the application of the inclusion and exclusion criteria, which may lead to inconsistent study-selection decisions. To mitigate this threat, predefined inclusion and exclusion criteria were applied throughout the screening process. A subset of studies was independently assessed by two reviewers at different screening stages, and inter-rater agreement was evaluated using Cohen’s Kappa \cite{p114}. Disagreements were discussed and resolved through consensus. These procedures were used to improve consistency in study-selection decisions and reduce the potential influence of individual reviewer judgment.\vspace{5pt}

\textbf{\textit{Stage 3:}} 
Bias or inconsistency during data extraction represents a potential threat to the reliability of the mapped findings. To reduce this risk, a structured data extraction form (see Table~\ref{tab:data_extraction}) was developed and consistently applied across the selected studies. The initial data extraction was conducted by the first and fourth authors, while ambiguities were discussed collaboratively among the authors. Following the recommendations in \cite{p111}, a subset of the extracted data was independently verified by the second and third authors to assess the accuracy and consistency of the extraction process. These procedures were used to reduce extraction errors and inconsistencies across the selected studies.\vspace{5pt}

\textbf{\textit{Stage 4:}} 
Subjective interpretation during coding and thematic classification may lead to inconsistent categorization of extracted data and primary studies. To mitigate this threat, the thematic analysis guidelines proposed in \cite{p115} were followed, and a systematic thematic analysis process was applied as described in the Data Extraction and Analysis subsection of the Research Methodology. Codes, subthemes, and themes were iteratively reviewed and refined to identify overlaps and inconsistencies, while the authors collaboratively participated in the refinement and naming of themes. These procedures were used to improve the consistency and traceability of the thematic classification.\vspace{5pt}

\textbf{\textit{Stage 5:}}Potential bias may arise when synthesizing findings across studies and interpreting recurring patterns in the extracted data. To reduce this threat, the qualitative findings were synthesized through the systematic thematic analysis process described in the Research Methodology, including iterative comparison, grouping, and refinement of codes, subthemes, and themes. The qualitative synthesis was complemented by descriptive quantitative information, such as study frequencies and distributions, where appropriate. These procedures were used to maintain consistency between the extracted evidence, thematic interpretations, and the reported findings.\vspace{5pt}

\textbf{\textit{Stage 6:}}
Subjective judgment in assigning quality-assessment scores may introduce inconsistency across the selected studies. To reduce this threat, predefined quality-assessment criteria and a consistent scoring scheme were applied to all included studies, as described in the “Quality Assessment of Selected Studies” subsection of the Research Methodology. The study- and criterion-level quality scores are also provided in the replication package to improve transparency and traceability. However, detailed per-study rationales for individual scores are not provided, which may limit the complete reproducibility of some quality-assessment decisions.

\subsection{External Validity}
External validity concerns the extent to which the findings may be applicable beyond the scope of the reviewed literature. This SMS specifically focuses on psychological safety and team performance in software development teams; therefore, the transferability of its findings to other industries and non-software teams may be limited. In addition, the review includes only peer-reviewed academic literature, which may exclude relevant evidence reported in industry and unpublished studies. The restriction to peer-reviewed literature was intended to maintain a scholarly evidence base \cite{p110}, while the quality of the included studies was further considered through the quality-assessment process described in the Research Methodology. Future research could complement the academic evidence with relevant industry-based studies and examine whether the identified patterns are also observed in other organizational and professional contexts.

%------------------Table 7--------------
\renewcommand{\arraystretch}{1.3}

\begin{table*}[t]
\centering
\caption{Threats to Validity and Mitigation Strategies}
\label{tab:validity}

\footnotesize
\setlength{\tabcolsep}{5pt}

\begin{tabular}{l l l}
\toprule

\rowcolor{gray!40}
\textbf{Validity Type} & \textbf{Threat} & \textbf{Mitigation Strategy} \\

\midrule

\multirow{4}{*}{Internal Validity} 
& Missing relevant studies & Multiple databases, pilot searches, and snowballing \\
& Study selection bias & Predefined inclusion/exclusion criteria and inter-rater agreement \\
& Data extraction bias & Structured extraction form and independent verification of a subset \\
& Theme classification bias & Systematic thematic analysis and collaborative theme refinement \\
& Data synthesis and interpretation bias & Iterative comparison and systematic synthesis across studies \\
& Quality-assessment bias & Predefined criteria, consistent scoring scheme, and documented scores \\

\midrule

\multirow{2}{*}{External Validity} 
& Limited to software teams & Future research in other domains \\
& Only academic studies & Inclusion of industry studies in future work \\

\midrule

Construct Validity 
& Search term limitations & Refined search strings and pilot testing \\

\midrule

Conclusion Validity 
& Interpretation bias & Cross-study comparison and systematic analysis \\

\bottomrule
\end{tabular}

\end{table*}
%----------------------end table 7-------------------

\subsection{Construct Validity}
Construct validity concerns whether the search and classification procedures adequately capture the concepts investigated in this SMS. A potential threat arises from variation in the terminology used to describe psychological safety and related human and team factors, which may result in relevant studies being missed or conceptually related studies being retrieved. To reduce this threat, the search strategy incorporated multiple relevant keywords and alternative terms and was refined through pilot searches following the recommendations in \cite{p111}. During study selection and thematic analysis, psychological safety was distinguished from related constructs such as trust, team climate, and communication-related factors rather than treating these concepts as equivalent. Nevertheless, differences in definitions and operationalizations across the primary studies may still introduce variation in how psychological safety is represented within the mapped evidence.

\subsection{Conclusion Validity}
Conclusion validity concerns whether the conclusions drawn from the synthesized evidence are adequately supported by the mapped studies. A potential threat arises from researcher interpretation during thematic analysis and synthesis. To reduce this threat, the findings were derived through systematic coding, theme development, and comparison of evidence across the selected studies \cite{p110}. Conclusions were based on recurring patterns in the mapped evidence rather than on individual studies alone. Nevertheless, the number of studies contributing to individual themes and subthemes varies, and thematic frequency should not be interpreted as evidence of causal strength or effect size. Therefore, the conclusions were formulated cautiously to reflect associations, recurring patterns, and reported practices rather than causal relationships.

%------------------Conclusion------------

\section{Conclusion}
This study presented a systematic mapping study (SMS) of psychological safety, related human factors, and team performance in software development teams. The study examined how psychological safety is conceptualized in software engineering, how it is associated with team-performance outcomes and team processes, which contextual and influencing factors are reported, what barriers affect its establishment and maintenance, and what strategies, practices, and interventions are reported for supporting it. A total of 112 primary studies published between 2006 and 06/2026 were selected and analyzed using descriptive and thematic analysis to address the six research questions.\vspace{5pt}

The \textit{demographic analysis} shows that research addressing psychological safety and related human factors in software engineering spans the period from 2006 to 06/2026, with publication activity increasing in recent years. This trend indicates growing research attention to the role of psychological safety and human factors in software development teams. The selected studies are distributed across multiple major publication sources, including IEEE, ACM, Springer, ScienceDirect, and Wiley, demonstrating that the topic has been investigated across a range of software-engineering research outlets. Overall, the publication trends suggest increasing scholarly interest in psychological safety, particularly in relation to contemporary software-development and team-based working environments.\vspace{5pt}

The \textit{key findings} show that psychological safety is conceptualized in software engineering through multiple interpersonal, social, communication, collaboration, learning, team-dynamic, and individual well-being dimensions. Regarding team performance, psychological safety is predominantly associated with productivity, software quality, innovation, and team satisfaction. It is also closely related to key team processes, particularly collaboration, communication, and decision-making. The findings further indicate that the relationship between psychological safety and team performance is context-dependent, with leadership, organizational culture, work environment, trust, and sociocultural and individual conditions represented as important contextual and influencing factors. At the same time, communication problems, cultural and trust-related conditions, structural and organizational constraints, leadership issues, diversity and inclusion challenges, and workload and resource pressures are reported as barriers to establishing and maintaining psychological safety. Finally, the literature reports a broad range of leadership, organizational, agile and process-related, learning and feedback, relational, well-being, and technology-related practices for supporting psychological safety, although evidence regarding their effectiveness and long-term sustainability remains comparatively limited.\vspace{5pt}

This \textit{study contributes} to software-engineering research by providing a systematic mapping of psychological safety, related human factors, and their associations with team performance. First, it organizes the existing evidence across six interconnected dimensions: conceptualizations of psychological safety, team-performance outcomes, team processes, contextual and influencing factors, barriers, and reported strategies and practices. Second, the thematic synthesis provides a structured view of recurring themes and subthemes across these dimensions, helping to consolidate evidence that is otherwise distributed across different software-engineering contexts and research perspectives. Third, the study identifies conceptual, contextual, and methodological gaps in the existing literature, including comparatively limited attention to decision-making, diversity and inclusion, workload and well-being, distributed and hybrid teams, emerging technologies, longitudinal research, and consistent measurement approaches. Together, these contributions provide a structured evidence base that can support future empirical research and the development and evaluation of context-sensitive approaches to psychological safety in software development teams.\vspace{5pt}

The findings provide several \textit{practical implications} for software organizations, managers, and team leaders seeking to support psychological safety. Organizations should consider creating team environments that encourage open communication, constructive feedback, participation, knowledge sharing, and respectful interpersonal interaction. Leadership and management practices should support inclusive participation, clear roles, trust, and opportunities for team members to raise concerns or express different viewpoints without fear of negative interpersonal consequences. Attention should also be given to organizational and structural barriers, including communication problems, excessive workload, resource constraints, exclusionary practices, and unclear responsibilities that may undermine psychological safety. Agile and process-related practices, continuous learning and feedback, relationship-building, and well-being support may also be considered as part of a broader approach rather than as isolated interventions. However, because the literature provides limited evidence regarding the comparative effectiveness and long-term sustainability of many reported practices, organizations should adapt and evaluate these approaches according to their specific team and organizational contexts.\vspace{5pt}

Overall, this systematic mapping study shows that psychological safety is a multidimensional and context-dependent phenomenon that is closely associated with team performance and team processes in software engineering. The mapped evidence highlights the importance of communication, collaboration, leadership, trust, organizational conditions, and individual and team-related factors, while also identifying multiple barriers and a broad range of reported practices for supporting psychological safety. However, the evidence remains uneven across research dimensions and contexts, particularly regarding decision-making, diversity and inclusion, workload and well-being, distributed and hybrid teams, and emerging technologies. Future research should therefore employ larger-scale empirical, longitudinal, and context-sensitive studies to examine how psychological safety develops and is associated with team functioning over time. Greater attention is also needed to consistent measurement approaches and to the systematic evaluation of reported strategies and interventions. Such research can strengthen the evidence base and provide a more robust understanding of psychological safety within contemporary software development teams.

\nocite{p1,p2,p3,p4,p5,p6,p7,p8,p9,p10,p11,p12,p13,p14,p15,p16,p17,p18,p19,p20,p21,p22,p23,p24,p25,p26,p27,p28,p29,p30,p31,p32,p33,p34,p35,p36,p37,p38,p39,p40,p41,p42,p43,p44,p45,p46,p47,p48,p49,p50,p51,p52,p53,p54,p55,p56,p57,p58,p59,p60,p61,p62,p63,p64,p65,p66,p67,p68,p69,p70,p71,p72,p73,p74,p75,p76,p77,p78,p79,p80,p81,p82,p83,p84,p85,p86,p87,p88,p89,p90,p91,p92,p93,p94,p95,p96,p97,p98,p99,p100,p101,p102,p103,p104,p105,p106,p107,p108,p109}

\section*{Statements and Declarations}

\textbf{Ethical Approval:}
Not applicable. This study is a systematic mapping study based exclusively on previously published literature and does not involve human participants or animals.\vspace{5pt}

\textbf{Competing Interests:}
The authors declare that they have no known competing financial interests or personal relationships that could have influenced the work reported in this paper.\vspace{5pt}

\textbf{Funding:}
The authors declare that no specific funding was received for this research.\vspace{5pt}

\textbf{Availability of Data and Materials:}
The dataset and supporting materials used in this study are publicly available through the replication package provided at the end of the Methodology section.\vspace{5pt}

\textbf{Authors' Contributions:}
Adnan Manzoor and Muhammad Azeem Akbar carried out the primary research activities, including conceptualization, methodology development, systematic search, study selection, data extraction, data analysis, visualization, and preparation of the original manuscript. Aakash Ahmad, Jussi Kasurinen, and Arif Ali Khan reviewed and contributed to the refinement of the methodology
and results. Muhammad Hamza, together with the other co-authors, contributed to the review and refinement of the remaining sections of the manuscript. All authors reviewed and approved the final manuscript.\vspace{5pt}

\textbf{Use of Generative AI:}
During the preparation of this manuscript, the authors used generative AI tools for language and grammar improvement and for assistance in creating graphical illustrations. The authors reviewed and verified all AI-assisted content and take full responsibility for the final content of the manuscript.\vspace{5pt}

\textbf{Availability of Data and Materials:}
The complete dataset used in this study, including the selected primary
studies, extracted data, initial coding, open codes, themes, and subthemes,
is available in the accompanying replication package at
Zenodo~\cite{dataset2026}.

\bibliographystyle{IEEEtran}
\bibliography{references}

\appendices
\section{Inter-Rater Agreement (IRA)}
\label{app:ira}
To ensure the reliability and consistency of the study selection process, inter-rater agreement was measured using Cohen’s Kappa statistic~\cite{p123}. Two reviewers independently applied the inclusion and exclusion criteria during the relevance screening, title screening, abstract screening, and full-text screening stages. A random sample of approximately 10\% of the studies at each stage was selected and independently reviewed by both reviewers.

Cohen’s Kappa ($\kappa$) was calculated to assess the level of agreement between the reviewers beyond chance, as defined in Equation~\ref{eq:kappa}. The Kappa value was computed using observed agreement ($P_o$) and expected agreement ($P_e$). The resulting values were interpreted using standard agreement level guidelines~\cite{p124}. Any disagreements between the reviewers were resolved through discussion until a final consensus was reached.

\begin{equation}
\kappa = \frac{P_o - P_e}{1 - P_e}
\label{eq:kappa}
\end{equation}

Where:
\begin{equation}
P_o = \frac{a + d}{n}
\end{equation}

\begin{equation}
P_e = \left(\frac{a + b}{n} \cdot \frac{a + c}{n}\right) + \left(\frac{c + d}{n} \cdot \frac{b + d}{n}\right)
\end{equation}

Where:
\begin{itemize}
\item $a$ = number of studies both reviewers agreed to include
\item $d$ = number of studies both reviewers agreed to exclude
\item $b$ and $c$ = number of disagreements between reviewers
\item $n$ = total number of studies reviewed
\end{itemize}

The inter-rater agreement results obtained from Cohen's Kappa analysis indicate moderate to substantial agreement across the different screening stages, as presented in Table~\ref{tab:kappa}. These results suggest that the inclusion and exclusion criteria were applied with 
reasonable consistency throughout the study selection process. Any disagreements that arose during the screening stages were discussed and resolved through mutual consensus. This process helped reduce potential selection bias and enhanced the reliability of the study selection procedure.

Any disagreements that arose during the screening stages were discussed and resolved through mutual consensus between the reviewers. This process helped reduce potential selection bias and enhanced the reliability and validity of the study selection procedure.
%---------------------GQM----------------------------------------
\section{\textbf{Goal--Question--Metric (GQM) Approach}\hspace{0pt}}
This study adopts the Goal–Question–Metric (GQM) followed the approach of Basili et al~\cite{p119} to systematically define the research goal, derive research questions, and identify the data required to address them. The GQM paradigm ensures that the research process is explicitly driven by the study objectives, thereby enhancing methodological rigor and analytical coherence.\vspace{0.5cm}

\textbf{Goal:}
The goal of this study is to investigate the role of psychological safety and its influence on team performance in software development teams. It aims to provide a structured overview of existing research on psychological safety, including its relationship with team performance, team processes, moderating factors, challenges, and strategies in software development environments.

\textbf{Research Questions:} To achieve this goal, the study addresses a set of six research questions see (Table~\ref{tab:rq}), covering key dimensions of psychological safety in software engineering.

\textbf{Metrics:}
To answer these research questions, specific data items were extracted from the selected studies. Each research question is linked to specific data extraction items (see Table~\ref{tab:data_extraction}).

The GQM framework establishes a clear traceability link between the research goal, research questions, data extraction, and analysis. This alignment supports construct validity, reduces irrelevant data collection, and enhances transparency and replicability. It also facilitates systematic synthesis by enabling consistent mapping of evidence across studies, supporting clear and structured presentation of the findings.\vspace{0.5cm}

\end{document}